# Impact of Space Weather on Climate and Habitability of Terrestrial Type Exoplanets

*Airapetian, V. S.[1,2], Barnes, R.[3], Cohen, O.[4], Collinson, G. A.[1], Danchi, W. C.[1], Dong, C. F.[5], Del Genio, A. D.[6], France, K.[7], Garcia-Sage, K.[1], Glocer, A.[1], Gopalswamy, N.[1], Grenfell, J. L.[8], Gronoff, G.[9], Güdel, M.[10], Herbst, K.[11], Henning, W. G.[1], Jackman, C. H.[1], Jin, M.[12], Johnstone, C. P.[10], Kaltenegger, L., Kay, C. D.,[11], Kobayashi, K.[14], Kuang, W.[1], Li, G.[15], Lynch, B. J.[16], Lüftinger, T.[10], Luhmann, J. G.[16], Maehara, H.[17], Mlynczak, M. G.[9], Notsu, Y.[18], Ramirez, R., M.[19], Rugheimer, S.[20], Scheucher, M.[21], Schlieder, J. E.[1], Shibata, K.[22], Sousa-Silva, C.[23], Stamenković, V.[24], Strangeway, R. J.[25], Usmanov, A. V.[1,26], Vergados, P.[24], Verkhoglyadova, O. P.[24], Vidotto, A. A.[27], Voytek, M.[28], Way, M. J.[6], Zank, G. P.[15], Yamashiki, Y[29].*

[1]Sellers Exoplanet Environments Collaboration, NASA/GSFC, Greenbelt, MD, USA; [2]Department of Physics, American University, Washington DC, USA; [3]University of Washington, Seattle, Washington, USA; [4]Lowell Center for Space Science and Technology, University of Massachusetts, Lowell, MA, USA; [5]Department of Astrophysical Sciences, Princeton University, Princeton, NJ, USA; [6]NASA Goddard Institute for Space Studies, New York, NY, USA; [7]Laboratory for Atmospheric and Space Physics, University of Colorado, 600 UCB, Boulder, CO 80309, USA; [8]Department of Extrasolar Planets and Atmospheres (EPA), Institute for Planetary Research, German Aerospace Centre (DLR), Rutherfordstr. 2, 12489 Berlin Adlershof, Germany; [9]NASA/LaRC, Hampton, VA, USA; [10]University of Vienna, Dept. of Astrophysics, Türkenschanzstr. 17, 1180, Vienna, Austria; [11]Christian-Albrechts-Universität zu Kiel, Institute for Experimental and Applied Physics, Leibnizstr. 11, 24118 Kiel, Germany; [12]SETI Institute, Mountain View, CA 94043, USA; [13]Carl Sagan Institute, Cornell University, Ithaca, NY, USA; [14]Department of Chemistry, Yokohama National University, Yokohama, Japan; [15]Center for Space Plasma and Aeronomic Research (CSPAR), University of Alabama in Huntsville, Huntsville, AL 35899, USA; [16]Space Sciences Laboratory, University of California at Berkeley, Berkeley, CA, USA; [17]Subaru Telescope Okayama Branch Office, NAOJ, Asakuchi, Okayama 719-02, Japan; [18]Department of Astronomy, Kyoto University, Kitashirakawa-Oiwake-cho, Sakyo-ku, Kyoto, Japan; [19]Earth-Life Science Institute, Tokyo Institute of Technology, 2-12-1, Tokyo 152-8550, Japan; [20]Centre for Exoplanet Science, University of St. Andrews, School of Earth and Environmental Sciences, Irvine Building, North Street, St. Andrews, KY16 9AL, UK; [21]Zentrum für Astronomie und Astrophysik, Technische Universität Berlin, D-10623 Berlin, Germany; [22]Astronomical Observatory, Kyoto University, Sakyo, Kyoto 606-8502, Japan; [23]Massachusetts Institute of Technology, Dept. of Earth, Atmospheric, and Planetary Sciences, Cambridge, MA, 02139, USA; [24]NASA Jet Propulsion Laboratory, California Institute of Technology, Pasadena, CA, USA; [25]University of California, Los Angeles, CA USA; [26]University of Delaware, DE, 19716, USA; [27]Trinity College Dublin, Dublin, Ireland; [28]NASA Headquarters, Washington, DC, USA;[29]Graduate School of Advanced Integrated Studies in Human Survivability (GSAIS), Kyoto University, Kyoto, Japan.

*Corresponding author: NASA/GSFC/SEEC, Greenbelt, MD, 20771 and American University, DC; phone: 1-301-286-4014; fax:1-301-286-1752; vladimir.airapetian@nasa.gov

**Abstract**

The search for life in the Universe is a fundamental problem of astrobiology and modern science. The current progress in the detection of terrestrial type exoplanets has opened a new avenue in the characterization of exoplanetary atmospheres and in the search for biosignatures of life with the upcoming ground-based and space missions. To specify the conditions favorable for the origin, development and sustainment of life as we know it in other worlds, we need to understand the nature of global (astrospheric), and local (atmospheric and surface) environments of exoplanets in habitable zones around G-K-M dwarf stars including our young Sun. Global environment is formed by propagated disturbances from the planet-hosting stars in the form of stellar flares, coronal mass ejections, energetic particles, and winds collectively known as astrospheric space weather. Its characterization will help in understanding how an exoplanetary ecosystem interacts with its host star, as well as in the specification of the physical, chemical and biochemical conditions that can create favorable and/or detrimental conditions for planetary climate and habitability along with evolution of planetary internal dynamics over geological timescales. A key linkage of (astro) physical, chemical, and geological processes can only be understood in the framework of interdisciplinary studies with the incorporation of progress in heliophysics, astrophysics, planetary and Earth sciences. The assessment of the impacts of host stars on the climate and habitability of terrestrial (exo)planets will significantly expand the current definition of the habitable zone to the biogenic zone and provide new observational strategies for searching for signatures of life. The major goal of this paper is to describe and discuss the current status and recent progress in this interdisciplinary field in light of presentations and discussions during the NASA Nexus for Exoplanetary System Science (NExSS) funded workshop "Exoplanetary Space Weather, Climate and Habitability" and to provide a new roadmap for the future development of the emerging field of exoplanetary science and astrobiology.

# 1.0 Introduction

The recent explosion of the number of planets detected around other stars (exoplanets, currently over 4000) by space and ground-based missions created a great leap in the progress of the fields of exoplanetary science and astrobiology. These observations provide a boost to scientifically addressing one of the major questions of modern science "Are we alone in the Universe?". Although the answer to this question is unknown, the Kepler Space Telescope's discoveries of terrestrial-type (rocky) exoplanets within Circumstellar Habitable Zones (CHZs, the regions where standing bodies of liquid water can be present on the exoplanetary surface) around main-sequence stars have provided an important step in addressing this difficult question. An ongoing issue relates to how well the classical CHZ as calculated by Kasting et al. (1993) and revisited by Kopparapu et al. (2013) relates to the actual physico-chemical conditions required for the origin, development and support of life as we know it within so-called "Biogenic Zones" (Airapetian et al. 2016) or "Abiogenesis Zones" (Rimmer et al. 2018).

With growing numbers of NASA (National Aeronautics and Space Administration) and ESA (European Space Agency) exoplanetary missions such as the Transiting Exoplanet Survey Satellite (TESS), the upcoming James Webb Space Telescope (JWST), Characterizing ExOPlanet Satellite (CHEOPS), the PLAnetary Transits and Oscillations of stars (PLATO), the Atmospheric Remote-sensing Infrared Exoplanet Large-Survey (ARIEL) missions, in the relatively near term we will be better equipped with high quality observations to move from the phase of exoplanetary discovery to that of physical and chemical characterization of exoplanets suitable for life.

It is currently unknown if or when life may have begun on Earth, and possibly Mars, Venus and exoplanets, or how long those planetary bodies could remain viable for life. The principal cause of this is a lack of understanding of the detailed interaction between stars and exoplanets over geological timescales, the dynamical evolution of planetary systems, and atmospheric and internal dynamics. In the last few years, there has been a growing appreciation that the atmospheric chemistry, and even retention of an atmosphere in many cases, depends critically on the high-energy radiation and particle environments around these stars (Segura et al., 2005; Domagal-Goldman et al., 2014; Rugheimer et al., 2015; Airapetian et al. 2017a).

Recent studies have suggested that stellar magnetic activity and its product, astrospheric space weather (SW), the perturbations traveling from stars to planets, in the form of flares, winds, coronal mass ejections (CMEs) and energetic particles from planet hosting stars, may profoundly affect the dynamics, chemistry and exoplanetary climate (Cohen et al. 2014; Airapetian et al. 2016; 2017b; Garcia-Sage et al. 2017; Dong et al. 2018).

The question of impact of stars on exoplanets is complex, and to answer it we must start with the host star itself to determine its effect on the exoplanet environment, all the way from its magnetosphere to its surface. To understand whether an exoplanet is habitable at its surface, not only do we need to understand the changes in the chemistry of its atmosphere due to the penetration of energetic particles and their interaction with constituent molecules, but also the loss of neutral and ionic species, and the addition of molecules due to outgassing from volcanic and tectonic activity. These effects will produce a net gain or loss to the surface pressure and this will affect the surface temperature, as well as a net change in the molecular chemistry. Therefore, due to the complexity of the problem, we should work with a set of interlinked research questions, all of which contribute pieces to the answer, with contributions from various disciplines involved in each topic.

To quantify the effects of stellar ionizing radiation including soft X-ray and Extreme Ultraviolet, EUV (100 – 920 Å) later referred to as XUV fluxes from superflares detected by the Kepler mission on exoplanetary systems, specifically, on close-in exoplanets around low luminosity M dwarf stars, we need to examine what do we know about the impact of space weather from our own star on Venus, Mars and Earth, the only inhabited planet known to us.

How can we apply lessons learned from the extreme space weather events to understand how exoplanet are affected by their host stars? Can heliophysics science with its methodologies and models developed to describe the effects of solar flares, CMEs as the factors of space weather, on Venus, Earth and Mars be expanded to address the extreme conditions on exoplanets around young solar like stars and close-in exoplanets around active F, G, K and M dwarfs? These questions are of critical importance as the major factors of habitability of exoplanets.

In this paper we present the roadmap to study various aspects of star-planet interactions in a global exoplanetary system environment with a systematic, integrated approach using theoretical, observational and laboratory methods combining tools and methodologies of four science disciplines: astrophysics, heliophysics, planetary and Earth science as presented in Figure 1. The components of the presented roadmap had been discussed during the NExSS sponsored Workshop Without Wall "Impact of Exoplanetary Space Weather on Climate and Habitability" and recent white papers submitted to the US National Academy of Sciences for Exoplanet Science Strategy, Astrobiology Science Strategy and Astronomy and Astrophysics (Astro2020) calls (see Airapetian et al. 2018a, b; 2019a and the link at

https://nexss.info/community/woskshops/workshop-without-walls-exoplanetary-scape-weather-climate-and-habitability").

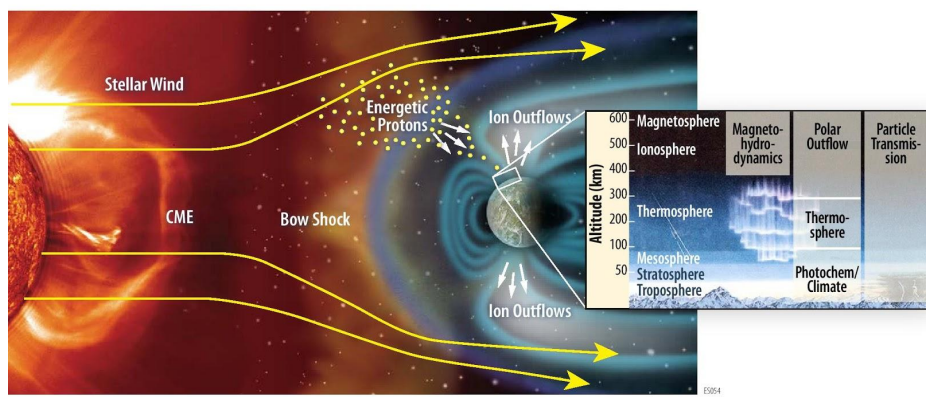

*Figure 1. Schematic view of the complex exoplanetary SW system that incorporates the physical processes driving stellar activity and associated SW including stellar flares, CMEs and their interactions with an exoplanetary atmosphere driven by its internal dynamics. While stellar winds and CMEs affect the shape of an exoplanetary magnetosphere, XUV and energetic particles accelerated on CME-driven shocks enter the atmosphere. The combined effects of XUV, stellar winds and CMEs drive outflows from the exoplanetary atmosphere. These processes are controlling factors of exoplanetary climate and habitability.*

We describe recent progress and challenges in understanding the nature of solar and stellar magnetic activity and associated space weather processes from the modern Sun, the young Sun (at the time when life started on Earth) and from other cool (K through M spectral type) stars. We discuss the physical processes that drive the interaction of space weather with the Earth, Mars and Venus and the implications for exoplanets around active stars including internal dynamics that drive outgassing. This review paper consists of Introduction, 9 sections.

Section 2 discusses the structures of global solar corona and the solar wind, properties of solar flares, CMEs and SEPs and their observational signatures.

In Section 3, we review recent observations and modeling efforts of coronae, winds and superflares from young solar type planet hosting stars.

In Section 4, we present a review of current observations and theoretical models of atmospheric erosion and chemical changes in the atmospheres of modern Earth, Mars and Venus caused by XUV emission from solar flares, CMEs and SEP events.

Section 5 highlights the current understanding of the effects of solar and stellar XUV driven emission and the dynamic pressure exerted by solar and stellar winds, CMEs on atmospheric escape processes from early Earth, Mars, Venus and terrestrial type exoplanets including exoplanets around Proxima Centauri and TRAPPIST 1.

Section 6 discusses the impact of SEP events on atmospheric chemistry of early Earth and SEP driven surface dosages of ionizing radiation of terrestrial type exoplanets.

Section 7 reviews the impact of space weather in the form of XUV fluxes and SEP events on climate-related processes, exoplanetary habitability and the properties of atmospheric biosignatures from terrestrial type exoplanets orbiting K-G-M stars.

Section 8 discusses interior dynamics of terrestrial-type planets as a function of their chemical composition, mass, size and explores the potential of volcanic and tectonic activity which plays a key role for atmospheric evolution. It also discusses current efforts in magnetohydrodynamic modeling of exoplanetary dynamos and their effects on habitability.

In Section 9, we present observational strategies for detection of habitable environments on terrestrial type exoplanets around different main-sequence stars and review progress in recent observations of exoplanets by the Kepler Space Telescope, HST and ground-based telescopes which employ transit spectroscopy, direct imaging, radial velocity and gravitational lensing. We also discuss a roadmap to detect biosignatures using stellar interferometers and the next generation ground based and space telescopes: TESS, CHEOPS, JWST, PLATO 2.0, ARIEL, the E-ELT, LUVOIR, ORIGINS, LYNX and HabEx.

Section 10 discusses future prospects and provides recommendations for the next steps in understanding of various aspects of star-planet interactions, the ways they affect planetary habitability and observational strategies to detect habitable worlds in the coming decade.

**2.0 Drivers and Signatures of Space Weather from The Sun**

Our Sun is a major source of energy for much of life on Earth. Our central star was formed from a collapsing protostellar cloud 4.65 billion years ago. The Solar system formed from the protoplanetary disk left after the Sun's birth was bombarded by a vast amount of energy in the form of electromagnetic radiation, solar wind, magnetic clouds, shock waves and energetic electrons and protons. Recent heliospheric missions including the Solar and Heliospheric Observatory (SOHO), the Solar Dynamic Observatory (SDO) and the Solar Terrestrial Relations Observatory (STEREO) have provided a wealth of information about our magnetic star. This has helped to recover statistical information about the spatial and temporal relationship between eruptive events occurring in the solar corona and provided clues to the physical mechanisms driving their underlying processes.

The primary output from the Sun is in the energy flux in the form of electromagnetic and mass emissions, ultimately powered by the thermonuclear reactions in the solar interior that convert hydrogen to helium and amplified by the magnetic dynamo that generates and transports magnetic fields to surface and the atmosphere. The steady or quiescent electromagnetic emission in the visible and infrared bands support life on Earth, while solar XUV flux creates the ionosphere around Earth and affects its upper atmospheric chemistry.

Electromagnetic emission can also be transient, in the form of solar flares at almost all wavelengths. Solar flares cause transient disturbances in Earth's ionosphere. The mass emission occurs in the form of steady two-component solar wind with speeds in the range 300-900 km/s and as CMEs that have speeds ranging from <100 km/s to >3000 km/s (e.g., Gopalswamy 2016). Fast CMEs (faster than the magnetosonic speed in the corona and interplanetary medium) drive magnetohydrodynamic (MHD) shocks that are responsible for the copious acceleration of electrons, protons, and heavy ions commonly referred to as SEPs. Energetic protons are known to significantly impact Earth's atmosphere. CMEs are magnetized plasmas with a flux rope structure, which when impacting a magnetized plasma can lead to geomagnetic storms that have serious consequences in the planetary magnetosphere/ionosphere/atmosphere. Solar wind magnetic structures can also result in moderate magnetic storms. Thus, SEPs and magnetic storms associated with CMEs are considered to be severe space weather consequences of solar eruptions on the technological infrastructure of the modern world (Schrijver et al. 2015). Flares and CMEs are formed within closed solar magnetic regions (active regions), while the solar wind originates from open field regions. Magnetic regions on the Sun also modulate the amount of visible radiation emitted by the solar plasma, because of a combination of sunspots and the surrounding plages (e.g., Solanki et al. 2013).

**2.1 Origin and Patterns of Solar Magnetic Activity: Global Solar Corona**

The outermost layer of the Sun, the solar corona, is the hottest atmospheric region of the Sun heated to ~ 1-2 MK, which is by a factor of 200 hotter than the solar photosphere. This suggests that the solar corona is heated from the lower atmosphere of the Sun driven by the surface (photospheric) magnetic field supplying and dissipating its energy in the upper atmospheric heating. Two possible mechanisms of heating include upward propagating magnetic waves (in the form of Alfvén waves) generated by photospheric convection and nanoflares, magnetic reconnection driven explosions releasing energy of ~ $10^{20}$ ergs (De Pontieu et al. 2014; Ishikawa et al. 2017). The coronal heating varies across the solar surface forming a diffuse corona (see left panel of Figure 2) and active regions (white concentrated regions above sunspots).

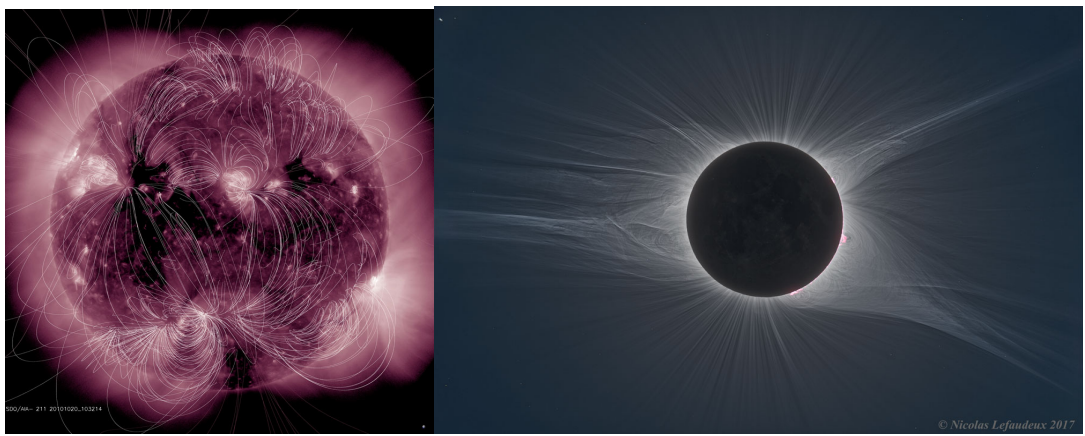

*Figure 2. Left panel: NASA/SDO image of the magnetic Sun in 211Å band with superimposed magnetic field lines interconnecting active regions (NASA/SDO); Right panel: global solar corona highlighted during "Great American" solar eclipse of August 17, 2017 (Copyright: Nicolas Lafaudeux (https://apod.nasa.gov/apod/ap180430.html).*

The left panel of Figure 2 shows the SDO image in *211Å* band (representative T ~ 2.5 MK) (https://sdo.gsfc.nasa.gov/gallery/main/item/37). The superimposed coronal magnetic field lines show that the magnetic field in the solar corona is organized into small (up to a few tens of km as resolved by ground-based solar telescopes) and large scale (comparable to the solar radius) structures connecting active regions, with their complexity varying with the solar cycle. The solar corona represents a magnetically controlled environment, and thus plasma structures observed on the solar active regions trace magnetic field line (bright regions on the left pane of Figure 2). Right panel of Figure 2 shows the solar corona during the solar eclipse of August 17, 2017. Its polar regions show open rays that are solar coronal holes, the regions of the open magnetic flux with lower density and temperature plasma that appear as dark regions in X-ray and EUV images (see the lower hole in the left panel of Figure 2). The open magnetic flux and coronal structures evolve with the phase of the solar cycle (from solar minimum of magnetic activity to solar maximum) running through the polarity flip every 11 years. While global magnetic field during minimum of activity can be relatively well represented by dipole field (close to the global field represented in the left panel of Figure 2), it becomes less organized with the presence of quadrupole and other multipole components (Kramar et a. 2014; 2015).

During solar maximum coronal holes migrate from low latitudes toward the equator with the progression of the solar cycle, while during solar minimum they can be mostly found in the polar regions (McIntosh et al. 2014). Solar coronal streamers near solar maximum are mostly located near the polar regions (see the right panel of Figure 2) of the Sun and can be described with higher magnetic multipole moments associated with coronal active regions in addition to the dipole component of the global magnetic field. During solar minimum, solar streamers are formed near the equator and are not associated with coronal active regions. They can be described by mostly dipolar field component inclined 10 degrees to the solar rotation axis with the heliospheric current concentrated in the heliospheric current sheet above the dipolar magnetic field at ~2.5 $R_\odot$ (Zhao & Hoeksema 1996).

Static solar corona extends into interplanetary space as the supersonic outflow known as the solar wind first predicted by Eugene Parker (1958). The solar wind forms a background for propagation of coronal disturbances including coronal mass ejections (CME) and associated solar energetic particle (SEP) events, and thus constitutes the major component of space weather. Remote sensing and in-situ measurements demonstrate that near solar minimum the solar wind has a bi-modal structure. The fast wind represents a high-speed (up to 800 km/s at 1 AU) low density (1 - 5 $cm^{-3}$) outflow emanating from the lower solar corona around polar regions and is associated with unipolar coronal holes. The slow wind (~350-400 km/s) is a factor of 3-10 denser and forms above the low latitude regions associated with large-scale equatorial structures known as coronal streamer belt with the mass loss rate of 2 x $10^{-14}$ $M_\odot/yr$ (McComas et al. 2007). Driven by the solar rotation, the high and low wind components form alternating streams moving outward into interplanetary space in an Archimedean spiral. At distance around 1 AU or farther away from the Sun, the high-speed streams eventually overtake the slow-speed flows and form regions of enhanced density and magnetic field known as co-rotating interaction regions (CIRs). These compressed interstream regions play an important role in space weather as a trigger of geomagnetic storms.

Thus, it is important to understand the physics of the dynamic solar wind as a major factor of the impact of the associated variable space weather on the Earth's magnetosphere. Historically, the first thermally driven coronal solar wind model was developed by Eugene

Parker (Parker, 1958). It could satisfactory describe the slow solar wind component but failed to explain the fast wind component. In order to explain the fast wind component, an additional momentum term is required (Usmanov et al. 2000; Ofman 2010; Airapetian et al. 2010; Airapetian and Cuntz 2015). Recently developed data driven global MHD models can successfully reproduce the overall global structure of the solar corona, the solar wind and incorporate the physical processes of heating and acceleration of the solar wind with the initial state and boundary conditions directly derived from observations (van der Holst 2014; Oran et al. 2017; Reiss et al. 2019).

**2.2 Origin and Patterns of Solar Magnetic Activity: Transient Events**

Observations and characterization of solar activity in the form of varying numbers of sunspots on the its disk have been performed since 1600s. Wilhelm Herschel (1801) recognized periods of low and high sunspot activity and correlated the low sunspot activity with high wheat prices in England. Decades later, Schwabe (1843) in his pursuit of a hypothetical planet closer to the Sun than Mercury, discovered a 10-year periodicity in the sunspot number; in 1850, the periodicity was confirmed by Rudolf Wolf and refined to be about 11 years. This discovery has long-lasting implications because the impact on planets accordingly waxes and wanes. The discovery of magnetic fields in sunspots by Hale (1908) led to the identification of 22-year magnetic cycle of the Sun (Hale cycle). Sunspots typically appear in pairs with leading and following spots having opposite magnetic polarity in a given hemisphere; the polarity is switched in the other hemisphere. This pattern is maintained over the 11-year sunspot cycle (also known as the Schwabe cycle). In the new cycle, the polarity of the leading and following polarities are switched in both hemispheres. The polarity switching is referred to as the Hale-Nicholson law. Other periodicities are also known, especially the Gleisberg cycle with a periodicity in the range 80-100 years (see e.g., Petrovay 2010). Finally, solar activity has grand minima and maxima over millennial time scales that do not seem to be periodic. The well-known example is the Maunder Minimum discovered by Eddy (1976), when the sunspot activity almost vanished during 1640 – 1715 AD. Sunspots were observed for only about 2% of the days in this interval. Estimates show that the Sun has spent <20% of the time in grand minima and <15% of the time in grand maxima in the Holocene (see Usoskin 2017 for a review).

The magnetic field associated with sunspots is known as the toroidal component of the solar magnetic field. The other component is the poloidal magnetic field, which peaks during the minimum phase of a solar cycle. The polarity of the poloidal field reverses during the maximum phase of the solar cycle (Babcock and Babcock, 1955; Babcock 1959). Figure 3 shows the toroidal (low latitude) and poloidal (high latitude) components using longitudinally averaged photospheric magnetic field and the microwave brightness temperature, which is a proxy to the magnetic field strength. The data correspond to part of solar cycle 22 (before the year 1996), whole of cycle 23 (1996 - 2008) and most of cycle 24 (after the year 2008). Note that the toroidal field starts building up when the poloidal field starts declining and vice versa. The poloidal field development is clearly connected to the evolution of the sunspot fields (local peaks at low latitudes) via plumes, which consist of the eastern parts of sunspot regions moving toward the poles as a consequence of the Joy's law.

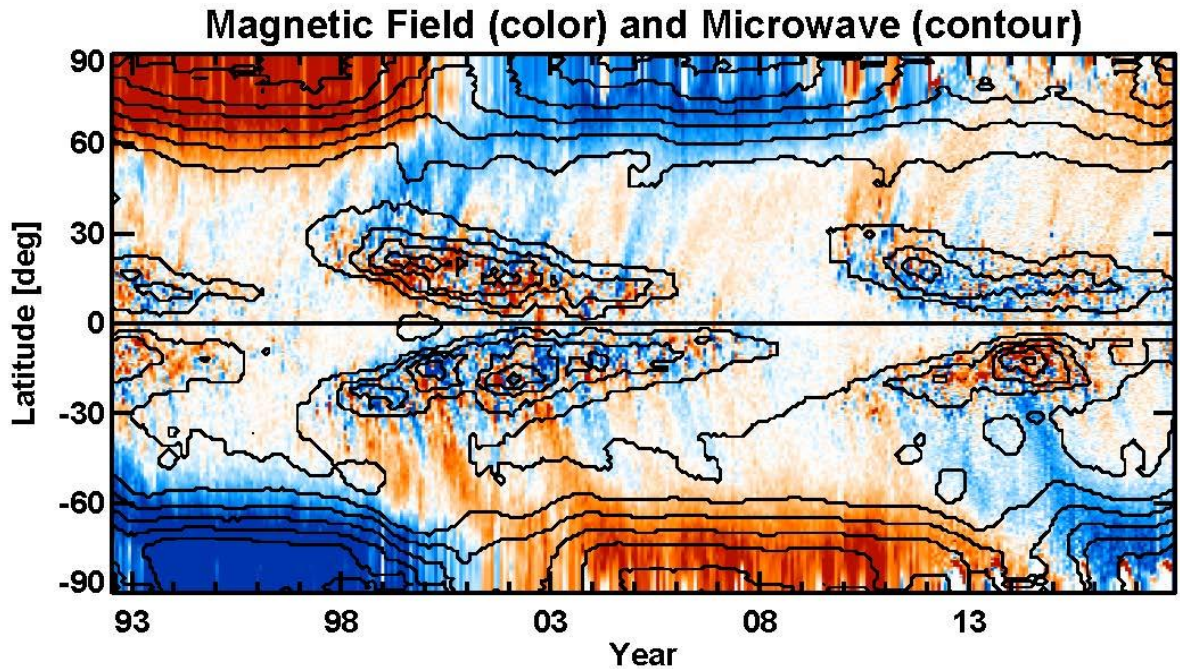

*Figure 3. The toroidal and poloidal components of the solar magnetic field (blue- negative; red - positive). The contours represent microwave brightness temperature at 17 GHz obtained from the Nobeyama Radioheliograph (contour levels: 9400, 9700, 10000, 10300, 10600, 10900, 11200, 11500, 11800, 12100 K). The field distribution between +/- 30 degree latitude represents the toroidal field, while that poleward of ~60 degree latitudes represents the poloidal field. Data updated from Gopalswamy et al. (2016).*

The sunspot cycle can thus be explained by a magnetic dynamo model in which the poloidal and toroidal fields mutually generate each other (see e. g., Charboneau 2010 for a review) in the presence of differential rotation, convective motion beneath the surface, and meridional circulation. Filaments (also known as prominences when appearing at the solar limb) are another important phenomena occurring generally in the mid latitudes but move toward the poles during the maximum phase. Filaments usually mark the polarity inversion lines in magnetic regions either in sunspot regions or in bipolar magnetic regions without sunspots. The disappearance of the polar crown filaments roughly marks the time of polarity reversal at the poles.

The extension of sunspot regions into the outermost region of the Sun, solar corona is known as an active region. Some active regions can be without sunspots (the quiescent filament regions). In X-rays and Extreme UV (EUV), active regions appear as a collection of loops, which are thought to be magnetic in nature. Solar eruptions, a collective term representing energy release in the form of flares and coronal mass ejections (CMEs), occur in active regions. The energy release happens in the form of heating, particle acceleration, and mass motion. Accelerated particles from the corona travel along magnetic field line toward the Sun produce various signatures of flares from radio to gamma-ray wavelengths. Non-thermal particles precipitating into the photosphere cause enhanced optical emission, which was originally recognized as solar flares by Carrington (1859) and Hodgson (1859). Chromospheric signatures of solar flares are readily observed in H-$\alpha$ in the form of flare ribbons and post-eruption arcades. Soft X-rays are sensitive indicators of flare heating and the intensity of flare X-rays can vary over six orders of magnitude. Non-thermal electrons propagating away from the Sun produce various types of radio bursts from decimetric to kilometric wavelengths. Type III bursts generally indicate electrons propagating from the flare site into open magnetic field lines. Another source of particle acceleration is the shocks formed ahead of CMEs that can form very close to the Sun (0.2 Rs above the solar surface (Gopalswamy et al. 2013) and survive to 1 AU and beyond. Electrons accelerated at the shock front produce type II radio bursts (Gopalswamy 2011). Type II radio bursts are the second most intense class in the 30MHz to 30m kHz band (metric band). Their frequency varies in time with the frequency drift from 200 MHz to 30 MHz in about 5 minutes in the solar corona, and in interplanetary space from 30 MHz to 30 kHz in 1-3 days. There is strong evidence that such emission during type II events is generated near the first and second harmonics of the local plasma frequency upstream of the shock via the plasma mechanism: non-thermal electrons accelerated at the shock generate Langmuir waves at the local plasma frequency that get converted into electromagnetic radiation at the fundamental and second harmonic of the local plasma frequency (Cairns 2011). The same shock accelerates protons to very high energies and hence are thought to be the primary source of SEP events observed in the interplanetary space.

CMEs and flares can be ultimately traced to magnetic regions on the Sun. The CME kinetic energy is typically the largest among various components of the energy release (e.g., Emslie et al. 2012). CME kinetic energies have been observed to have values exceeding $\sim 10^{33}$ erg. The only plausible source of energy for eruption is likely to be magnetic in nature (e.g., Forbes 2000). It is thought that the closed magnetic field lines in the active region store free energy when stressed by photospheric motions and the energy can be released by a trigger mechanism involving magnetic reconnection.

Flares and CMEs generally are closely related: flares represent plasma heating while CMEs represent mass motion resulting from a common energy release. This is especially true for large flares and energetic CMEs. Small flares may not be associated with CMEs and occasionally large CMEs can occur with extremely weak flares, especially when eruptions occur outside sunspot regions (Gopalswamy et al. 2015). Even X-class flares can occur without CMEs, but in these cases, no metric radio bursts or non-thermal particles observed in the interplanetary medium, suggesting that the only thing that escapes is the electromagnetic radiation (Gopalswamy et al. 2009). Flares without CMEs are known as confined flares as opposed to eruptive flares that involve mass motion.

## 2.3 Space Weather: Solar Flares, CMEs and SEPs

Solar eruptions are one of the major sources of space weather at Earth and other planets. Solar flares suddenly increase the X-ray and EUV input to the planetary atmosphere by many orders of magnitude. They are also sources of energetic electrons and protons accelerated at the flare reconnection sites in the solar corona forming short-lived SEP events known as impulsive events with maximum particle energy of ~ 10 Mev (Kallenrode et al. 2003). Another type of energetic particles observed in interplanetary (IP) space, gradual SEP events, last over 1 day, which are accelerated to energies over a few GeV per nucleon. These events are associated with CME driven shocks forming in the outer corona at $\geq 2$ $R_{sun}$. Particles are accelerated via diffusive shock acceleration mechanism based on Fermi I acceleration via multiple scattering of particles on plasma turbulent homogeneities as they cross the shock front (Zank et al. 2000).

On the other hand, CME and associated SEPs events have a much longer-term impact from the time the shock forms near the Sun to times well beyond shock arrival at the planet. In the case of Earth this duration can be several days. SEPs precipitate in the polar region and participate in atmospheric chemical processes. When the shock arrives at Earth, a population of locally accelerated energetic particles known as energetic storm particles are encountered. If a geomagnetic storm ensues after the shock (due to sheath and/or CME) then additional particles are accelerated within the magnetosphere. Thus, the ability of a CME to accelerate SEPs and to cause magnetic storms are the most important consequences of space weather. Fast CME driven IP shocks are associated with narrow band metric radio burst emissions (1-14 MHz) and broad band radio emissions (< 4 MHz) called IP type II events.

The continuous CME observations over the past two decades with the simultaneous availability of SEP observations, IP shock observations, and IP type II event data over the past two decades have helped us characterize CMEs that cause SEP events and geomagnetic storms. Figure 4 shows a cumulative distribution of CME speeds observed by SOHO coronagraphs. The average speed of all CMEs is ~400 km/s. All populations of CMEs marked on the plot are fast events (over 600 km/s): metric type II radio bursts (m2) due to CME-driven shocks in the corona at heliocentric distances <2.5 $R_\odot$; magnetic clouds (MC) that are interplanetary CMEs with a flux rope structure; interplanetary CMEs lacking flux rope structure (ejecta, EJ); shocks (S) ahead of interplanetary CMEs detected in the solar wind; geomagnetic storms (GM) caused by CME magnetic field or shock sheath; halo CMEs (Halo) that appear to surround the occulting disk of the coronagraph and propagating Earthward or anti-Earthward; decameter-hectometric (DH) type II radio bursts indicating electron acceleration by CME-driven shocks in the interplanetary medium; SEP events caused by CME-driven shocks; ground level enhancement (GLE) in SEP events indicating the presence of GeV particles. It must be noted that MC, EJ, GM, and Halo are related to the internal structure of CMEs in the solar wind, while the remaining are all related to the shock-driving capability of CMEs and hence particle acceleration. Note that SEP-causing CMEs have an average speed that is four times larger than the average speed of all CMEs. CMEs causing magnetic storms have an average speed of ~1000 km/s, which is more than two times the average speed of all CMEs. It must be noted that these CMEs originate close to the disk center of the Sun and hence are subject to severe projection effects. The projected speeds are likely to be similar to those of SEP-causing CMEs. SEP-causing CMEs forms a subset of CMEs that produce DH type II bursts because the same shock accelerates electrons (for type II bursts) and SEPs. SEPs in the energy range 10-100 MeV interact with various layers of Earth's atmosphere, while the GeV particles reach the ground. Note that only about 3000 CMEs belong to the energetic populations ranging from metric type II bursts to GLE events, suggesting that only ~15% of CMEs are very important for adverse space weather, while the remaining 85% form significant inhomogeneities in the background solar wind.

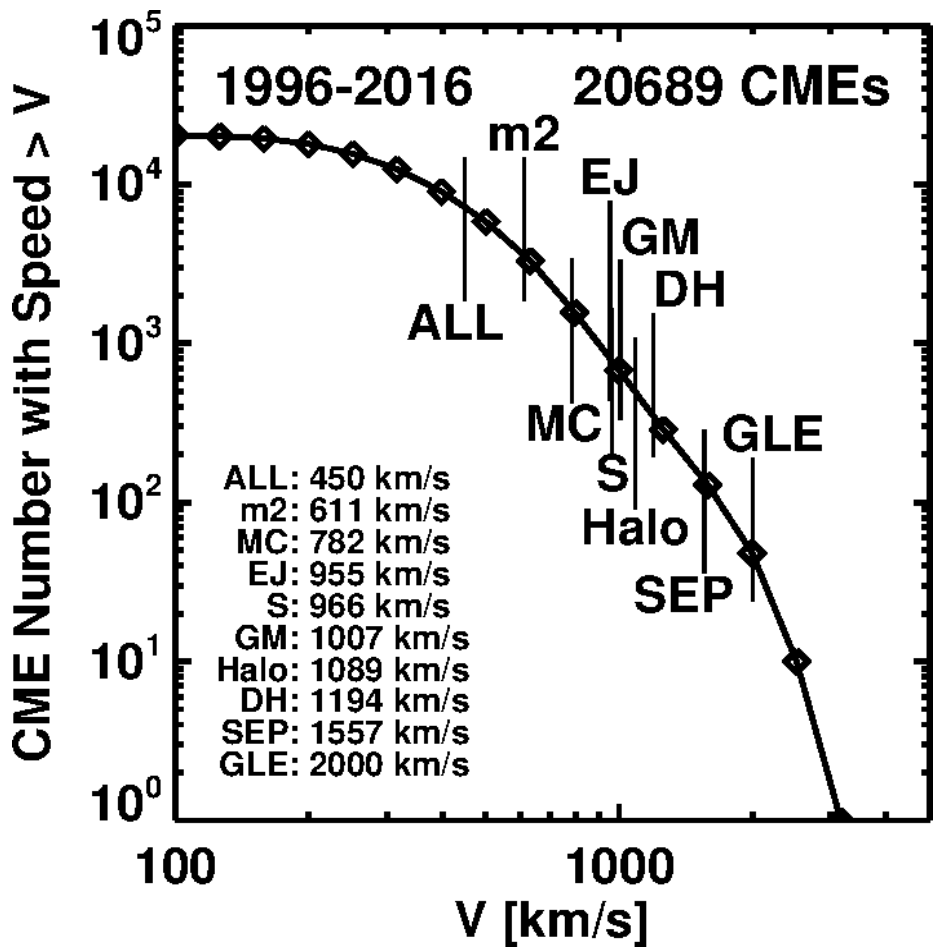

*Figure 4. Cumulative distribution of CME sky-plane speeds (V) from the SOHO coronagraphs during 1996-2016. More than 20,000 CMEs cataloged at the CDAW Data Center (https://cdaw.gsfc.nasa.gov) have been used to make this plot. The average speeds of CME populations associated with various coronal and interplanetary phenomena are marked on the plot. Updated from Gopalswamy (2017).*

Another point to be noted is that CMEs from the Sun do not have speeds exceeding ~4000 km/s. This limitation is most likely imposed by the size of solar active regions, their magnetic content, and the efficiency with which the free energy can be converted into CME kinetic energy. Gopalswamy (2017) estimated magnetic energy up to ~$10^{36}$ erg can be stored in solar active regions and a resulting CME can have a kinetic energy of up to $10^{35}$ erg. Such events may occur once in several thousand years. Note that the highest kinetic energy observed over the past two decades is ~4 x 10$^{33}$ erg.

Figure 5(a) shows the relation between sunspot number, numbers of large SEP events, major geomagnetic storms, major X-ray flares, and fast CMEs. There is a general pattern in which fast CMEs and major flares occur more frequently during the maximum phase of solar activity. Accordingly, the space weather events also occur in correlation with the sunspot number. There are clearly periods when there is discordant behavior between CMEs and flares, but there is generally a better correlation between CMEs, SEPs, and magnetic storms.

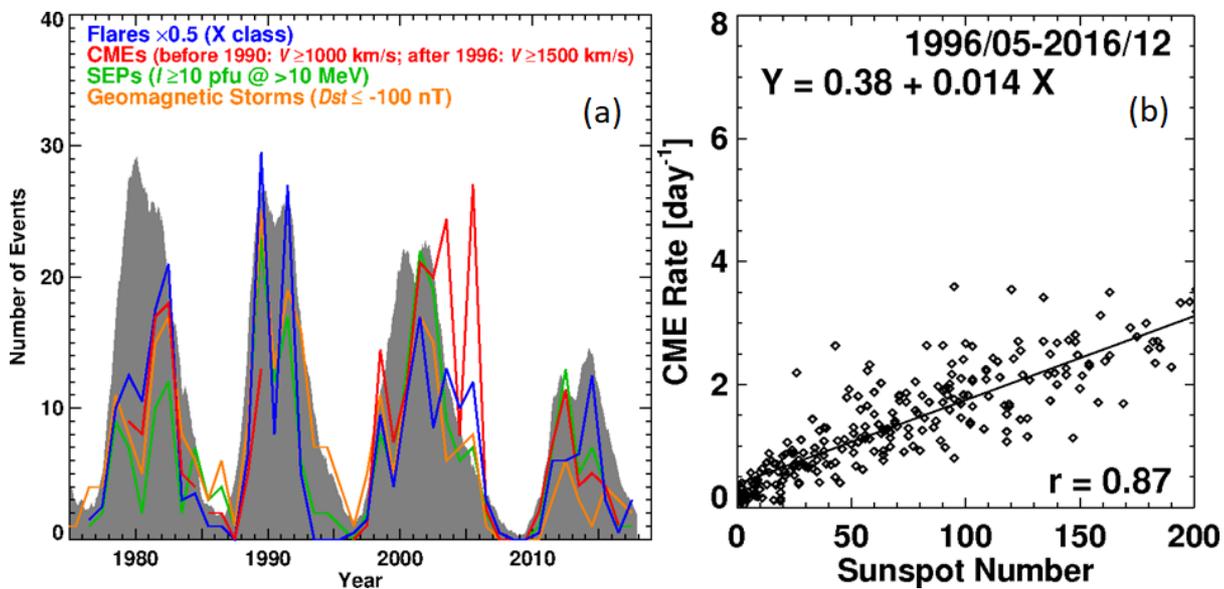

*Figure 5. (a) Solar-cycle variation of X-class soft X-ray flares, fast CMEs, large SEP events, and major geomagnetic storms. All the vent types generally follow the solar cycle represented by the sunspot number (gray). (b) Scatter plot between the daily CME rate and sunspot number during 1996 to 2016.*

The close connection between sunspot number and space weather events can be understood from the fact that large magnetic energy required to power the CMEs can be stored only in sunspot regions. Figure 5(b) shows that the CME rate is higher during high sunspot number, with a correlation coefficient of 0.87. The correlation is high, but not perfect. In fact, Gopalswamy et al (2010) showed that during the rising and declining phases of the solar cycle, the CME rate - sunspot number correlations are very high (r ~0.90), but it is slightly lower (~0.70) during the maximum phase. This is because CMEs also originate from non-spot magnetic regions (quiescent filament regions), which are very frequent in the maximum phase.

**3.0 Space Weather from Active Stars**

Recent X-ray and UV missions including CHANDRA, XMM-NEWTON, the Hubble Space Telescope and the Kepler Space Telescope have opened new windows onto the lives of stars resembling our Sun at various phases of evolution. This has provided a unique opportunity to infer the magnetic properties of planet hosting stars.

The evolutionary history of the Sun and active stars can be reconstructed by studying atmospheric signatures of solar-like stars at various phases of evolution. Physical properties such as rotation velocity and magnetic activity of young solar-like stars strongly depend on age (see, e.g., Shaviv 2003; Cohen et al. 2012; Wood et al. 2014; Johnstone et al. 2015; Airapetian and Usmanov 2016). This section describes recent observational and modeling efforts in reconstructing XUV emission and stellar wind properties of active planet hosting stars as their output plays a crucial role in dynamics of exoplanetary atmospheres.

### 3.1. Coronal Properties of Active Stars

In general, young Sun-like stars have higher rotation velocities, a higher magnetic activity as well as significantly higher mass loss rates (see, e.g., Güdel et al. 1997; Güdel and Naze 2009; Wood et al., 2002; Cleeves et al., 2013). Furthermore, observations of young Sun-like stars have shown the signatures of large magnetic spots that are concentrated at higher latitudes (Strassmeier, 2001) than the sunspots observed on the current Sun. It is also known that the rotation velocity of a star is correlated with its magnetic activity specified by X-ray flux (Güdel 2007). Rapidly rotating young solar-type stars show stronger surface magnetic field (a few hundreds of G) and two-to-three orders of magnitude greater XUV flux than the modern Sun according to solar analogue data (Ribas et al. 2005).

While the long-term evolution of the Sun's bolometric radiation is quite well understood from calculations of the Sun's internal structure and nuclear reactions (e.g., Sackmann & Boothroyd 2003), the evolution of the high-energy radiation is less clear. This emission shortward of approximately 150-200 nm originates in magnetically active regions at the stellar chromospheric, transition region, and corona (in order of increasing height and temperature). The heating of the plasma in these regions may be caused by the dissipation of magnetic energy as it is observed on the Sun (see Section 2.1).

The long-term evolution of the magnetically induced high-energy radiation is related to solar/stellar rotation rate with chromospheric emission also increasing with the rotation rate. As stars age, they spin down due to a magnetized wind and coronal mass ejections (Kraft 1967 Weber & Davis 1967). In stellar astronomy, the activity-rotation-age relations became established with ultraviolet (see, e.g., Zahnle & Walker 1982) and X-ray observations (Pallavicini et al. 1981, Walter 1981) from space. It consists of three important relations.

First, stellar activity, for example as expressed by the total coronal X-ray luminosity, follows a decay law with increasing rotation period $P_{rot}$ of the form $L_X \propto P_{rot}^{-2.7}$ for a given stellar mass on the main sequence. Because X-ray flux is driven by the magnetic field generated by the stellar differential rotation and convection, this correlation suggests that the internal magnetic dynamo strongly relates to the surface rotation period (e.g., Pallavicini et al. 1981, Walter 1981, Maggio et al. 1987, Güdel et al. 1997, Ayres 1997, Wright et al. 2011). However, for rotation periods as short as a couple of days (depending on stellar mass), the X-ray luminosity saturates at $L_{X,\,sat} \sim 10^{-3}\,L_{bol}$ (see, e.g., Wright et al. 2011). The cause for this saturation is not well understood, but could be related to saturation of surface magnetic flux, or some internal threshold of the magnetic dynamo. A unified relation for the activity-rotation relation was presented by Pizzolato et al. (2003) for all spectral types on the cool main sequence.

Second, activity decays with stellar age, t. The decay laws can be reconstructed from open cluster observations in X-rays. On average, one finds $L_X \propto t^{-1.5}$ for solar analogs (Maggio et al. 1987, Güdel et al. 1997). Obviously, this is related to stellar spin-down.

The third ingredient, the relation between stellar spin-down and age (Skumanich 1972) is also well studied based on co-eval cluster samples. On average, a relation between the stellar rotation period and the stellar age, t, follows as $P_{rot} \propto t^{0.6}$ (e.g., Ayres 1997 for solar analogs), and is compatible with the above two relations.

The activity decay law has often been used for exoplanetary loss calculations based on simple fits to observed trends using limited samples. For X-rays, the "Sun in Time" sample was used with ages between ~100 Myr and ~5 Gyr (Güdel et al. 1997), and this was complemented with the corresponding decay laws for the extreme-ultraviolet and far ultraviolet radiation (Ribas et al. 2005) and also near-UV (Claire et al. 2012). Generally, the decay in time is steeper for higher-energetic radiation, which therefore also decays by a larger factor over a given time. Using the average regression laws referred to above, X-rays decrease by a factor of ~1000-2000 over the main-sequence life of a Sun-like star, EUV by a few hundred, and UV by factors of a few tens (e.g., Ribas et al. 2005). However, such a statistical approach could be flawed because in reality the rotation behavior of young stars (ages less than a few hundred Myr for solar analogs) is highly non-unique. Cluster samples show a wide dispersion in rotation periods for ages up to a few hundred Myr, after which they gradually converge to a unique, stellar-mass dependent value (Soderblom et al. 1993). This convergence is attributed to the feedback between the magnetic dynamo and angular momentum loss in a magnetized wind. The wide dispersion of $P_{rot}$ for young stars instead reflects the initial conditions for rotation starting after the protostellar disk phase (e.g., Gallet & Bouvier 2013).

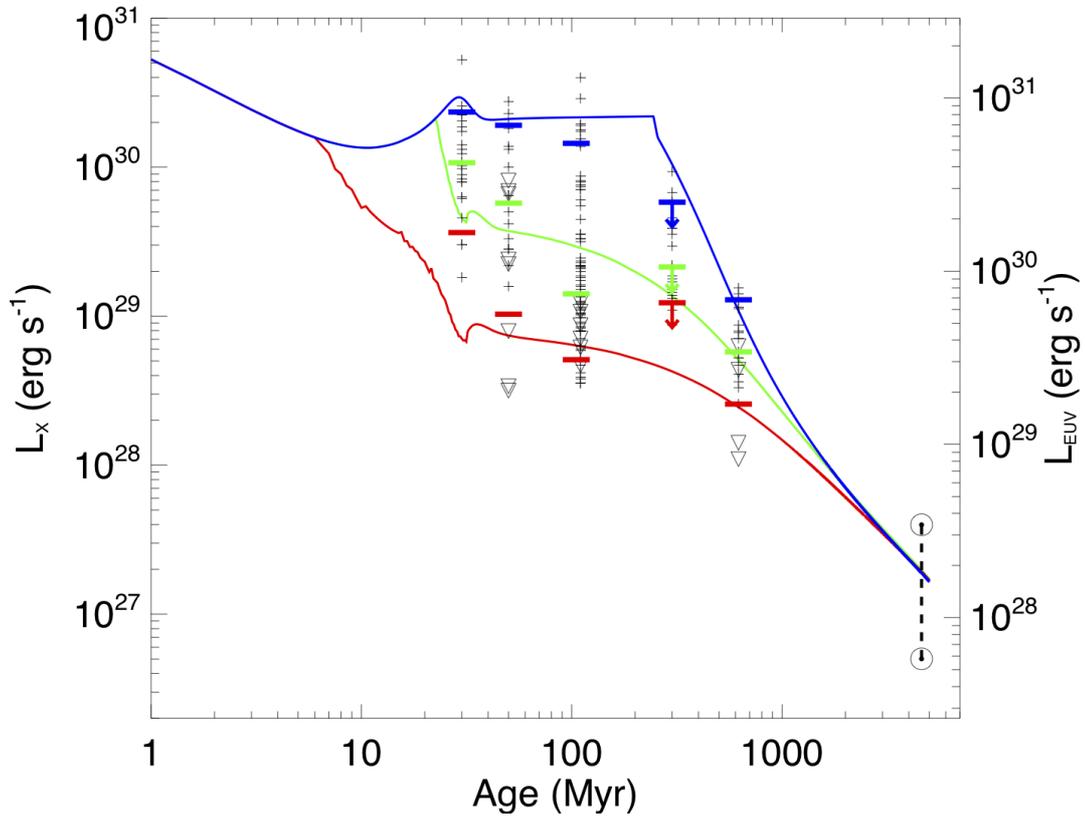

*Figure 6. Tracks of $L_X$ for a $1M_\odot$ star our calculated from rotation tracks using an observed rotation period distribution after the protostellar disk phase. The red, green, and blue tracks refer to the 10th, 50th, and 90th percentiles of the rotation period distribution. The + signs and the $\nabla$ symbols are observed values of $L_X$ or, respectively, their upper limits, from several open clusters at the respective ages. The solid horizontal lines show the 10th, 50th, and 90th percentiles of the observed distributions of $L_X$ at each age calculated by counting upper limits as detections. The two solar symbols at 4.5 Gyr show the range of $L_X$ for the Sun over the*

*course of the solar cycle. The scale on the right y-axis shows the associated $L_{EUV}$ (from Tu et al. 2015).*

An evolutionary decay law for high-energy radiation therefore needs to be accounted for the dispersion of rotation periods. Observationally, a wide distribution of $L_X$ in young clusters was in fact known from early cluster surveys (e.g., Stauffer et al. 1994). A proper analysis of the problem was laid out in studies by Johnstone et al. (2015a,b) and Tu et al. (2015), in which a solar-wind model was adapted to stars at different activity levels and different magnetic fluxes, fitting distributions of $P_{rot}$ in time from various clusters. Translating rotation to high-energy radiation, Tu et al. (2015) reported the finding illustrated in Figure 6. This figure shows that, depending on whether a solar analog starts out as a slow or fast rotator after the disk phase, the X-ray evolutionary tracks first diverge (i.e., $L_X$ of a slow rotator rapidly decays with time while that of a fast rotator does not) and then converge again after several hundred Myr when the rotation periods converge. The nearly constant X-ray luminosity for fast (but spinning-down) rotators is due to the saturation effect.

The distribution of high-energy radiation is broadest in the range of a few tens to a few hundreds of Myr, precisely the range of interest for proto-atmospheric loss, the formation of a secondary atmosphere, a crust and a liquid water ocean on Earth, and the earliest steps toward the formation of life. A consistent study of atmospheric evolution therefore needs to account for the uncertainty of early stellar high-energy evolution. To the present day, we do not know the evolutionary track in $L_X$ which the Sun has taken. The first clues on the Sun as a slow rotator come from the modeling of loss of moderate volatiles such as sodium and potassium from the surface of the Moon (Saxena et al. 2019).

We should also add that the hardness of high-energy radiation, specifically, the X-ray hardness (the relative amount of "harder" to "softer" radiation) decreases with decreasing X-ray surface flux, because more active stars are dominated by hotter coronal plasma (Johnstone & Güdel 2015). This has consequences for the irradiating spectra because the penetration depth of radiation in an atmosphere depends on photon energy.

While stellar X-ray emission can be well characterized by space missions including CHANDRA, XMM-Newton, Swift, NICER and MAXI, most of stellar XUV emission remains hidden from us because most of the EUV flux longer that 40 nm is absorbed by interstellar medium even from the closest stars. The XUV stellar spectrum is crucial for understanding the exoplanetary atmospheric evolution and its impact on habitable worlds as it drives and regulates atmospheric heating, mass loss and chemistry on Earth-like planets, and thus is critical to the long-term stability of terrestrial atmospheres.

The stellar XUV emission can be reconstructed using empirical and theoretical approaches. Empirical reconstructions have already provided valuable insights on the level of ionizing radiation from F, K, G and M dwarfs (Cuntz & Guinan 2016; France et al. 2016; 2018; Loyd et al. 2016; Youngblood et al. 2016; 2017). The latter approach is based on analysis of HST-STIS and COS based stellar far-ultraviolet (FUV) observations as proxies for reconstructing the EUV flux from cool stars, either through the use of solar scaling relations (Linsky et al. 2014; Youngblood et al. 2016) or more detailed differential emission measure techniques (e.g., Louden et al. 2017). Using these datasets the authors found that the exoplanet host stars, on average, display factors of 5–10 lower UV activity levels compared with the non-planet-hosting sample. The data also suggest that UV activity-rotation relation in the full F–M star sample is characterized by a power-law decline (with index $\alpha \approx -1.1$), starting at rotation periods 3.5 days. France et al. (2018) used N V or Si IV spectra and knowledge of the star's bolometric flux to estimate the intrinsic stellar EUV irradiance in the 90–360Å band with an accuracy of roughly a factor of $\approx 2$. The data suggest that many active K, G and most of "quiet" M dwarfs generate high XUV fluxes from their magnetically driven chromospheres, transition regions and coronae. Another approach is based on semi-empirical non-LTE modeling of stellar spectra using radiative transfer codes (Peacock et al. 2018). This model has recently been applied to reconstruct the XUV flux from TRAPPIST-1 constrained by HST Ly-alpha and GALEX FUV and NUV observations.

Airapetian et al. (2017a) used the reconstructed XUV flux from one of the quiet M dwarf, M1.5 red dwarf, GJ 832 to compare with the XUV fluxes from the young (0.7 Gyr) and current Sun. Figure 7 shows the reconstructed spectral energy distribution (SED) of the current Sun at the average level of activity (between solar minimum and maximum with the total flux, $F_0$ (5 - 1216 Å) = 5.6 erg/cm$^2$/s; yellow dotted line), the X5.5 solar flare occurred on March 7, 2012 (blue line), the young Sun at 0.7 Gyr (yellow solid line) and an inactive M1.5 red dwarf, GJ 832 (red line). (Airapetian et al. 2017a). The XUV flux from the young Sun, and GJ 832 are comparable in magnitude and shape at wavelengths shorter (and including) the Ly-alpha emission line. This suggests that the contribution of X-type flare activity flux is dominant in the "quiescent" fluxes from the young Sun including other young suns and inactive M dwarfs, which should play a critical role in habitability conditions on terrestrial type exoplanets around these stars.

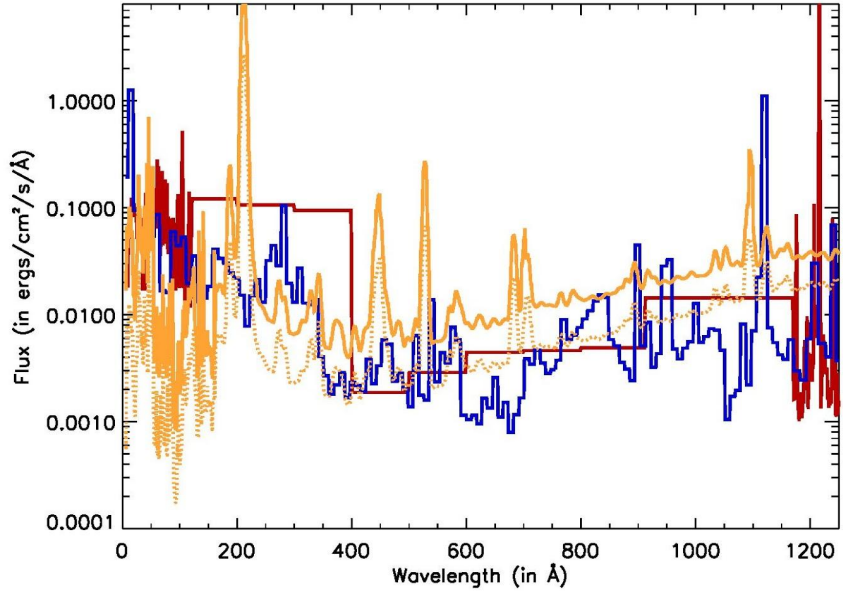

*Figure 7. Spectral energy distribution in the young (yellow dotted), current Sun (yellow solid) as compared to the X5.5 solar flare (blue) and M dwarf (red) (Airapetian et al. 2017a)*

Theoretical models of XUV emission from planet hosting stars are based by multi-dimensional MHD simulations of stellar coronae and winds driven by the energy flux generated in the chromosphere constrained by FUV emission line fluxes and ground-based spectropolarimetric observation enabled reconstruction of stellar surface magnetic fields are in early phase of development. The first data driven 3D MHD model of global corona of a young solar twin star, $k^1$ *Ceti*, has recently been developed by Airapetian et al. (2019b). Two different techniques are mostly used to reconstruct stellar magnetism, namely the Zeeman broadening technique (ZB, Johns-Krull 2007) and the Zeeman Doppler Imaging technique (ZDI, Donati & Brown 1997). The ZB technique measures Zeeman-induced line broadening of unpolarized light (Stokes *I*). This technique is sensitive to the *total* (to large- and small-scale), unsigned surface field. The ZDI technique recovers information about the large-scale magnetic field (its intensity and orientation) from a series of circularly polarized spectra (Stokes *V* signatures). These techniques have their limitations. While ZB provides a measurement of the total field, it does not provide a measurement of the field polarity. On the other hand, while ZDI provides measurement of the field polarity, it does not have access to small-scale fields. For this reason, these techniques are complementary to each other (see, eg, Vidotto et al 2014).

Using ZDI magnetic maps, Vidotto et al (2014) demonstrated that the average, unsigned large-scale magnetic field of solar like stars decay with age and with rotation (see also Petit et al 2008; Folsom et al 2018) as, respectively,

$$<|B|> \sim t^{-0.655 \pm 0.045}$$

$$<|B|> \sim P_{rot}^{-1.32 \pm 0.14}.$$

These empirical trends provide important constraints on the evolution of the large-scale magnetism of cool stars. In particular, the magnetism-age relation (see Figure 8) presents a similar power dependence empirically identified in the seminal work of Skumanich (1972) and the magnetism-rotation relation suggests that a linear dynamo of the type B ~ 1/$P_{rot}$ is in operation in solar-like stars. These empirical relations contain significant spread partially caused due to magnetic field evolution on years, and sometimes only months (see Figure 8).

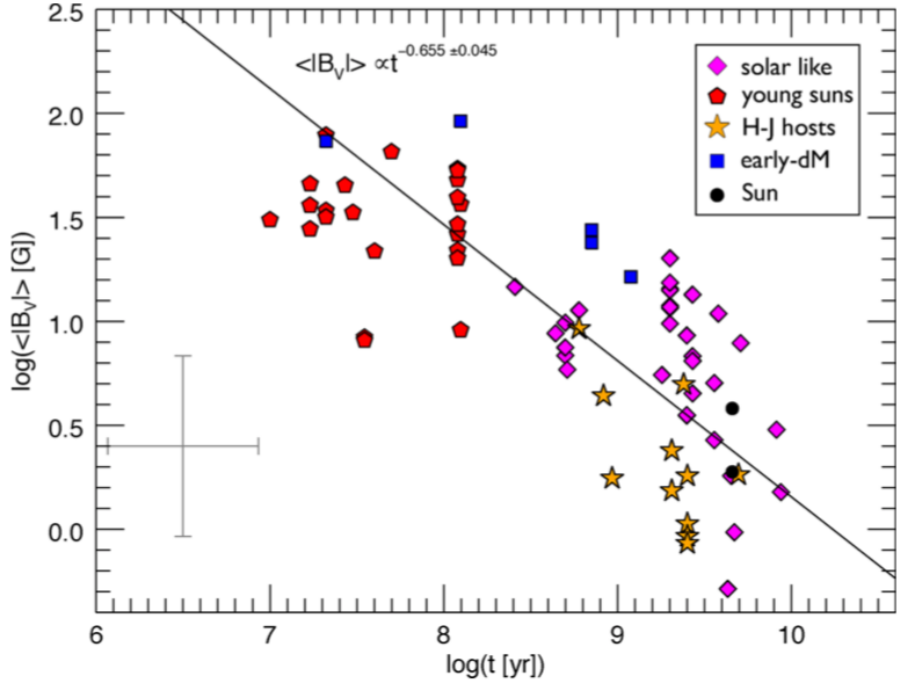

*Figure 8: Empirical relation between the unsigned average stellar magnetic field (derived using the Zeeman Doppler Imaging technique) and age. Figure from Vidotto et al (2014).*

Given that ZDI allows one to obtain the topology of the surface magnetic field, See et al (2015) investigated the energy contained in the poloidal and toroidal components of the stellar surface field. They found that the energy in these components are correlated, with solar type stars more massive than 0.5 $M_\odot$ having $<B_{tor}^2> \sim <B_{pol}^2>^{1.25 \pm 0.06}$.

**3.2 Winds from Active Stars**

Stellar winds represent an extension of global stellar corona into the interplanetary space and are fundamental property of F-M dwarf stars. Stellar coronal winds are weak and no reliable detection of a wind from another star other than the Sun was reported. The empirical method of detection of stellar wind relies on the observations of HI Ly-$\alpha$ absorption from the stellar astrosphere forming due to the wind interaction with the surrounding interstellar medium (Wood 2018). The mass loss rates from young GK dwarfs are well correlated with X-ray coronal flux as $\dot{M} \propto F_X^{1.29}$ reaching the maximum rates of 100 times of the current Sun's rate. However, this relation fails for stars with the greater X-ray flux suggesting the saturation of surface magnetic flux (see Section 2.1).

Recent measurements of surface magnetic fields and X-ray properties of the coronae from solar-type stars paved a way for inputs to heliophysics based multi-dimensional models of the stellar coronae and winds using "Star-As-The-Sun" approach. This approach suggests the availability of stellar inputs and boundary conditions to be used for heliophysics MHD code. The first MHD wind models from young solar-type stars resembling our Sun in its infancy were developed by Sterenborg et al. (2011); Airapetian & Usmanov (2016) and De Nascimento et al. 2016; Ó Fionnagáin et al. 2018. Specifically, the young Sun's wind speed at 1 AU was twice as fast, five times hotter with the mass loss rate of > 50 times greater as compared to the current solar wind properties (see Figure 9).

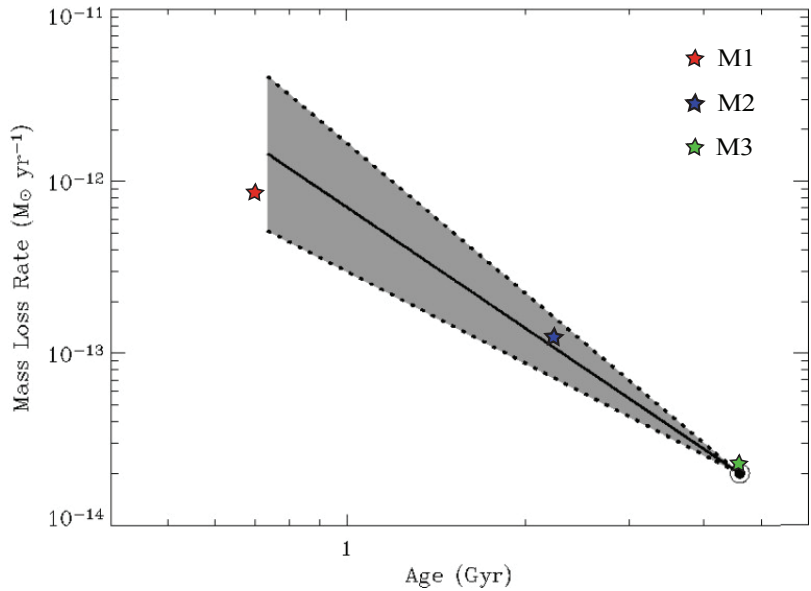

*Figure 9. The model and empirical mass loss rates shown as the shadow region (Wood et al. 2005) from the evolving Sun. Red, blue and green stars show the model mass loss rates for the 0.7 Gyr 2.2 Gyr and 4.65 Gyr old Sun (Airapetian and Usmanov 2016).*

The young Sun's wind model was recently extended to describe the formation of global solar corona by incorporating the observationally derived stellar magnetograms and the chromospheric parameters of the best-known young Sun's proxy, $k^1$ Cet, into the data driven 3D MHD thermodynamic model (Airapetian et al. 2019b). Figure 10 shows the drastic change in the topology of global magnetic field of the star in 11 months (2012.9 (left) and 2013.8 (right)). If the shape of global stellar corona in the left panel resemble a dipole-like magnetic field, global coronal field tilts at 45 degrees and becomes complex. The model predicts that the variation of X-ray flux by a factor of 2 over 11 months. The coordinated spectropolarimetric observations that provide stellar magnetic field with TESS, HST and X-ray observations (XMM-Newton and NICER mission) are currently in place to provide epoch specific model inputs and outputs that are required to check the model predictions. Recent spectropolarimetric observations of another active star, K dwarf, 61 Cyg A, show that the star's magnetic field has undergone full dipole flip during the magnetic cycle (Boro Saikia et al. 2015). Such data driven models are required in order to characterize the range of variations of ionizing radiation fluxes and their time scale in order to asses the impact of the stellar high energy radiation onto habitability of exoplanets.

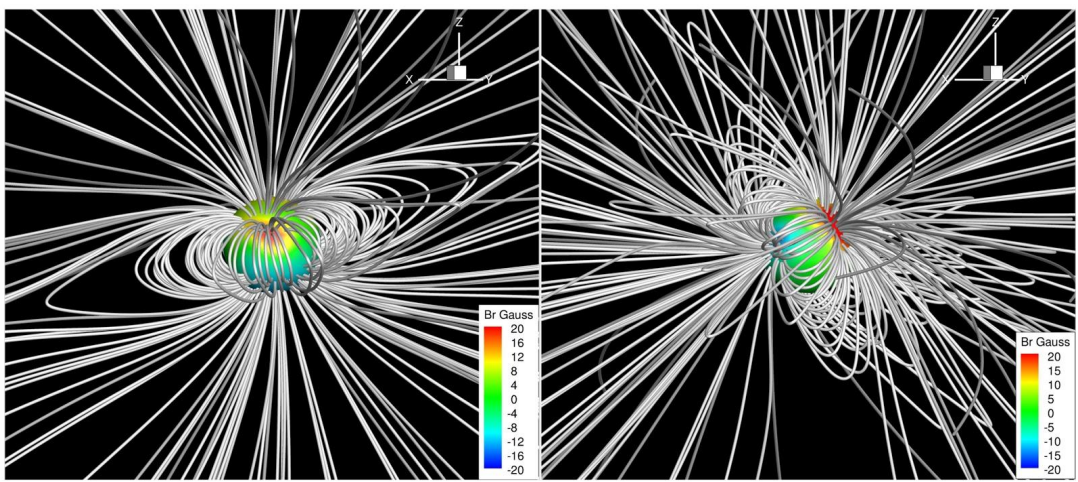

*Figure 10. The evolution of the 0.7 Gyr Sun proxy's global magnetic field over the course of 11 months, at 2012.9 (left) and 2013.8 (right) (from Airapetian et al. 2019b).*

### 3.3 Superflares from Active Stars

Recent observations of stars of F, G, K and M spectral types in X-ray, UV and optical bands provide evidence of stellar flares with energy release 10 to 10,000 times larger than that of the largest solar flare ever observed on the Sun and referred to as superflares (Schaefer et al. 2000; Zhou et al. 2011; Walkowicz et al. 2011; Maehara et al. 2012; 2015; 2017). The energy of the smallest detected solar flare (a nanoflare which was recently detected by Focusing Optics X-ray Solar Imager (FOXSI) mission, Ishikawa et al. 2017) is 13 orders of magnitude smaller than the largest stellar superflare from active stars. Many lines of evidence suggest that solar flares are powered by magnetic reconnection of coronal magnetic fields emerging from the solar convective zone into the solar corona (Shibata and Magara 2011).

This raises a fundamental question i.e. whether powerful stellar flares on magnetically active stars are also driven by magnetic energy release? If they are, then can we extrapolate statistics of white-light solar flares and its energy partition into different modes of energy release (direct heating, non-thermal energy of electrons and ions and kinetic energy of flows) to 5 orders of magnitude? Because our understanding of the nature of solar flares is far from complete, we need to rely on statistical patterns observed for solar and stellar flares. Characterization of the frequency of occurrence of flares with energy may serve as one of such important discriminators. The measurements of solar hard X-ray (HXR) bursts obtained by the Solar Maximum Mission (SMM), radio and optical bands suggest that the frequency of occurrence of flares, dN, within the energy interval E, E+δE follows a power law with flare energy as

$$\frac{dN}{dE} = k_0 E^{-\alpha}$$

with power law index, α, varying between -1.13 to -2 (Crosby et al. 1993). The data suggest that flare energy distribution in the optical band corresponds to α = -1.8. This can be directly compared to recent optical observations by the Kepler mission of thousands of white-light flares on hundreds of G, K and M dwarfs (Maehara et al. 2012). The power-law index for superflares from both short- and long-cadence data, band corresponds to α = −1.5 ± 0.1 for flare energies from $4 \times 10^{33}$ erg to $10^{36}$ erg (Maehara et al. 2015).

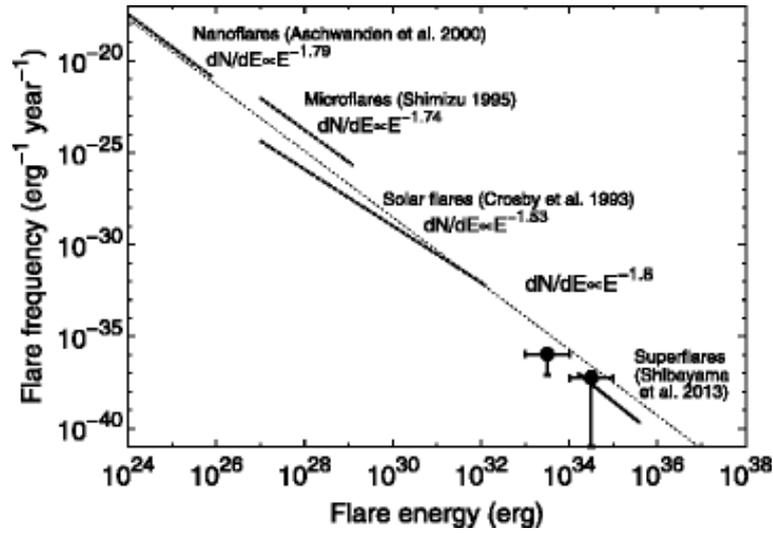

*Figure 11. The power-law distribution of frequency of occurrence of solar and stellar flares.*

Figure 11 shows the frequency distribution for a range of flare processes: from the smallest solar flares to the most powerful stellar superflares. Filled circles indicate the frequency distribution of superflares on Sun-like stars (G-type main sequence stars with $P_{rot} > 10$ days and $5,800 < T_{eff} < 6,300$ K) derived from short-cadence data. Horizontal error bars represent the range of each energy bin. Bold solid line represents the power-law frequency distribution of superflares on Sun-like stars taken from (Shibayama et al. 2013). Dashed lines indicate the power-law frequency distribution of solar flares observed in hard X-ray (Crosby et al. 1993), soft X-ray (Shimizu 1995), and EUV bands (Aschwanden et al. 2000). Frequency distributions of superflares on Sun-like stars and solar flares are roughly on the same power-law line with an index of α = -1.8 (thin solid line) for the wide energy range between $10^{24}$ and $10^{35}$ erg. Shibayama et al. (2013) found that superflares from young (0.7 Gyr) solar-like stars with the energy of $10^{34}$ erg occur at the rate of 0.1 event per day, which suggests that the frequency of superflares with energies ~ $10^{33}$ erg (see Figure 11) is ~ 10 events/day. The event of the comparable energy was observed on September 1-2, 1859 by British astronomers Richard

Carrington (Carrington 1859) and Richard Hodgson (Hodgson 1959) and was coined as the Carrington event. Recent studies suggest that our Sun had produced powerful flares associated CMEs with the energy at least 3-10 times stronger than the Carrington event. Possible solar superflare events could be associated extreme SPE events that produce cosmogenic radionuclides including the enhanced content of carbon-14, $^{14}$C, detected in tree rings, beryllium-10, $^{10}$Be, and chlorine-36, $^{36}$Cl, measured in both Arctic and Antarctic ice cores (Mekhaldi et al. 2015). These data suggest that solar superflares events were accompanied with hard energy SEP protons occurred in A.D. 774 to 775, A.D. 993 to 994 and 660 BC (Miyake et al. 2012; 2013; O'Hare et al. 2019). The proton fluence associated with such powerful events would be equivalent to solar flares with the energy of ~ $10^{34}$ ergs. The occurrence frequency of these events (two events in 220 years) is roughly comparable to the average occurrence frequency of superflares on Sun-like stars with the energy of $10^{33}$ to $10^{34}$ erg.

As the Sun evolved to the current age, its flare frequency was reduced dramatically to 1 event per 70, 500 and 4,000 years for the flare bolometric energies of $10^{33}$, $10^{34}$, and $10^{35}$ erg, respectively. Also, data suggest that the frequency of superflares strongly depends on the rotation period, $P_{rot}$ (Notsu et al. 2013; Maehara et al. 2015; Davenport 2016). These frequencies are lower than the upper limits derived from radionuclides in lunar samples (Schrijver et al. 2012).

The frequency of stellar superflares also correlates with the effective temperatures of the star. Kepler data indicate that main sequence stars with lower temperature exhibit more frequent superflares. The frequency of superflares on K- and early M-type stars ($T_{eff}$=4000-5000K) is roughly 1 order of magnitude higher than that on G-type stars (e.g., Candelaresi et al. 2014).

As described above, the flare frequency depends on the rotation period. On the other hand, Kepler data suggest that the bolometric energy of the largest superflare on solar-type stars does not depend on the rotation period (e.g., Maehara et al. 2012; Notsu et al. 2013). This implies that the slowly-rotating solar-type stars could exhibit superflares with the energy of $10^{34}$ -$10^{35}$ erg. Most of superflare stars show periodic brightness variations with the amplitude of ~1% due to the rotation of the star which suggest that the stars showing superflares may have large starspots on their surface. According to Shibata et al. (2013), the energy of the largest superflares is roughly proportional to $A_{spot}^{3/2}$ ($A_{spot}$ is the starspot area), and large starspots with the area of > 1% of the solar hemisphere ($> 3 \times 10^{20}$ cm$^2$) would be necessary to produce superflares with the energy of $\geq 10^{34}$ erg.

Existence of starspots on Kepler superflare stars are also supported from the follow-up spectroscopic observations (Notsu et al. 2015; Karoff et al. 2016). A recent statistical analysis of starspots based on Kepler data performed suggests that the magnetic activity pattern in terms of the frequency-energy distribution of solar flares is consistent with superflares detected on more active solar-like stars, which implies that the same magnetically driven physical processes of the energy release within coronal active regions are responsible for the origin of solar and stellar flares (Maehara et al. 2017). Recent estimates of lifetimes and emergence/decay rates of large starspots on young solar type stars suggest that sunspots and large starspots share the same underlying processes (Namekata et al. 2018).

## 3.4. Coronal Mass Ejections from Active Stars

As discussed in Section 2.2, large solar flares are associated with powerful CMEs. Because eruptive events on solar-types stars are also driven by magnetic field energy, this begs the important question as to whether we can use correlations between X-ray flare fluxes and CME parameters to derive the properties of stellar CMEs, estimate their rates of occurrence and derive their observational signatures. Sub-section 3.4.1 discusses recent efforts in modeling

CMEs on other stars from first principles and the model predictions of their properties. Subsection 3.4.2 presents the summary of recent searches of CMEs and its methodology.

### 3.4.1 MHD Models of Stellar CMEs

Sudden eruptions on the Sun and solar-type stars occur due to the rapid release of free magnetic energy stored in the sheared and/or twisted strong fields typically associated with sunspots and active regions (e.g. Fletcher et al. 2011; Shibata & Magara 2011; Kazachenko et al. 2012; Janvier et al. 2015). Large flares are often accompanied by CMEs (Gopalswamy et al. 2005). Regardless of the specific details of the ideal or resistive instabilities that initiate the catastrophic, run-away eruption of CMEs, the relationship between the eruptive flare and its resulting CME is well understood (Forbes 2000; Zhang & Dere 2006; Lynch et al. 2016; Török et al. 2018; Welsch 2018; Green et al. 2018).

One such model for the initiation and launching of solar CMEs into the corona in global and local active region environments is the magnetic breakout model (Antiochos et al. 1999; DeVore & Antiochos, 2008; Lynch et al. 2008, 2009; Karpen et al. 2012; Masson et al. 2013). Lynch et al. (2016) have demonstrated that even in bipolar streamer distributions, the overlying, restraining closed flux is removed in a breakout-like way through an opening into the solar wind, thus enabling the eruption of low-lying energized flux. Therefore, the evolution and interaction between the low-lying energized and overlying restraining fields are extremely important aspects of modeling eruption processes in solar and stellar coronae to correctly estimate the CME ejecta properties and energetics. Figure 12 illustrates recent ARMS 3D simulation results by Lynch et al. (2019) of a massive halo-type (width of $360^0$, see Figure 12) energetic CME eruption based on the observationally derived $k^1$ *Cet* magnetogram. The entire stellar streamer-belt visible on this figure is energized via radial field-preserving shearing flows and the eruption releases ~7 × 10$^{33}$ erg of magnetic free energy in ~10 hours (Lynch et al. 2008; 2019). Magnetic reconnection during the stellar flare creates the twisted flux rope structure of the ejecta and the ~2000 km/s eruption creates a CME-driven strongly magnetized shock.

The maximum increase in total kinetic energy during the eruption was ~ 2.8 × 10$^{33}$ erg – on the order of the 1859 Carrington flare event. The meridional planes plot the logarithmic current density magnitude showing the circular cross-section the CME (Vourlidas et al. 2013) and representative magnetic field lines illustrate 3D flux rope structure.

CME initiation process from an active region (compact CMEs) suggests that a solar-- like star with the large-scale dipolar magnetic field of 75 G would confine impulsive coronal eruptions with energies less than 3 x 10$^{32}$ ergs. Thus, only Carrington type flare eruptions can eject compact CMEs from these stars. These results also imply that strong (~ 600 - 1000 G) dipole fields on young active M dwarfs would confine CMEs with E < 3 x 10$^{34}$ ergs, and thus, should occur at much lower (at least 10,000 times) frequency that the flare frequencies that are typically ~ 1 event per hour.

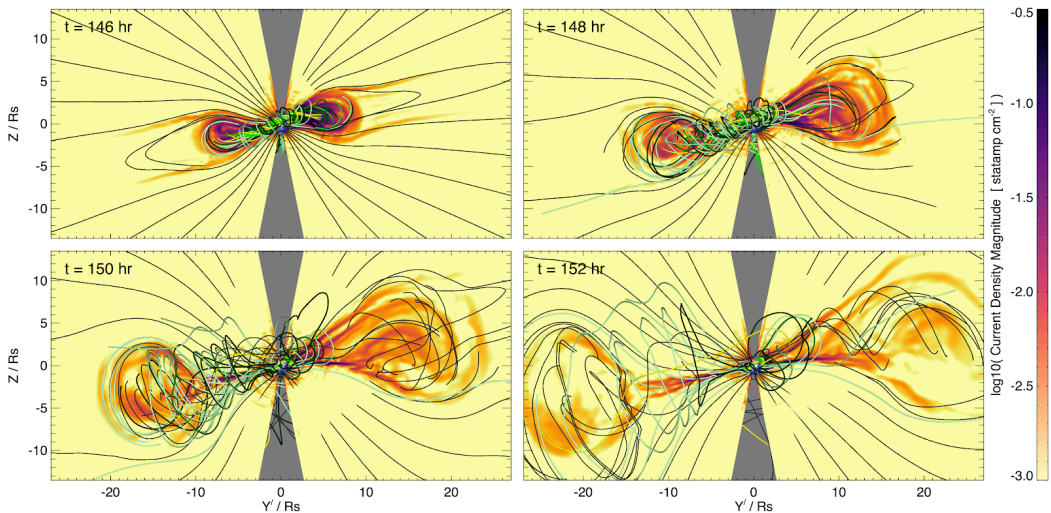

*Figure 12. ARMS 3D simulation of a massive (Carrington scale) stellar coronal eruption initiated from the k$^1$ Cet magnetogram. The contour planes show current density magnitude in the erupting flux rope (Lynch et al. 2019).*

### 3.4.2 Search for Stellar CMEs.

If we apply the empirical relationships solar energetic flares and associated CMEs (see Section 2.2), then stellar superflares should be associated with fast and massive CMEs from magnetically active stars. Can we observationally detect these large stellar CMEs? The three major observational techniques for detecting the signatures of stellar CMEs include a) Type II radio bursts; b) Doppler shifts in UV/optical lines, and c) continuous absorption in the X-ray spectrum. Type II radio bursts occur at the low frequency of < 30 MHz. Within few minutes the burst frequency shifts to < 1 MHz. The emission below 8-15 MHz cannot be detected from the Earth due to the reflection from the terrestrial ionosphere (Davies 1969). Also, high velocity outflows generated during powerful stellar eruptions could also be signatures of CMEs. A detection of blue-shifted hydrogen and Ca II chromospheric lines from the spectrum of the young M dwarf star, AD Leo, during a powerful superflare event (with the energy of > $10^{34}$ erg) could be possibly associated with the massive CMW with the speed up to 5,800 km/s (Houdebine et al. 1990). Another signature of stellar CMEs could be associated with the excess amount of absorption of X-ray emission as for example for observed early in the flare from UX Ari (Franciosini et al. 2001). This absorption requires 5 times greater column density in neutral hydrogen that attributable to the interstellar medium on the line of sight. Favata & Schmitt (1999) discussed X-ray absorption observed during a superflare from Algol star as a signature of a giant CME. Recently, Moschou et al. (2018) have extended their suggestion to a stellar CME model that can fit the statistical correlations of CME mass with X-ray flare flux. The aforementioned and other possible observational signatures of stellar flares and CMEs are summarized in Table 1.

|  | **Observational Signature** | **Seen in Sun?** | **Seen in Other Stars?** |
|---|---|---|---|
| Flare | Nonthermal hard X-ray emission | Y | Unknown |
|  | Incoherent radio emission | Y | Y |
|  | Coherent radio emission | Y | Y |
|  | Far UV (FUV, 920 - 1200 Å) emission lines | Y | Y |
|  | Hot blackbody optical-UV emission | Y | Y |
|  | Coronal emission lines and continuum | Y | Y |
|  | Optical/UV chromospheric emission lines | Y | Y |
| CME | Coronagraph measurements via Thomson scattering of photospheric photons off coronal electrons | Y | N |
|  | Radio type II bursts | Y | Possibly |
|  | High velocity outflows from escaping material | Y | Possibly |
|  | Scintillation of background radio sources | Y | Possibly |
|  | Coronal dimmings | Y | Possibly |
|  | $N_H$ increases in X-ray flare spectra | N | Possibly |
|  | Pre-flare dips prior to impulsive phase | Unknown | Possibly |

|  | Effect of CMEs on stellar environment | Y | Possibly |
|--|--|--|--|
|  | Flare/CME connections | Y | Y |

*Table 1: Comparison of observational signatures used to study flares and coronal mass ejections on the Sun and in aggregate across other stars.*

It is evident that there is significant overlap between observational signatures of flares on the Sun and other stars, even if on closer examination the specific bandpasses and observing methods differ (e.g. spatially and temporally resolved solar observations versus spatially integrated temporal stellar observations). Osten (2016) discusses the observational perspective on stellar flares and what is known about stellar flaring events throughout the age of the Solar system. The comparison for coronal mass ejections shows significantly less overlap; this is due primarily to the much lower contrast of the ejected material compared to the stellar disk, and the need to disentangle any CME-related emission from flare-related emission in order to attribute a signature entirely to a CME. To date, while there have been suggestions of ejections of CMEs in solar-type stars, these have generally been attributed to specific circumstances, such that the wider applicability is unknown. Osten & Wolk (2017) described the framework for finding and interpreting stellar CMEs, including a discussion of the observational signatures listed in the CME table above.

As discussed in the previous section, given that only energetic CMEs can be ejected into the astrosphere, the frequency of CMEs on magnetically active stars should be reduced the by at least 1-2 orders of magnitude. Also, the expected radio frequency from shocks associated with energetic CMEs and superflares from magnetically active stars should be in the order of 10 MHz or lower (Lynch et al. 2019). This suggests that extended low-frequency observations are required to hunt for elusive stellar CMEs.

## 4.0 Impact of Space Weather Effects on Modern Earth, Venus and Mars

Distinguishing the different evolutionary pathways of Earth and its rocky neighbors, as well as for Earth-like planets are central tasks in (exo)planetary science. Much is to be learned by comparing the response of a rocky planet which features a magnetosphere (the Earth) to changes in stellar output compared with the responses of our weakly magnetized neighboring planets Venus and Mars (Brain et al., 2016). Did Venus' climate evolve via a runaway greenhouse from earlier, potentially habitable conditions (Way et al., 2016) into its current extreme state (Bullock and Grinspoon, 2001)? What role the Sun played in its climate evolution (e.g. Lammer et al, 2012, Chassefiere et al 2010)? Is Venus a magma ocean planet preserved from its early phase (Hamano et al. 2013)? How and to what extent escape processes driven by stellar output drove atmospheric loss of the Martian atmosphere into its current cold, dry, dusty state is a key theme in Martian science (Jakosky and Phillips, 2001)?

Chemical responses of planetary atmospheres to space weather can be broadly split into two mechanisms - firstly due to enhanced radiation (e.g. EUV) which enhances e.g. photolytic destruction and secondly due to interaction of gas-phase species with high energy particles. Space plasma interaction responses can similarly be classified as either atmospheric ion erosion or sputtering of neutral atmosphere by the energized ions that are not on escape trajectories but instead deposit their gained energies into the upper atmosphere. But first it is useful to briefly consider the situation at the Earth for comparisons.

### 4.1 Modern Earth

Our planet, like all the other terrestrial planets in our Solar system, is exposed to the XUV and particle emissions from the quiescent and active Sun. The latter includes the solar wind plasma with its embedded magnetic fields. The extremes of these external influences occur in the forms of solar flares, CME driven plasma and magnetic field enhancements proceeded by shocks, and the related solar energetic particles. In addition, galactic cosmic rays diffuse into the inner heliosphere and into the planetary atmospheres. Each of these has specific atmospheric (and related surface) influences, with those affecting Earth the most potentially consequential for both our technological society and the biosphere. Such effects, which can also have evolutionary time scale influences of importance, can generally be classified into two main categories: Atmospheric erosion or loss to space that can become significant on long time scales, and atmospheric chemistry alterations, which can have both long and short-term impacts.

### 4.1.1 Space Weather Drivers of Ionospheric Outflows

The contemporary flux of outflowing ions at high latitudes at the Earth is a combination of a light ion (Hydrogen and Helium ion) dominated 'polar wind' and heavier ion outflows composed mainly of oxygen ions associated with auroral activity (Yau and André, 1997) that is of most interest here. The estimated integrated oxygen ion outflows are of the order $10^{24}$ s$^{-1}$ for low geomagnetic activity and $10^{26}$ s$^{-1}$ for high activity. For the Earth, the solar wind does not interact directly with the ionosphere because of the presence of a planetary scale magnetic field. Nevertheless, the solar wind does interact with the Earth's magnetosphere, primarily through magnetic field reconnection. This allows solar wind momentum and energy to be coupled into the magnetosphere, and affect the high latitude ionosphere, mainly in the auroral zone and cusp region (e.g. see Moore et al., 2010 and references therein). The latter is the more direct point of entry for solar wind-related energy fluxes, be they particle or electromagnetic energy fluxes. This was explored by Strangeway et al. (2005) using data from the Fast Auroral Snapshot (FAST) Small Explorer. FAST acquired data from a high inclination elliptical orbit with apogee around 4000 km altitude. Strangeway et al. (2005) derived scaling laws that related Poynting flux (electromagnetic energy flux) and precipitating auroral electron fluxes (particle energy) to ion outflows. This analysis was extended by Brambles et al. (2011) to include the effects of Alfvén waves, which are an additional source of electromagnetic energy flux. The connections explored by Strangeway et al. (2005) and Brambles et al. (2011) are illustrated in Figure 13. Moore and Horwitz (2007) also explored these connections with numerical simulations as a means of understanding and interpreting the global picture of the various processes at work that lead to the heavy ion energization and escape.

The upper part of Figure 13 shows parameters measured above the ionosphere (in green), as well as the type of connection: correlated (magenta arrows) or possibly causal (open blue arrows). As an example of a possible causal relation, Alfvén waves could contribute to the Poynting flux into the ionosphere (left-hand path) or to enhanced precipitating electron fluxes (right-path). It is known that some form of heating is required for upwelling ions to escape (the escape energy for oxygen ions at the Earth is ~ 10 eV, at least an order of magnitude larger than the temperature of upwelling ions). Since the Poynting flux is carried by field-aligned currents (FACs), it is also possible that the FACs could contribute to heating via a current driven instability, while the precipitating electrons may also be a source of free energy for wave instabilities.

The lower half of Figure 13 shows how the incoming energy fluxes can result in outflows. For the left-hand path the electromagnetic energy results in Joule dissipation that heats the ions in the ionosphere, increasing the scale-height, so that more ions upwell to altitudes were waves can heat the ions, allowing them to escape. The right-path corresponds to electron precipitation that increases the ionospheric electron temperature. The resultant ambipolar electric field also increases the ionospheric scale height. Moore and Horwitz (2007) explored these connections, using numerical simulations to understand and interpret the global picture of the various processes at work that lead to the heavy ion energization and escape.

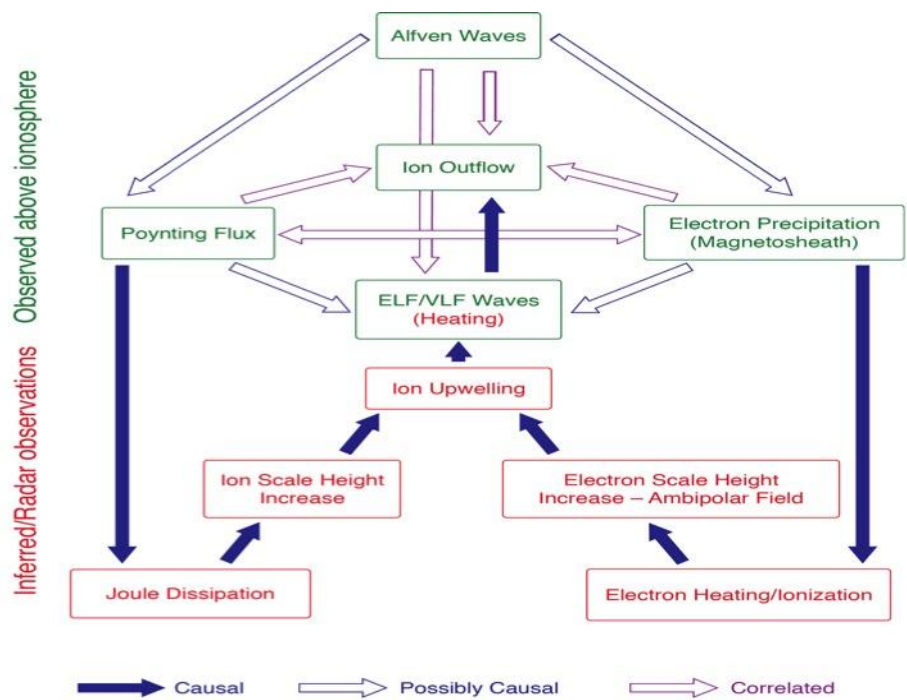

*Figure 13. Connections between input energy fluxes and outflowing ions as explored by in situ measurements above the ionosphere (after Strangeway et al. 2005, with the addition of Alfvén waves as an energy source).*

Much like the solar wind interaction with the unmagnetized planets, the ultimate energy source to drive the ionospheric outflows at the Earth is the solar wind. But what is less clear is how efficiently that energy is coupled to the ionosphere. The planetary magnetic field inhibits any direct interaction with the ionosphere, but also makes the planetary obstacle much larger, with an effective radius of the order 10 Earth radii. At the same time, the amount of energy available to drive outflows depends on the highly variable reconnection efficiency. The Earth's magnetic field also results in any energy from the solar wind being mainly directed to the polar ionosphere. Whether or not the outflowing ions ultimately escape the magnetosphere depends on their transport through the magnetotail lobes. The ions could escape downtail, or be convected into the plasmasheet where they are then transported sunward. Kistler et al. (2010) have reported escaping fluxes of the order $10^{24}$ $s^{-1}$ far downtail, and they argue that most of the outflowing ions are transported to the plasmasheet. But even then some of the ions could escape through transport to the dayside magnetopause. Fuselier et al. (2017) has shown that ions of ionospheric origin are transported to the dayside magnetopause, although the mass density of these ions is a small fraction of the magnetosheath plasma mass density.

The question for planetary atmospheric erosion through ionospheric escape is how do the ionospheric rates for the magnetized and unmagnetized planets vary as a function of the drivers, such as the solar wind dynamic pressure and solar EUV, and what role does a planetary scale magnetic field have in enhancing or inhibiting ionospheric outflow and loss? It may be purely coincidence that planetary outflow rates for Venus, the Earth and Mars are of the same order of magnitude. But determining if this is the case requires developing a detailed understanding of the various processes involved, and including such processes in models that allow us to explore parameter regimes including those corresponding to the early Solar system (see Section 5.1 and 5.2).

**4.1.2 Space Weather Effects on the Chemistry of Ionosphere-Thermosphere System.**

SEP events represent protons and heavy ions with energies from MeV to tens of MeV for gradual events lasting for days and from tens of MeV to several hundred MeV for impulsive events lasting for hours. These types of events are often associated with large solar flares and CMEs. At Earth, similarly energetic particles occupy the radiation belts, which are perturbed (both enhanced and depleted) as a part of the related geomagnetic activity. The Energetic Electron Precipitation (EEP) events are divided into three categories based on their origin of precipitation and energy: a) low energy electrons (LEE, < 30 KeV), b) medium energy electrons (MEE, 30–300 KeV), and c) high energy electrons (HEE, 300 KeV to several MeV). The LEEs reside in the magnetosphere, whereas MEE and HEE are trapped in the outer Van Allen radiation belt, from where they precipitate into the atmosphere during geomagnetic storms and sub-storms (*Sinnhuber et al.*, 2012). The LEEs affect the atmospheric ion chemistry > 90 km, the MEEs affect the atmospheric ion chemistry between 70 and 90 km (*Egorova et al.*, 2011), and the HEEs affect atmospheric chemistry < 70 km at middle and high latitudes (*Mironova et al.*, 2015). Finally, the Galactic Cosmic Rays (GCRs) originate from outside of our Solar system and consist of high-energy protons with energies ranging from ~1 MeV up to ~$10^{15}$ MeV (*Mironova et al.*, 2015). They are the main source of ionization between 3 km and 40 km. Due to geomagnetic field cutoffs that reduce the access of the lower energy galactic cosmic rays to low-to-mid latitudes, the maximum ionization occurs at around 10 km at the poles and gradually decreases towards the equator (*Usoskin et al.*, 2010). These cutoffs also affect the access of solar energetic particles to the atmosphere. The ionization rate by GCRs is about three orders of magnitude compared to the rest of the EPP events.

EPP events affect the middle-to-upper atmosphere at middle-to-high latitudes by modulating the local atmospheric ion chemistry by increasing the local ionization rate (*Wissing and Kallenrode*, 2009). EPPs produce odd hydrogen ($HO_x$ = H, OH, $HO_2$), odd nitrogen ($NO_x$ = N, NO, $NO_2$) as well as chlorine compounds, which destroy the upper stratospheric and mesospheric ozone ($O_3$) through a series of catalytic chemical reactions. These processes have been extensively studied using observations, theoretical approaches and modeling, e.g., (*Rusch*

*et al.,* 1981; *Solomon et al.,* 1981; *Jackman et al.,* 2005, 2006, 2011; *Verronen et al.,* 2006, 2007, 2011; *Seppala et al.,* 2006; *Damiani et al.,* 2008, 2010; *Verkhoglyadova et al.,* 2015, 2016). The lifecycle of $HO_x$ is very short (minutes to hours) in the middle atmosphere compared to the $NO_x$ lifecycle (months to years). This has further repercussions on the Earth's thermal properties that in turn influence global circulation properties, as we detail below.

In the mesosphere, below 80 km, $HO_x$ species can be formed through positively charged $O_2^+$ ions that are produced during EPP. These ions interact with water molecules via $O_4^+$ to create heavier hydrated $O(H_2O)$ ions, which eventually recombine with an electron to convert one water molecule into two odd hydrogen species (OH and H) (*Rusch et al.,* 1981) that destroy $O_3$ via: $OH + O_3 \rightarrow HO_2 + O_2$. Aside from ion reactions, $HO_x$ compounds are also produced via water vapor photolysis ($H_2O + h\nu \rightarrow H + HO$) and oxidation ($H_2O + O(^1D) \rightarrow 2OH$). Dissociation and dissociative ionization of $N_2$ during EPP events, also produce $NO_x$ species via a series of ion chemistry reactions. Firstly: $N_2^+ + O \rightarrow NO^+ + N(^2D)$. The positive $NO^+$ then reacts with an electron to produce ground state nitrogen and atomic oxygen via: $NO^+ + e^- \rightarrow N(^4S) + O$, as well as an excited nitrogen atom and molecular oxygen via: $NO^+ + e^- \rightarrow N(^2D) + O$, which further produce $NO_x$ species via: $N(^2D) + O_2 \rightarrow NO + O$. Mesospheric $O_3$ is primarily destroyed via: $NO + O_3 \rightarrow NO_2 + O_2$, although $O_3$ depletion via $NO_x$ at mesospheric altitudes is not as effective as via $HO_x$. Contrary to $HO_x$, the longer-lived, mesospherically-produced $NO_x$ species can descend into the upper stratosphere (*Solomon et al., 1982; Russell et al.,* 1984; *Randall et al.,* 2007) through the polar vortex during high latitude winter (in the absence of sunlight, when the $NO_x$ lifetime is particularly long), destroying upper stratospheric $O_3$ during spring via: $NO + O_3 \rightarrow NO_2 + O_2$. Meteorologically, the depletion of mesospheric $O_3$ and $H_2O$ vapor during EPP events reduces the local heating rate that alters the thermal wind balance, which changes the filtering of mesospheric gravity waves. This causes weaker air ascent and a reduced adiabatic cooling. Evidently, EPP events influence the background atmospheric conditions, which are directly related to numerous atmospheric phenomena such as, mesospheric cloud formation (during summer solstice), circulation, and dynamics.

In the upper stratosphere, only very high-energy EPP events directly produce $HO_x$ and $NO_x$ species to catalytically destroy $O_3$. In the middle and lower stratosphere, where direct SEP production of $HO_x$ and $NO_x$ species is rare, the primary source of $O_3$ depletion proceeds mainly via $NO_x$ species that are dynamically transported downwards from the mesosphere lower thermosphere region. The barrier to mixing at the edge of the polar vortex acts as a containment vessel for $NO_x$ species. During the season of the polar vortex, cold temperatures favor heterogeneous chemical reactions on the surface of polar stratospheric clouds, which are extremely important in ozone depletion during winter. One of the most important chemical pathways through which ozone is destroyed inside the polar vortex is the liberation of highly reactive chlorine and bromine forms from their reservoir species via: $ClONO_2 + HCl \rightarrow Cl_2 + HNO_3$. While chlorine is being released into the atmosphere, the reactive odd nitrogen species ($NO_x$) are locked up as non-reactive $HNO_3$ and cannot deactivate chlorine. Thus, the chlorine is free to destroy ozone at an extremely efficient rate via the $Cl_x$ catalytic cycles. After the polar vortex starts dissipating at the beginning of spring, in the presence of sunlight, the unreactive nitrogen reservoir species (e.g., $HNO_3$, $N_2O_5$) are transformed into NOx which is released into mid and lower latitudes in the stratosphere destroying the middle and lower stratospheric $O_3$ via the catalytic $NO_x$ cycles. The *World Meterological Organization* (WMO) (2014) assessment provides an in-depth overview of processes affecting ozone in Earth's atmosphere. Since $O_3$ regulates the vertical thermal structure of the stratosphere globally, fluctuations in its concentration induce temperature gradients, which alter the zonal, meridional, and vertical winds that drive global circulation and dynamics. Now that we have established the link between EPP events and the physical-chemical pathways through which they could affect atmospheric composition and thermodynamics, we provide a quantitative description of their impact in the terrestrial atmosphere over short and long timescales.

In the mesosphere, satellite observations (*Jackman et al.*, 2005) and model simulations (*Jackman et al.*, 2008) showed >30% decrease in $O_3$ in the polar cap regions (> 60° geomagnetic latitude) during the October–November 2003, during strong solar proton events (SPEs). *Andersson et al.* (2014), using satellite observations, showed as high as 90% mesospheric $O_3$ depletion between 60 and 80 km at geomagnetic latitudes 55–65° N/S, during EEP events. In the stratosphere, at high latitudes, modeling studies showed 10-30% $O_3$ reduction due to SPEs (*Semeniuk et al.*, 2011) and EEP events (*Rozanov et al.*, 2005). It was found in the late 20$^{th}$ century that $O_3$ was the primary driver of the observed poleward shift of the summer tropospheric jet, which has been linked to changes in tropospheric and surface temperatures, clouds and cloud radiative effects, and precipitation at both middle and low latitudes (*Previdi and Polvani*, 2014). *Rozanov et al.* (2005) found a 2.0 K cooling in the polar middle stratosphere in response to energetic electron precipitation. *Calisto et al.* (2012) simulated 3.0 K cooling at 60 km in a possible SPE Carrington-like scenario. EPP events introduce temperature gradients at stratospheric altitudes between the equator and the poles in both hemispheres, which cause acceleration of the meridional winds. To achieve thermal wind balance, the zonal winds accelerate (*Limpasuvan et al.*, 2005) by up to 5.0 m/s (~ 20% of the background wind) (*Calisto et al.*, 2012). Stronger zonal winds lead to a more stable polar vortex (*Seppälä et al.*, 2013) (defined as a large-scale low-pressure area that rotates counter-clockwise in the NH and clockwise in the Southern hemisphere (SH) with its base in the upper troposphere extending up to the upper stratosphere). The polar vortex vertically transports $HO_x$ and $NO_x$ and other chemical species (including greenhouse gases such as, water vapor) and affects the mesospheric (*Baumgaertner et al.*, 2011) and tropospheric (*Limpasuvan et al.*, 2005) wind speed patterns.

Although all the studies mentioned above provide an example of how EPP events could influence the terrestrial atmosphere over short-term timescales (days to months), recent studies have started investigating long-term effects of the EPP events. *Andersson et al*. (2014) showed that the integrated effect of short-term events occurring in high frequency leads to long-term effects after observing 34% mesospheric $O_3$ variations over a solar cycle between 2002 and 2012. They identified the continuous LEE precipitation as the missing link in the Sun-Climate system. Whether the longer-term trend of frequency, strength, and scale of temporary changes have any impact on atmospheric composition and thermodynamics still remains a question under investigation (*NRC*, 2012). Future detailed analyses are required in order to define the coupling processes through which EPP events may influence the terrestrial atmosphere over short and long timescales. Efforts towards this investigation are currently being made by the Intergovernmental Panel on Climate Change (IPCC), which has recognized the possible role of EPPs in global atmospheric variability, and efforts are already underway to include detailed EPP-ozone ion chemistry in the next Coupled Model Intercomparison Project Assessment Report Phase 6 (CMIP6) (*Matthes et al.,* 2016). EPP is a very important factor in the dynamics of Earth's middle atmosphere and is known to induce thermodynamic effects there. We suggest that the particle radiation environment can also influence the dynamics of exoplanetary atmospheres, and ultimately can affect atmospheric escape, hence the evolution of an exoplanetary atmosphere and its long-term variability. There is a strong need to develop pathways to estimate EPP input based on specific stellar environment and extend our knowledge of stellar flares and CME-like eruptions. To estimate EPP effects we need to adapt an atmospheric ionization model to exoplanetary middle atmospheres by incorporating flexible atmospheric chemistry modules suitable for target exoplanets. Establishing such dynamical exoplanetary atmosphere models will aid our understanding of potential biogenic zones and habitability of exoplanets.

For the past 16 years (and counting), NASA's TIMED satellite has been observing Earth's thermosphere and mesosphere, offering unique, unprecedented views of the response of the atmosphere to space weather events. Of particular interest has been the observations of infrared radiative cooling of the atmosphere by the nitric oxide (NO) and carbon dioxide ($CO_2$) molecules by the SABER instrument (*Mlynczak*, 1996, 1997; *Russell et al.*, 1999). Shortly after the launch of the TIMED mission in December 2001, SABER observed the first of many geomagnetic storm events in April, 2002. *Mlynczak et al.* (2003, 2005, 2007, 2010, 2015, 2016) demonstrated the role of infrared radiation from NO as a "natural thermostat," effectively serving as the conduit by which geomagnetic storm energy is lost from the atmosphere, allowing it to return to its pre-storm state in a matter of a few days. *Mlynczak et al.* (2017) have shown that for a number of strong storms observed by SABER, approximately 2/3 of the storm energy is radiated in the mid-infrared band by NO molecules, with the other 1/3 radiated by $CO_2$ molecules. The strongest of these infrared, storm-generated signals approach 3 terawatts (globally integrated total). Around younger, more active stars, the total infrared signal may be significantly higher. The infrared signatures of NO, $CO_2$ (and also hydroxyl, OH, and $O_2(^1D)$, electronically excited state of the $O_2$ molecule) are essentially "beacons of life" from exoplanets around young stars that may be readily observed on future space-based and ground-based telescopes (Airapetian et al. 2017b). Observations of these infrared emissions confirm the existence of nitrogen, carbon, oxygen (both atomic and molecular), and hydrogen within an exoplanet's atmosphere. The relative abundances of these in an exoplanet could be compared to those on Earth, providing powerful evidence for habitability and the probability of life.

**4.2. Modern Venus and Mars**

One of the striking phenomena observed by spacecraft orbiting Venus and Mars is the apparently continuous escape of atmospheric ions. In particular, oxygen ions, which are often taken to be a proxy for water, are observed with planetary escape rates of the order $10^{24}$ - $10^{25}$ s$^{-1}$ (Brain et al., 2017), similar to the oxygen ion escape rates for Earth. This is of particular interest because both of these planets lack a global-scale magnetic field, and so the solar wind can interact directly with the planetary upper atmosphere and ionosphere. It has been argued that one of the primary controlling factors for the ion loss rates in the direct solar-wind - ionosphere interaction is the solar wind dynamic pressure (Strangeway et al., 2010), and this has been recently verified through MAVEN observations at Mars (Dubinin et al., 2017). Another controlling factor appears to be solar EUV flux (e.g., Lundin et al., 2007; Dubinin et al., 2017).

**4.2.1 Space Weather at Modern Venus**.

Of all other planets, Venus is in many ways the most Earth-like, with comparable surface gravity, a substantial atmosphere, and once orbited within the habitable zone of its star (Way et al. 2016). The inner edge of the habitable zone moved 4% farther out than Venus's orbital distance at 0.72 AU about 1 Gyr ago. This ignited conditions for a runaway greenhouse on Venus, when the planet lost its oceans within a few tens of millions of years via photodissociation and subsequent escape of hydrogen from the planet at the time when the Sun was 8% less bright than today (Kasting 1988; Kasting et al. 1993; Kasting et al. 2014). This specifies the conditions for the inner edges of habitable zones for Earth-like planets (Kopparapu et al. 2013). However, Venus differs from Earth in another important respect: it has essentially no intrinsic magnetic dipole field (Smith et al. 1965, Phillips and McComas 1991), and thus the fundamental way it interacts with the Sun is very different from Earth's interaction. Without a magnetic field, the primary obstacle to the solar wind at Venus is the thick, conductive ionosphere. Interplanetary magnetic field (IMF) advecting with the solar wind penetrates the ionosphere and induces a current within, resulting in the generation of an induced magnetosphere (see, e.g. Zhang et al. 2008, and references therein). This induced magnetic field is an obstacle to the fast solar wind, and thus a bow shock and magnetosheath are generated upstream of the ionosphere (Russell et al., 1979). The induced Venusian magnetosphere is an order of magnitude smaller than the terrestrial intrinsic magnetosphere, with a sub-solar stand-off distance of the bow shock of ~1.5 Venus radii ($R_V$ = 6,052 km) (Slavin et al., 1980), as opposed to the ~15 Earth radii ($R_E$ = 6,371 km) (Fairfield, 1971). Inside the bow shock is a turbulent region of shocked solar wind, which for consistency with other planets we shall refer to as the Magnetosheath (but is sometimes referred to as an "ionosheath" in the literature). The boundary between the magnetosheath and ionosphere is specified by the pressure balance between the incident solar wind pressure and the ionospheric (thermal plasma plus magnetic) pressure and is typically referred to as the "ionopause". At solar maximum, the ionospheric pressure typically dominates and fully stands off the solar wind, in which case the ionosphere is largely "unmagnetized" (unlike Mars). However, when solar EUV flux is low or the solar wind dynamic pressure high, the interplanetary magnetic field (IMF) is able to penetrate and enter the ionosphere which becomes "magnetized" (e.g. Angsmann et al. 2011).

Observations (e.g. Strangeway and Crawford, 1995, Crawford et al., 1998) and simulations (Omidi et al., 2017) of the Venusian interaction with the solar wind have revealed that, as at Earth, upstream from the Venusian bow shock lies a turbulent region magnetically connected to the bow shock known as the foreshock. Observations revealed the Venusian foreshock to be a miniaturized version of Earth's, full of the same wave-generating and particle-energizing plasma phenomena (e.g. Collinson et al. 2014). Without a magnetic field for protection, it has been hypothesized that such foreshock phenomena can directly impact the upper ionosphere.

The effects of transient space weather events such as Interplanetary Coronal Mass Ejections (ICMEs), flares, and Corotating Interaction Regions (CIRs) are much different at Venus than at Earth. Our understanding of the impact of space weather at Venus can be broken down into a few categories related to the effects of enhanced solar wind pressure, increased planetary ion energization and outflow, and effects of related particle precipitation into the atmosphere.

The responses of the Venusian ionosphere to perturbations by the solar wind have been an active topic of research since the Mariner Probes. Between the Mariner 5 and Mariner 10 flybys of Venus the magnetopause was observed to be significantly compressed to 250 km from the planetary surface versus 500 km as usually observed during quite conditions, which was thought to be a result of changes in solar wind conditions (Bauer and Hartle, 1974). The ionospheric magnetization events observed on the Pioneer Venus Orbiter (PVO) (e.g. Luhmann and Cravens, 1991) occurred during periods following ICME impact when post-shock enhancements of solar wind dynamic pressure pass by. Theoretical analysis of the magnetization showed that when the ionopause altitude is lowered to the vicinity of the exobase, collisional diffusion of the sheath field into the ionosphere occurs. More recently Vech et al. (2015) used Venus Express observations to investigate the impact of 42 strong interplanetary coronal mass ejections on Venus. Most of these CMEs drove interplanetary shocks, which were found to significantly enhance the strength of the magnetic field in the magnetosheath. The bow shock position was found to be significantly expanded during passages of magnetic clouds. Both Collinson et al. (2015) and Vech et al. (2015) found that the dayside ionopause was not significantly depressed by ICME impact, although Vech et al. (2015) found that the nightside ionosphere was significantly disturbed. This observation was likely related to the known lower limit on ionopause compression of ~250 km (subsolar) observed on PVO (e.g. Phillips et al., 1982; Luhmann et al., 1986). Regardless of the solar wind pressure the ionosphere production continues unabated, thus maintaining a basic profile by virtue of photoionization and recombination in the atmosphere.

Based on PVO data neutral mass spectrometer (ONMS), Luhmann et al., (2007) suggested that the passage of an ICME also enhanced the outflow of $O^+$ suprathermal ions by as much as 2 orders of magnitude. On the other hand, case studies by Luhmann et al., (2008) and McEnulty et al., (2010) based on Venus Express observations suggested escaping energetic pickup ions were accelerated to higher energies at lower altitudes during ICMEs, but with no apparent increases in fluxes than at the energies and in the cases they studied. Edberg et al. (2011) made a statistical study of the impact of 147 CIRs and ICMEs, including the lower energy planetary ions, and found escape rates were enhanced by a factor of 1.9 during their passage. More recently, Collinson et al (2015) studied the impact of a slow IMCE on Venus (convecting with the solar wind with no interplanetary shock), concluding that even weak ICMEs with no driving shocks can increase atmospheric loss rates at Venus and suggesting that the orientation of the Interplanetary Magnetic Field (IMF) may be a factor in atmospheric escape rates. One important point to note about Venus is that, similar to Earth's case, energization via solar wind interaction-related processes is required to bring heavy ions like oxygen up to at least the escape velocity of ~10 km/s. While polarization electric fields and other mechanisms such as charge exchange with solar wind protons may affect its hydrogen budget, the solar wind interaction may be a key factor in determining Venus's oxygen (and therefore possibly water) loss to space.

Other studies of space weather effects at Venus have focused on atmospheric ionization and chemistry effects from the impact of high-energy particles, including those that come from interstellar space, and those accelerated in association with ICMEs. Nordheim et al., (2015) simulated the interaction of both galactic cosmic rays and solar energetic particles (SEPs) with the Venusian atmosphere. Such particles are the main ionization source in Venus' atmosphere below ~ 100 km (Borucki et al., 1982, Nordheim et al. 2015) with a peak in the ion concentration occurring at 60-65 km and ion production rates ranging from $1-10^5$ $cm^{-3}s^{-1}$ depending on the solar cycle phase. The latter study (Nordheim et al. 2015) discussed the chemical formation and propagation mechanisms for atmospheric ions induced by high energy particles. They suggested that the primary ions formed in Venus' thick, $CO_2$-dominated atmosphere are: $CO_2^+$, $CO^+$ and $O^+$, which are quickly converted into secondary ions and ion clusters. Diffuse auroral UV emission was detected on the nightside of Venus by PVO's ultraviolet spectrometer in association with the apparent passage of an ICME solar wind pressure enhancement (Phillips et al., 1986). More recently, ground-based observations of Venus' oxygen green line nightglow emissions by Gray et al., (2014) found that they were associated with the passage of ICMEs or the related SEP influxes, and are probably a result of enhanced solar wind electron precipitation.

Many details of the consequences of space weather at Venus still remain to be explored. For example, effects of particle precipitation on cloud nucleation has been considered for Earth but not for Venus, where the external energetic particle fluxes have relatively global access to the atmosphere. Similarly, the effects on atmospheric chemistry of the deep ionization related to both solar flare XUV enhancements and the energetic particles is not understood beyond the model calculations described above. In addition, it has been suggested that sputtering of the atmosphere by precipitating energetic ions at Venus can enhance the escape of heavy neutral species (Curry et al. 2015b). Yet sputtering under normal circumstances today is expected to be fairly weak in terms of its contribution to the exosphere. Whether this situation changes during major space weather events is unknown, in part because of the difficulty of detecting sputtering effects in the presence of all of the other processes (e.g. hot oxygen corona formation by dissociative recombination). It may be timely to revisit the archives of Venus mission data and other observations toward putting together a more comprehensive picture of space weather storms at Venus. One other consideration is that space weather impacts cannot be considered in isolation of other events that may occur at modern Venus, such as volcanic eruptions, comet impacts (Gillmann et al. 2016), or even normal weather including storms and lightning (Hart et al. 2018). As at Earth, there are many couplings between the lower and upper atmospheres that can affect our interpretations and the outcome(s) of space weather events.

### 4.2.2 Space Weather Effects at Modern Mars.

With the availability of results from the MGS, Mars Express, and MAVEN missions, Mars is now known to have a particularly complicated 'hybrid' solar wind interaction that exhibits both some characteristics of an induced magnetosphere like that of Venus, and magnetospheric phenomena as well. Its distributed, generally small scale crustal magnetic fields are strongest on one face of the southern hemisphere, making it an asymmetric magnetosphere as well. The outer part of the interaction, consists of the bow shock and magnetosheath and appears almost Venus-like in scale and morphology. The magnetospheric "obstacle" of Mars, including its magnetotail, is a combination of induced magnetosphere-like draped fields, and "closed" and "open" local crustal field topologies. They are in a constant state of reconfiguration driven by external variable solar wind conditions and the surface nonuniform crustal fields (Dubinin et al. 2018). Different types of planetary ion energization and outflows are associated with all these structures and their temporal variations that produce atmosphere losses with both Venus-like and Earth-like characteristics. It is in this complex setting that researchers have endeavored to understand Mars' space weather responses in some detail.

As at the Earth and Venus, photoionization dominates the dayside ionosphere production. Space weather introduces SEP ionization mainly above ~70 km, while GCRs produce enhancements below this altitude. Energetic particle-induced air shower events on modern Mars were simulated by *Gurtner et al.* (2005) using the GEometry ANd Tracking Version 4 (GEANT4) model, by Norman et al. (2014) using NAIRAS and by Gronoff et al. (2015) using the Planetocosmics model. The main responses of positive and negative ions on modern Mars has been reviewed (e.g. Haider et al., 2007). The entry of cosmic rays leads to strong production of $CO_2^+$ ions and dissociated products. Charge transfer reactions subsequently lead to formation of $O_2^+$ at high altitudes and more complex chemistry at lower altitudes due to electron capture and multi-species chemical reactions. The subsequent ion-neutral chemistry has been compared with the Earth's D-region (see e.g. *Molina-Cuberos et al., 2001*). Electron capture can lead to $O^-$ and $O_2^-$ which go on to form $CO^-_n$ and $NO^-_n$. The study by Molina-Cuberos et al. (2001) applied a column model with ion chemistry which suggested that hydrated ion clusters would dominate both the positive and negative ion budgets. More recently, 3D global models of the Martian atmosphere including neutral and ion chemical networks with coupled global dynamics and radiative transfer have been developed (e.g. Gonzalez-Galindo et al., 2013; Bougher et al., 2015), although these models currently lack space weather impacts.

Again, as at Earth and Venus, ion outflows related to space weather events are of interest. Estimates of average heavy ion (e.g. $O_2^+$, $O^+$) atmospheric loss rates planet-wide at Mars range from about $10^{24}$ s$^{-1}$ at solar minimum up to about $(1-3) \times 10^{25}$ s$^{-1}$ at solar maximum as estimated from Mars Express data and MAVEN, Mars Atmosphere and Volatile EvolutioN, data (e.g. Lundin et al., 2013; Nilsson et al., 2011; Groeller et al., 2014; Brain et al., 2015; Dong et al., 2015a). These rates are similar to those quoted above for Earth and Venus. But at Mars under normal solar and solar wind conditions, the escape of species like oxygen is instead dominated by photochemical loss of neutral atoms, not ion loss. Space weather may change this situation. Futaana et al. (2008) suggested that heavy ion outflow on Mars measured by Mars Express increased by about one order of magnitude during the intense flare event in December 2006. Recent results (Rahmati et al., 2017) from the MAVEN orbiter suggested an order of magnitude change in hydrogen escape rates with changing Martian season but with less varied response over season for oxygen (Heavens et al. 2018). Jakosky et al. (2015) analyzed MAVEN data for an Interplanetary Coronal Mass Ejection (ICME) event in March 2015 and used global modeling of the solar wind interaction during the event to infer that ion escape rates where globally enhanced by an order of magnitude during passage of the high dynamic pressure phase of the ICME (also see Curry et al, 2015; Dong et al., 2015b; Ma et al., 2017, Luhmann et al., 2017). Another, even stronger event affected Mars in September 2017 (Guo et al. 2018). As for the March 2015 event, enhanced escape of planetary ions is indicated, but other atmospheric responses were also observed on this occasion, including heating and expansion of the upper atmosphere by the related preceding solar flare, and a global UV aurora indicating an almost planet-wide influx of energetic particles. Earlier diffuse auroras had also been observed by the MAVEN ultraviolet spectrometer in association with the arrival of solar energetic electrons, suggesting this again occurred but with greater fluxes and hence intensities (Duru et al. 2015). In addition, the magnetospheric topology was inferred to become more 'open' to interplanetary space as the ICME passed, the result of the combination of more deeply penetrating draped magnetic fields into the dayside ionosphere and enhancements in reconnection between the crustal fields and the interplanetary field. Note that the former effect is Venus-like while the latter is Earth-like in some respects. The solar energetic particles were so intense at the higher energies (>100 MeV) that they were even detected on the ground by the RAD instrument on MSL, which also observed a Forbush Decrease in the galactic cosmic rays from apparent deflection of them away from Mars by the ICME fields. The September

2017 event provided a wealth of Mars space weather response studies that are documented in the literature in part (Gup et al. 2018) but still ongoing.

It is considered that modern Mars and Venus present laboratories to better understand the interactions of space weather events including ICMEs and high energy particles with $CO_2$ - dominated planetary atmospheres for conditions ranging from the cold, thin, dusty atmosphere of Mars to the hot, thick atmosphere of Venus. In particular, the comparisons of their space weather responses to those at Earth are critical for understanding exoplanetary $CO_2$ atmospheres with implications for e.g. abiotic $O_2$ sources and sinks in a biosignature context. Further comparative terrestrial planet studies uniquely provide the "ground truth" needed to interpret the growing observational information becoming available for exoplanetary systems. Future investigations would be wise to take the activity of the host star into account in developing the overall descriptions of the local planetary atmospheres.

**5.0 Space Weather Impact on (Exo)planetary Systems: Atmospheric Loss**

Extreme space weather conditions in the form of frequent energetic flares, CMEs and SEPs can significantly impact the upper atmospheres of exoplanets around active G-K-M dwarf stars. In this section, we will describe the current understanding of the effects of stellar XUV driven emission and the impacts of dynamic pressure exerted by solar, stellar winds and associated CMEs on physical processes of atmospheric escape from early Earth, Mars, Venus and terrestrial type exoplanets including exoplanets around Proxima Centauri and TRAPPIST 1.

**5.1 Atmospheric Escape: Effect of Stellar XUV flux**

The enhanced XUV fluxes encountered by the early Earth as well as close-in exoplanets around active young G-K-M stars has significant consequences for atmospheric mass loss. The energy deposited by this flux through the absorption of XUV radiation can drive a number of thermal, nonthermal and chemical escape processes forming ionosphere and thermosphere. The thermnal escape process is driven by the tempetrature at the exobase, the atmospheric layer, where the particle mean free path is comparable to the pressure scale height. In this layer, high fast particles from the tail of Maxwellian distribution moving with outgoing velocity exceeding the plant's escape velocity will escape into space (Jean's escape, Jeans 1925). Its escape rate is controlled by the Jean's escape parameter, $\lambda_c$, represented by the ratio of gravitational energy to the mean particle's thermal energy. When $\lambda_c < 1.5$, the uncontrolled escape of all gas species known to as the atmospheric "blowoff" does not depend on the exospheric temperature (Öpik, 1963). However, in reality the pressure gradient due to excessive heating should contribute to the escape rate. Tian et al. (2008) has applied an 1D hydrodynamic thermosphere model to study the effects on high XUV fluxes (up to 20 times the current Sun's flux, $F_0$) from the young Sun on the early Earth. He found that at the XUV fluxes below $5.3F_0$, photoionization driven heating increases the exobase temperature to ~ 8,000K moving it to 7,700 km with the bulk velocity of > 10 m/s. However, above this critical XUV flux, very high Jeans escape at the exobase drives the upward flow producing adiabatic cooling, and thus reducing the exobase temperature. Lichtenegger et al. (2010) expanded Tian's model to calculate the atmospheric escape due to ionization and solar wind pick up of th exospheric gas to be on the order of $10^7$g/s, thus removing the Earth's atmosphere in 10 Myr. First fully hydrodynamic upper atmospheric models were developed for hydrogen-dominated atmospheres (Murray-Clay et al. 2009; Owen and Mohanty 2016; Odert et al. 2019). Recent hydrodynamic outflow model developed for an Earth-like planet ($N_2 - O_2$ dominated atmosphere) suggests that at XUV fluxes 60 times the current Sun's flux, the neutral atmosphere undergoes extreme hydrodynamic escape at the mass loss rate of 1.8 x $10^9$ g/s (Johnstone et al. 2019). This suggests that atmospheres of Earth-like planets around active stars should be depleted at the time scale of Myrs. More studies with multi-dimensional fully thermodynamic atmospheric models are required to study the initiation of hydrodynamic escape in atmospheres with various levels of $CO_2$ and NO tnat serve as the major atmospheric coolants (Mlynczak et al. 2005; Glocer et al. 2018).

In non-thermal escape processes, escaping particles acquire energy through energy from nonthermal sources. Fir example, this may occur when energetic ion $A^+$ collides with a clod neutral particle, M, through the charge exchange process $A^+$ (*energetic*) + M(*cold*) –> A(*energetic*) + $M^+$(*cold*), for example H/$H^+$ or $O^+$/O charge exchange processes.

Lighter species, such as hydrogen, tend to escape more easily through thermal escape, but strong XUV fluxes can also stimulate the loss of light and heavy ions through ionospheric outflow and exospheric pick-up by stellar winds (Glocer et al. 2012; Airapetian et al. 2017a; Kislyakova et al. 2014; Dong et. al. 2017; Lammer et al. 2008; 2018). In this process, the incident XUV photons ionize the atmospheric neutral particles yielding ions and electrons. The electrons generated by photoionization, known as photoelectrons, are much less gravitationally constrained then the much heavier ions, and in the absence of collisions would be largely free to escape to space. This is prevented, however, by a polarization electric field which serves to restrain the free escape of electrons while simultaneously enhancing the escape of ions. The role of photoelectrons, and the associated polarization electric, in enhancing ionospheric outflow has been extensively studied in the literature (see e.g., Tam et al.1995, 2008; Khazanov et al. 1997, Su et al. 1998), but the impact has only recently been evaluated for systems with high XUV fluxes relative to the current solar case.

The role of large XUV fluxes for generating enhanced ion escape for close-in planets was recently evaluated by Airapetian et al. (2017a). This study used a hydrodynamic model of the $O^+$ ion escape coupled to a kinetic model of electron transport, the Polar Wind Outflow Model (PWOM)-Superthermal Electron Transport (STET) code (see Glocer et al, 2017 for model details). Starting with the XUV flux of the Sun for average levels of activity, $F_0$, they evaluated the impact of enhancing the XUV by factors of 2, 5, 10, and 20 on the outflowing $O^+$ ion flux from an Earth-like planet with enhanced thermospheric temperature. The resulting escape particle mass fluxes are shown in Figure 14. While $N^+$ ions were not included into the model, its escape rate is expected to be similar to $O^+$ escape rate, because its mass differs from oxygen ions only by ~ 12%. Indeed, satellite measurements of ion escape from Earth's ionosphere during geomagnetic storms driven by XUV fluxes and wind induced particle precipitation suggest that the contribution of escape rate due to $N^+$ ions is comparable to $O^+$ ions although caution should be exercised when using these observational results for our analysis because the mass peaks of $O^+$ and $N^+$ are not well resolved (Yau et al. 2007). Future missions, including the recently proposed ESCAPE mission, should resolve the question of $N^+$ loss rate as a function of various space weather factors. Assuming $N^+/O^+ \sim 1$ during large geomagnetic storms driven by flare and CME events, the total ion mass loss rate scales as

$$M \text{ (in g/s)} \sim 1.6 \times 10^4 \, F_{XUV} \quad \text{(in erg cm}^{-2}\text{ s}^{-1}\text{)}$$

This relation represents the low bound for the ion escape rate, because it does not consider the ionospheric heating due to precipitating electrons during storms. Such additional heating swells the ionosphere, and thus enhances the escape rate. XUV and electron precipitation driven ion escape fluxes should be quite large for many close-in exoplanets around M dwarfs or young K stars as their XUV fluxes are 100-400 $F_0$ and can retain such high XUV emission fluxes over 1 Gyr. This will make it difficult for such planets to retain their atmospheres over the geological timescales unless considerable atmospheres are outgassed from the interior via volcanic outgassing during later phases of planetary evolution. In order to develop the complete picture of atmospheric escape from close-in exoplanets, these effects should be studied with future multi-dimensional hydrodynamic and kinetic models with inclusion of XUV and electron precipitation induced heating.

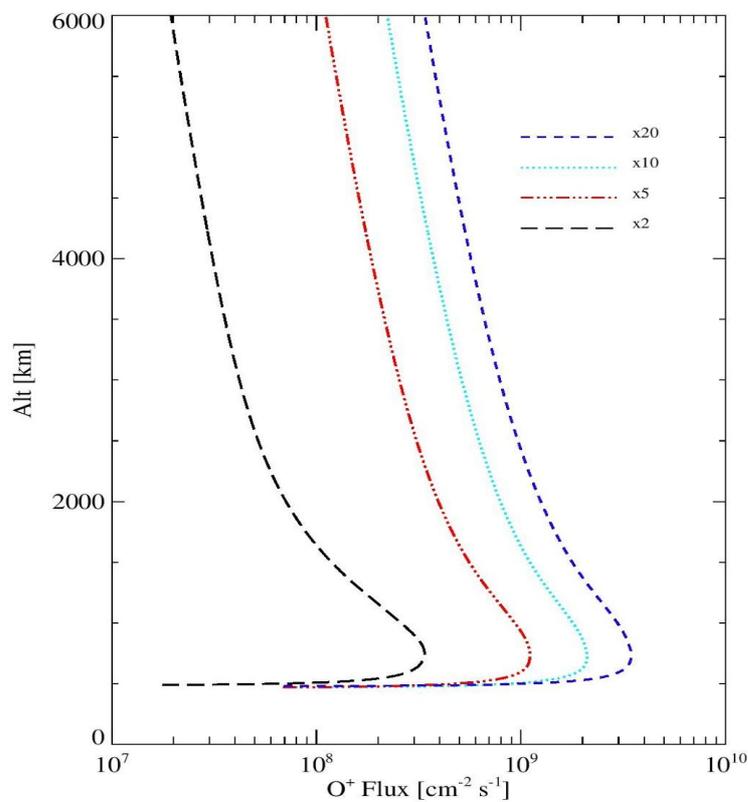

*Figure 14. The mass loss rate of oxygen ions from the Earth atmosphere due to XUV flux from the young Sun at $F_{XUV}=2\,F_0$ (long dash), $5\,F_0$ (dash-dot), $10\,F_0$ (dot), $20\,F_0$ (short dash) (from Airapetian et al. 2017a).*

**5.2. Impact of Space Weather on Atmospheric Chemistry of early Earth and Mars**

### 5.2.1. Dynamics of Early Atmospheres of Earth, Venus and Mars

The very first atmospheres ("proto-atmospheres") of Earth, Venus and Mars (excluding the initial silicate-containing envelope formed during accretion) consisted of a thick envelope of molecular hydrogen ($H_2$) with up to several hundreds of bars of surface pressure accreted directly from the protoplanetary disk. These proto-atmospheres were likely lost within a few million years after planetary formation due to rapid escape driven by strong EUV emissions by the active early Sun (e.g. Lammer et al. 2014; 2018). They were followed by a thick (with up to hundreds of bar surface pressure) steam atmosphere phase which arose due to volatile degassing and which lasted approximately from the magma ocean phase up to tens to hundreds of Myrs after formation of the crust (see e.g. Zahnle et al., 1988; Elkins-Tanton, 2012; Lammer et al. 2018; Nikolaou et al. 2019). This process is followed by formation of a secondary atmosphere of heavier gases, such as $N_2$-$CO_2$ rich environment as the planet cools and its surface solidifies (Elkins-Tanton 2008; Noack et al. 2014).

Comparative studies of the effect of the active young Sun on the atmospheric evolution of Earth, Venus and Mars were performed by Kulikov et al. 2006, Johnstone et al. 2018; 2019; Lammer et al. 2018; Tian et al. 2018, Airapetian et al. 2016; 2018, who emphasized that comparative atmospheric evolution could only be understood in the context of an evolving Sun. Specifically, flare driven XUV emission and the dense fast wind and associated frequent and associated CMEs from the young Sun could have been an important factor in setting habitability conditions on early terrestrial planets (Airapetian 2018). Its consequences on the magnetosphere and thermosphere of early Earth and Mars are the area of emerging research (Airapetian et al. 2016; 2018; Dong et al. 2017). As discussed in Section 3.2, the frequency of a Carrington-scale flare event from the young Sun at 0.7 Gyr was about 1 event per day that suggests that such events could have been accompanied by a super-CME of a comparable energy (Airapetian et al. 2016; Alvarado-Gomez et al. 2018). Recent study suggests that the reflection of CMEs from coronal holes of the young Sun deviated from radial path toward the equator, and thus increased the resulting frequency of "geoeffective" CMEs (with the southward directed z-component of the magnetic field, which is opposite to the Earth's magnetospheric field) that may hit the Earth and cause strong geomagnetic storms (Kay et al. 2017; Kay and Airapetian 2019).

As an energetic geoeffective Carrington-type CME propagating in the background of the solar wind from the young Sun (or an active star) moves toward the early Earth (or a young Earth-like planet), its dynamic pressure compresses and convects the planetary magnetospheric field inducing the convective electric field and associated ionospheric current. Thus, the shape of the Earth's magnetosphere is the result of the interaction with the dense and fast wind from the young Sun (Airapetian and Usmanov 2016). The distance of subsolar magnetosphere from the Earth's surface varies in times in response to the changing dynamic pressure from the solar wind and a CME. Its boundary known as the standoff distance is determined from the balance between the magnetic pressure of the Earth magnetosphere and the wind dynamic pressure. The larger the dynamic pressure, $\rho V_w^2$, the smaller is the standoff distance. The solar wind and a CME also compress the night-side magnetosphere and ignites magnetic reconnection at the night-side of the Earth's magnetosphere causing the magnetospheric storm as particles penetrate the polar regions of Earth. Also, the orientation of the magnetic field of the wind and CME as compared to that of the Earth's magnetospheric field controls the energy transfer from the wind to the planet, as it drives magnetic reconnection which in turn ignites particle acceleration and particle precipitation in the planetary ionosphere and thermosphere (e.g., Kivelson & Russell 1995; Tsurutani et al. 2006). This process is driven by electric fields that accelerate the ions against the neutrals resulting in current dissipation (Ohmic or Joule heating).

In addition, the particle precipitation in the upper atmosphere impacts the local ionization and modifies Joule heating processes (see e.g., Deng et al. 2011, with references therein) and atmospheric line excitation (i.e., auroral excitation, Schunk & Nagy 1980).

The simulations of the interaction of a Carrington-type CME with the early Earth's magnetosphere is shown in Figure 15. The left panel of Figure 15 shows that the magnetospheric standoff distance moves from ~ 10$R_E$ (right panel) to ~ 1.5 $R_E$ as the Carrington scale CME from the young Sun with the southward $B_z$ component of the magnetic field as a result of the combined effect of magnetic reconnection between two fields and the CME dynamic pressure (Airapetian et al. 2015; 2016; Panel b of Figure 15). The boundary of the magnetospheric open-closed field shifts to 36 degrees in latitude opening 70% of the Earth's magnetic field. The CME dynamic pressure drives large field aligned electric currents that dissipates in the ionosphere-thermosphere region (at ~ 110 km) of the Earth via ion-neutral frictional resistivity producing Joule heating rate reaching ~ 4 W/m$^2$. This is over 20 times larger than that observed during largest geomagnetic storms, and will thus increase the upper atmospheric temperature and ignite enhanced ionospheric outflow (Airapetian et al. 2017a). The models of associated atmospheric escape are in their infancy and require the extension of sophisticated tools to consider multi-fluid effects for consistent treatment of neutral species in the ionosphere-thermosphere dynamics developed for the current and early Earth, Mars and Titan (Smithro and Sojka 2005; Ridley et al. 2006; Tian et al. 2008; Glocer at al. 2012; Bell 2008; Johnstone et al. 2018).

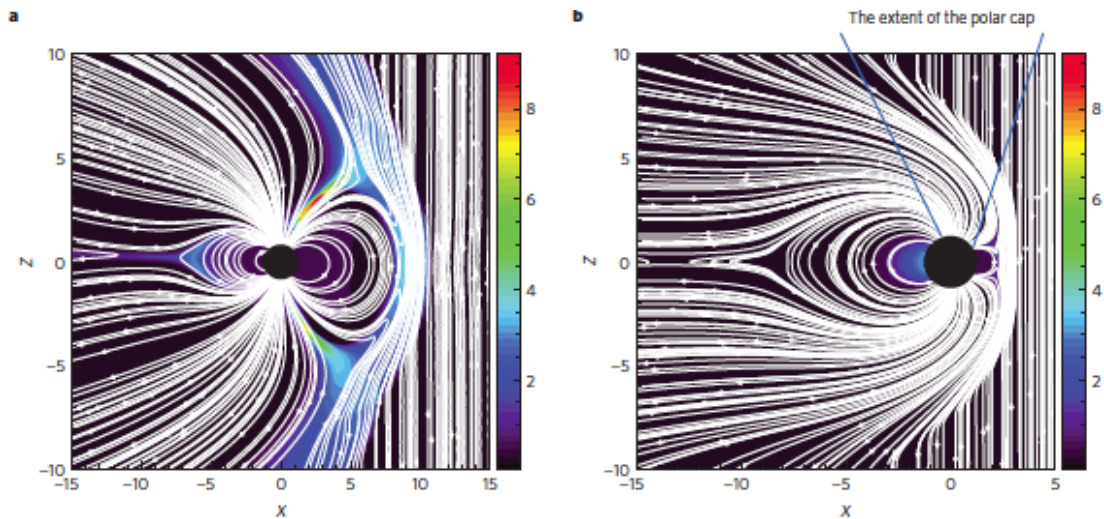

*Figure 15. The magnetic field lines and plasma pressure in the Earth's magnetosphere due to a CME event. a, Initial state. b, Final state. Magnetic field lines (white) and plasma pressure in nPa (color map). Axes represent distance from the Earth's center in units of Earth radius (Airapetian et al. 2016).*

**Early Earth**. Lammer et al. (2018) suggested that proto-Earth was formed via planetesimal accretion growing to 0.5-0.75 $M_{Earth}$ in about 10 Myr. The accretion phase is characterized by the formation of magma oceans surrounded with a nebular-captured $H_2$ - rich envelope (primary atmosphere) and delivery of volatiles via impactors formed in the outer solar system. High XUV driven from the young Sun heating via photoionization ignited hydrodynamic escape of the atmosphere (Erkaev et al. 2013; Owen and Mohanty 2016). As the steam atmosphere later collapses or erodes via escape processes, magma ocean will finally solidify, and thus promote the formation of Earth's oceans. Continued outgassing produced the secondary atmosphere could have been dominated by $H_2O$ and $CO_2$ with a minor amount of $N_2$ and traces of inert gases, hydrocarbons (e.g. $CH_4$), and sulfur-containing compounds or by more reduced atmosphere depending on the redox state of mantle (see e.g. Zahnle, 2007; Lammer et al. 2018; Schaefer and Fegley 2017). Numerous geochemical proxies exist which constrain the

Archaean atmosphere. For example, Rosing et al. (2010) suggested that magnetite siderite equilibria in banded iron minerals implied that atmospheric $CO_2$ concentrations did not exceed (~x3) the Present Atmospheric Level (PAL) on the early Earth, although this view is contested (Kasting et al., 2010). The evolution of $N_2$ is even more uncertain (Wordsworth, 2016; Lammer et al. 2018). Recent study of evolution of molecular nitrogen by Gebauer et al. (2019) suggests atmospheric pressure of early Archean Earth was less than 0.5 bar slowly rising in time due to diminishing effects of $N_2$ reduction via atmospheric escape and nitrogen fixation. Som et al. (2012) suggested that atmospheric pressure was less than ~ 2 bar during the late Archaean based upon an analysis of the size distribution of fossilized raindrop imprints. They revised this estimate to less than half of present levels in recent work (Som et al. 2016). Marty et al. (2013) argue that Archean $N_2$ levels were 0.5-1.1 bar, with 0.7 bars or less of $CO_2$. This is consistent with the evolutionary scenarios of $N_2$ atmospheric pressure at 2.7 Gyr (Gebauer et al. 2019). Such low atmospheric pressure makes it difficult to explain the warming of the Archean Earth via recently proposed scenarios (Goldblatt et al. 2009). To warm the atmosphere, a strong greenhouse gas is required. Airapetian et al. (2016) have recently suggested that a potent greenhouse gas, nitrous oxide ($N_2O$) can be efficiently formed via frequent energetic particle precipitation into the lower Archean atmosphere driven by SEP events from the young Sun (see also Section 6.5).

Oxygen isotope data in cherts could constrain paleo ocean temperatures (see e.g. discussion in Shields and Kasting (2007). Molybdenum (Anbar et al. 2007) data suggests early "whiffs" of atmospheric oxygen as a prelude to the Great Oxidation Rise which marked the end of the Archean (Lyons et al. 2014; Smit & Mezger 2017). Loss of mass independent fractionation of sulfur isotopes observed near the end of the Archaean could imply either a rise in atmospheric oxygen or/and a decrease in atmospheric methane (Zahnle et al. 2006). Finally, flow features preserved in ancient soil (paleosols) suggest the presence of liquid water despite the fainter early Sun - this issue is described in more detail in the habitability discussion in Section 6.3.

In summary, the Earth's atmosphere has undergone extremely strong and (in geological terms) rapid change both in composition and mass during its evolutionary history. More work is however required to constrain the detailed effects of space weather, especially the interaction of energetic XUV emission and high energy particles during the various phases of atmospheric evolution of our planet.

**Early Venus** - measurements of the Deuterium to Hydrogen (D/H) ratio taken from the clouds of Venus by the Pioneer Venus Mission (Donahue et al. 1982) suggest at least 150-fold enhancement compared with terrestrial water. This high fractionation of Deuterium suggests that the planet underwent considerable water loss during its evolution and that conditions on early Venus could have been habitable. Way et al. (2016) summarize possible scenarios of the water evolution on early Venus. They note that the early water inventory was likely lost within ~100 Myrs of years after formation due to atmospheric escape driven by XUV flux from the active young Sun (see also Chassefiere 1997 and Ramirez and Kaltenegger 2014). Since paleo estimates suggest the equivalent of 19-meter to 500-meter water covering the surface, this requires considerable water delivery during the late veneer (Donahue and Russell 1997). Wordsworth (2016) discusses possible mechanisms for the presently observed elevated amount of $N_2$ in Venus' atmosphere relative to that of Earth.

**Early Mars** - a wealth of geological proxies (e.g. geophysical flow features, clay deposition etc.) suggest that early Mars could have experienced habitable periods during its evolutionary history (Ramirez & Craddock 2018). Tian et al. (2009) had applied an upper atmospheric model to study the evolution of the Martian atmosphere and found that a dense primordial $CO_2$ atmosphere could not have been maintained. Later work by Erkaev et al. (2014) confirmed that of Tian et al. (2009) suggesting that several tens of bar of $CO_2$ (via catastrophic outgassing after the magma ocean solidified) could be quickly lost (within about 12 million years) due to extreme stellar-XUV driven escape arising from the active young sun. Outgassing and (possibly) delivery led to the subsequent build-up of up to ~1 bar atmospheric pressure (Kite et al., 2014) on early Mars which could have favored habitable conditions. Analogous to the early Earth, earlier numerical models mostly predicted surface temperatures below freezing on early Mars despite geoproxy data to the contrary (e.g. Craddock and Howard, 2002). To address this issue, the model study by Ramirez et al. (2014) suggested that the presence of 1.3-4 bars of CO2, 5-20% $H_2$ could enhance the efficiency of the early Martian greenhouse in order to attain habitable surface conditions. Von Paris et al. (2013) suggested up to 13K surface warming on adding 0.5 bar $N_2(g)$. Wordsworth et al. (2017) suggested that updating previously underestimated collision-induced absorption could lead to significant warming. Ramirez (2017) subsequently find that $H_2$ concentrations as low as 1% can sustain mean surface temperatures above freezing. Related model studies of early Mars-type environments investigated factors impacting habitability, e.g. clouds (Urata and Toon, 2013; Ramirez and Kasting, 2017; Kitzmann, 2017), the effects of surface ice (Ramirez 2017), and precipitation (von Paris et al., 2015). Detailed 3D model studies of the early Martian climate have been performed (Forget et al., 2013; Wordsworth et al., 2015) which investigate the response of surface habitability to varying e.g. atmospheric composition, mass and planetary orbital parameters.

Detailed studies of the effects of space weather upon the evolution of early Mars - in terms of e.g. enhanced EUV absorption and particle induced ionization during air-shower events - are to our knowledge rather lacking in the literature. In particular, studies which focus on the Noachian and Hesperian periods including space weather effects and assuming representative atmospheres i.e. $CO_2$ (~1-5 bar); $N_2$ (<0.5 bar); $H_2$ (<a few tenths of a bar) and $CH_4$ (<a few hundredths of a bar) (as quoted in the literature and discussed above) would be desirable.

**5.2.2 Impacts of Space Weather on Close-In Exoplanets**

**Effect on Habitability of Proxima b.** Proxima Centauri constitutes the third and smallest member of the triple star system Alpha Centauri which, with only 1.3pc (4.4 light-years) distance from Earth, are the closest stars to our Solar system. The only known planet in this star system is Proxima Centauri b (Anglada-Escudé et al. 2016), hereafter Proxima b, with a minimum mass of 1.27 M and an orbital period of 11.2 days i.e. clearly located in the CHZ (following Kopparapu et al. 2013). The planet is only inferred from radial velocity measurements and the probability of Proxima b of transiting is very small (Kipping 2017). Little is known about Proxima b's upper mass limit, although it could be rocky based on planetary population statistics from the Kepler mission (Weiss and Marcy 2014). It is also still unclear from the data whether there could be more planets around this M5.5V red dwarf star with its effective temperature $T_{eff}$ of 3050 K (Anglada-Escudé et al. 2016; Barnes et al. 2016). Proxima Cen is a highly active star with flaring events of ~$10^{30}$ erg once a day and low energy events (~$10^{28}$ erg) every hour (Walker 1981; Davenport 2016). This compares to the largest recorded solar SEP events e.g. SEP 1989 and the Carrington event (Atri 2016). Proxima Cen's surface magnetic field (B ~ 600G) is orders of magnitudes larger than on our Sun (Reiners and Basri 2008). Additionally, the Total Stellar Irradiation (TSI) was found to vary significantly by about 17% over Proxima Cen's rotation period of 83 days (Wilson et al. 1981), which might have a crucial impact on the atmospheric evolution of Proxima b which receives an average TSI of 65% of the Earth (Anglada-Escudé et al. 2016).

There are numerous factors e.g. planetary mass, orbital evolution, and atmospheric properties, which influence habitability (Kasting et al. 1993; Wordsworth et al. 2010; Pierrehumbert 2011; Kopparapu et al. 2013). Since Proxima b is non-transiting, the key parameter ranges determining the state of Proxima b's atmosphere are large. Proxima b could have formed either at its current semi-major axis or maybe beyond the snow line, and for both cases there are parameter ranges for which it could still possess a significant amount of water (Ciesla et al. 2015; Mulders et al. 2015; Carter-Bond et al. 2012). Given that the presence of water is, by definition, essential for the study of habitability, numerous model studies have shown scenarios for which Proxima b could support periods of liquid water, depending on e.g. the $H_2$, $O_2$ and $CO_2$ evolution of its atmosphere (Luger et al. 2015; Barnes et al. 2016; Meadows et al. 2016, Ribas et al. 2016). While assuming the best-case scenario of a rocky, low-mass planet, Turbet et al. (2016) and Boutle et al. 2017 have shown in 3D Global Climate Model (GCM) studies that water is more likely to be present if Proxima b is tidally-locked, at least at the sub-stellar point. Del Genio et al. 2017 found the possibility of an even broader region of liquid water with their 3D-GCM studies including a dynamic ocean.

Can the environments of such close-in terrestrial type planets within CHZ from their parent stars be hospitable to life? This answer requires the conditions of space weather around red dwarf stars including quiescent and flare driven XUV fluxes and their properties of stellar winds.

The XUV emission from Proxima Centauri was recently presented by Garcia-Sage et al. (2017). They used the reconstructed XUV fluxes that appear to be over 2 orders of magnitude greater at the planet location than that received by the Earth to evaluate the associated ion escape. The calculated escaping $O^+$ mass flux from Proxima Cen b appears to be high as presented in Figure 19, consistent with the results of Airapetian et al. (2017). They also discussed the impact of thermospheric temperature and polar cap area on the escape fluxes. A hotter thermosphere is obviously expected to result in stronger outflows, but a larger polar cap area (the area of open magnetic flux connected to the stellar wind) also results in more net mass flux of ionized particles lost to space. Figure 16 shows the connection between these quantities and the mass loss rate. As expected, the thermospheric temperature and the polar cap size have strong implications for the mass loss rate at Proxima Cen b and the ability for this exoplanet to retain its atmosphere on geological time scale.

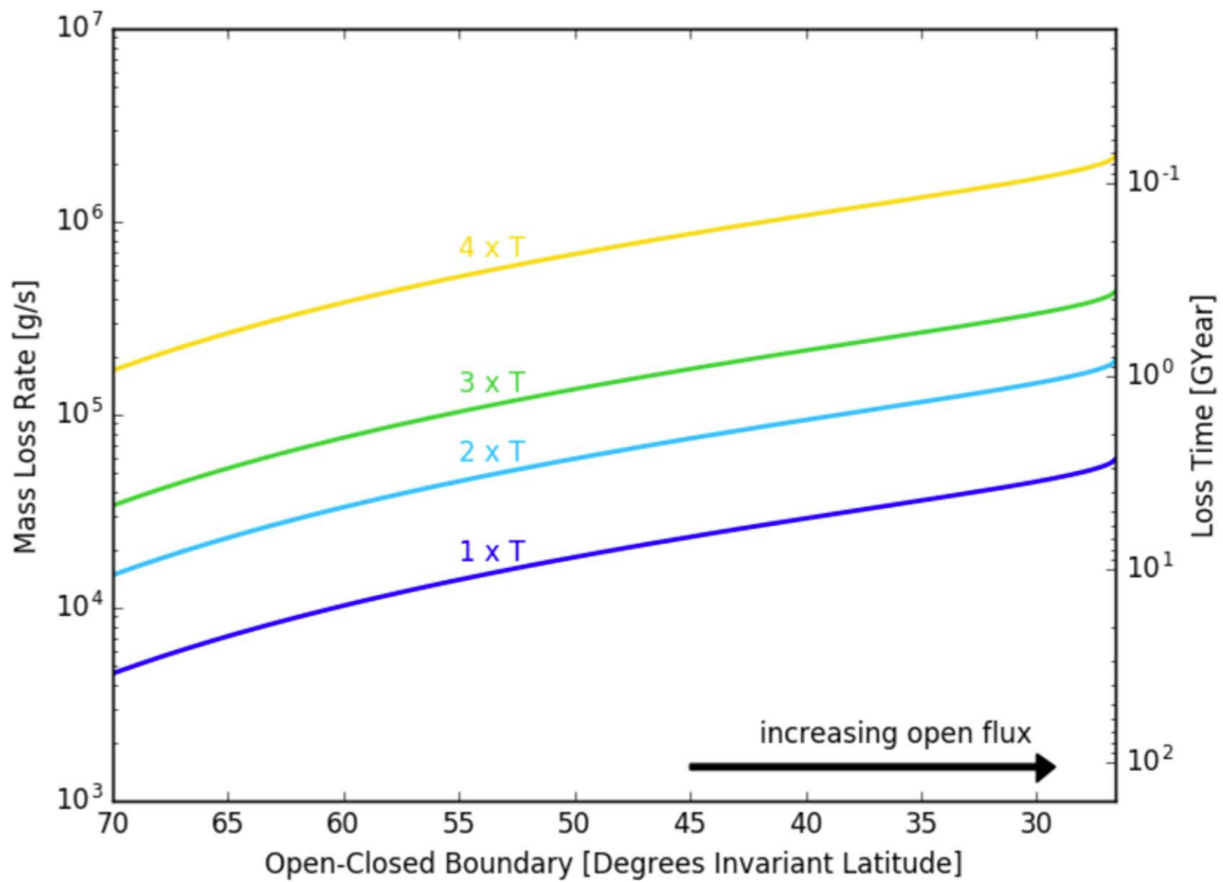

*Figure 16. The ion escape rate calculated for 4 different neutral thermospheric temperatures of Proxima b driven by XUV fluxes from Proxima Cen (Garcia-Sage et al. 2017).*

In summary, due to its proximity to the star, Proxima Cen b resides in an extremely hostile and extreme space environment that is likely to cause high atmospheric loss rates. If it is not clear whether the planet can sustain an atmosphere at all, it is less likely that the planet is habitable, even though it resides in the CHZ around the star.

The effects of stellar wind from Proxima Centauri on its exoplanet was recently characterized by Garraffo et al. (2016). They have theoretically reconstructed the stellar wind from Proxima Centauri and modelled the space environment conditions around Proxima Cen b. The model of the stellar coronal wind was driven by magnetic field observations (ZDI map) of the star GJ 51, which was used as a proxy for Proxima Cen for which ZDI observations are not available. Following the limited information about the magnetic field of Proxima Cen (Reiners and Basri 2008), they used two cases of average stellar magnetic field of 300 G and 600 G. They found that the stellar wind's dynamic and magnetic pressure along the orbit of Proxima Cen b are extremely large, up to 20,000 times that of the solar wind at 1AU. In addition, the planet experiences fast and very large variations in the ambient pressure along its orbit. As a result, the magnetosphere surrounding the planet undergoes huge variations within the relatively short period of the orbit, which potentially drive strong currents in the upper atmosphere and result in extreme atmospheric heating (see Figure 17).

The combination of extreme XUV fluxes and the ambient wind pressure, along with the fast and strong variations, potentially make the atmosphere of Proxima Cen b highly vulnerable to strong atmospheric stripping by the stellar wind, and atmospheric escape as a result of the additional strong heating. The simulations suggest that Proxima Cen b may reside in the sub-Alfvenic stellar wind at least part of the orbit, which removes the magnetopause completely, and exposes the planetary atmosphere to direct impact by stellar wind particles.

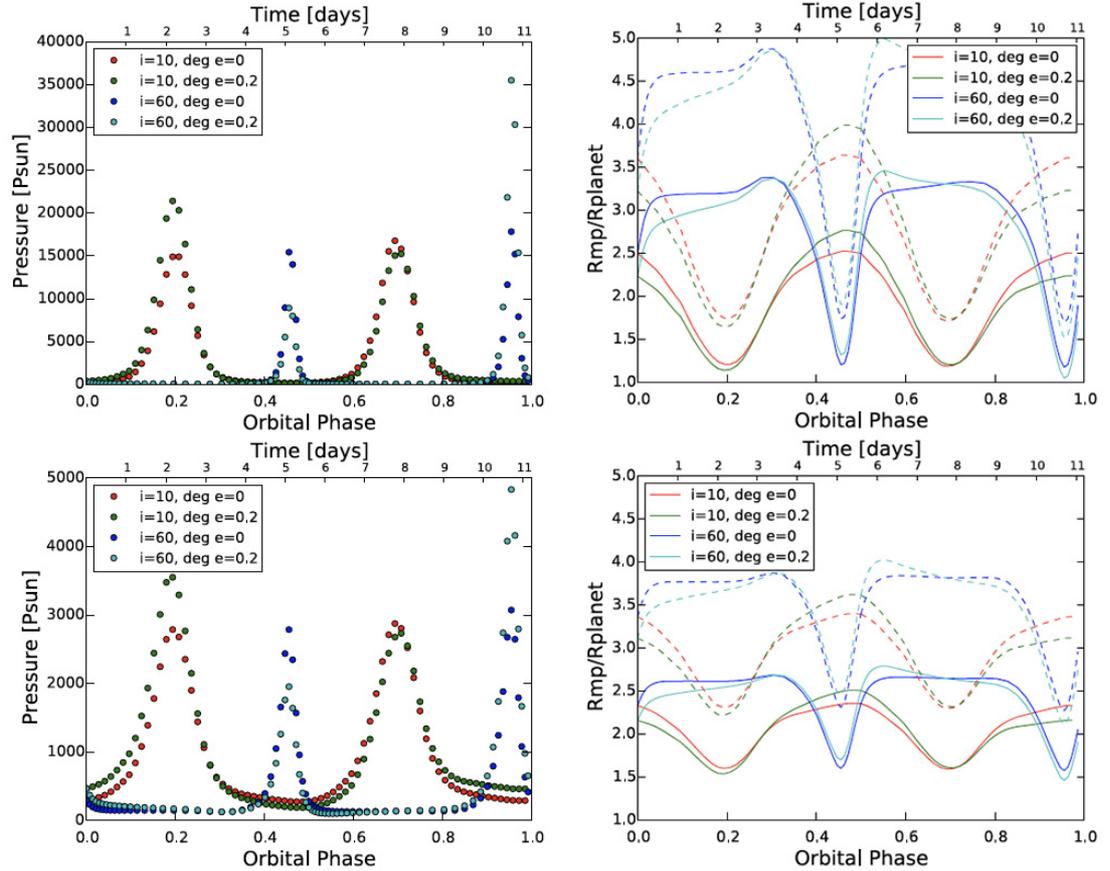

*Figure 17. Left: Dynamic pressure along the possible orbits of Proxima Cen b normalized to typical dynamic pressure of the solar wind at 1AU for stellar field maximum of 600G (top) and mean of 600G (bottom). Right: Same but plots show the magnetosphere standoff distance (from Garraffo et a. 2016).*

**Effect on Habitability of the TRAPPIST-1 System.** The recently discovered TRAPPIST-1 planetary system (Gillon et al. 2016) is 39 light-years away and contains a so-called ultra-cool magnetically active M8.2V dwarf star, surrounded by seven terrestrial type exoplanets three of which, d, e, and f, could be located within the CHZ and have orbital periods of 4, 6, and 9 days, respectively (Kopparapu et al. 2017). TRAPPIST-1 has a low $T_{eff}$ of around 2560K and (similar to Proxima Cen) exhibits strong surface (~600G) magnetic fields driving intensive chromospheric activity (Gizis et al. 2000; Reiners & Basri 2010) accompanied by frequent

(0.38 /day) stellar flares with energies between $10^{30}$-$10^{33}$ ergs (Vida et al. 2017). Unlike Proxima Cen, TRAPPIST-1 is a known transiting planetary system, which constrains the possible planetary parameters significantly. TRAPPIST-1b and c receive a greater stellar flux than Venus and are therefore suggested to have undergone a runaway greenhouse, or might still be in the runaway phase (Bourrier et al. 2017). Recent numerical studies by Wordsworth et al. (2017) have shown that abiotic $O_2$ build-up is possible for lose-in planets within CHZ, which makes them interesting targets for biosignature studies. On the other hand, the lower stellar flux on TRAPPIST-1f and g makes abiotic $O_2$ build-up unlikely, hence they are excellent targets for atmospheric characterization follow ups.

Model studies by Barstow and Irwin (2016) estimated that Earth's present-day ozone levels should be detectable with JWST on planets b, c, and d with observations of 30 to 60 transits. Recent hydrodynamic escape studies suggest that planets accreted wet (many Earth oceans of water), although this is generally debated at this point (*Tian and Ida 2015*). *Bolmont et al. (2016)* and *Bourrier et al. (2017)* estimate that the inner planets likely lost most of their water, whereas planets d, e, f, g, and h could have retained all but a few Earth oceans, depending on the age of the system. Furthermore, three-dimensional global climate models of the TRAPPIST-1 system, which are able to consider tidal-locking, have controversially shown that the accumulation of greenhouse gases could be difficult for all planets in the system (Yang et al. 2013). This is because $CO_2$ likely freezes out on the night side, and the high EUV flux rapidly photodissociates $NH_3$ and $CH_4$. In the case that $CO_2$ can eventually build up, planets e, f, and g could sustain liquid water (*Turbet et al. 2017*).

However, we should be cautious with aforementioned results that do not account for the $O^+$ escape rates due to XUV fluxes and orbital environments as discussed in the previous subsection (Garcia-Sage et al. 2017; Garaffo et al. 2017; Airapetian et al., 2017a). These factors may prevent the build-up of $O_2$ and subsequent formation of ozone in atmospheres of these planets and need to be modelled with the impacts of space weather effects from their host star. Another important impact on atmospheric escape may come from interaction of stellar winds with exoplanets. Specifically, strongly magnetized wind from TRAPPIST 1 introduces a strong dynamic pressure up to a factor of 1000 the solar wind pressure on Earth on hypothetical exoplanetary magnetospheres. As exoplanets orbit the star, the wind pressure changes by an order of magnitude and most planets spend a large fraction of their orbital period in the sub-Alfvenic regime (Garraffo et al. 2017). Recently, Dong et al. (2018) studied the atmospheric escape from the TRAPPIST-1 planets and discussed its implications for habitability. They modeled the stellar wind of TRAPPIST-1 by adopting the Alfvén Wave Solar Model (AWSoM; van der Holst et al. 2014) and simulated the star-planet interaction and the concomitant atmospheric ion loss by employing the BATS-R-US multi-species MHD model (Dong et al, 2017). Figure 18 (left panel) shows the total ion escape rate as a function of the semi-major axis for cases with both maximum (solid curve) and minimal (dashed curve) total pressure over each planet's orbit. An inspection of Figure 18 (left panel) reveals that 18e overall escape rate declines monotonically as one moves outwards, from TRAPPIST-1b to TRAPPIST-1h. Hence, taken collectively, this may suggest that TRAPPIST-1h ought to be most "habitable" planet amongst them, when viewed purely from the perspective of the specified mechanism of atmospheric loss, while XUV driven ion escape has not been consistently modelled in this study. Also, it must be recalled that the presence of liquid water on the surface is a prerequisite for habitability, and TRAPPIST-1h is not expected to be conventionally habitable (Gillon et al., 2017). It seems likely that TRAPPIST-1g will, instead, represent the best chance for habitable planet in this planetary system to support a stable atmosphere over long periods.

The ionospheric profiles for the TRAPPIST-1 planets in the HZ are not sensitive to the stellar wind conditions at altitudes ≤ 200 km (Dong et al. 2018 and see Figure 20, right panel).

This is an important result in light of the considerable variability and intensity of the stellar wind, since it suggests that the lower regions (such as the planetary surface) may remain mostly unaffected from under normal space weather conditions.

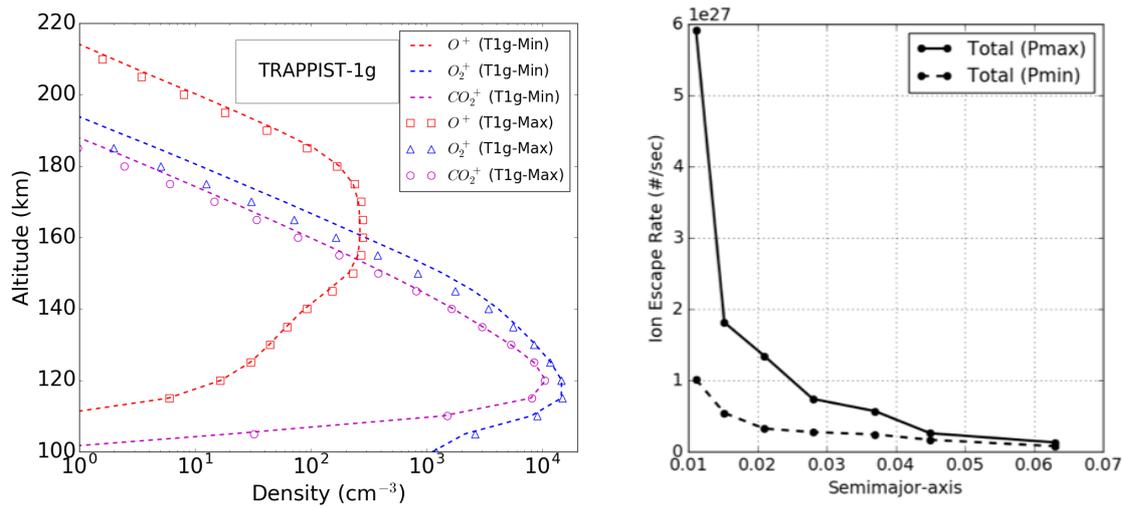

*Figure 18: (Left) Total atmospheric ion escape rate as a function of the semi-major axis for cases with both maximum (solid curve) and minimal (dashed curve) total pressure over each planet's orbit. (Right) The ionospheric profiles along the substellar line for TRAPPIST-1g for the cases of (i) maximum and (ii) minimum total pressure over its orbit (Dong et al., 2018). The seven distinct points on each curve represent the seven planets of the TRAPPIST-1 system.*

We should note that if the aforementioned exoplanets within CHZ would acquire large amount of water forming water-planets, their atmospheres should be composed of mostly $H_2O$ with small amounts of other volatiles delivered by comets and asteroids (Kaltenneger et al. 2013). How efficiently atmospheric escape mechanisms can remove the oceans from these planets? To make firm predictions, both the wind and XUV associated losses need to be modelled in further studies using multi-fluid hydrodynamic and kinetic models (Glocer et al. 2018).

### 5.2.3 Trends in the Transiting Exoplanet Population Likely Driven by Space Weather

The recent completion of the NASA *Kepler* mission has left a legacy dataset that will provide results for many years. In its prime mission, Kepler detected thousands of exoplanets, with sizes smaller than Earth to larger than Jupiter, transiting a wide variety of stars. A key result enabled by *Kepler* was the first statistical sample of small (R ≤ 4 $R_\oplus$), short period (P < 100 days), exoplanets with measured radii which allows for population studies. Spectroscopic measurements of the stars hosting these planets, along with precision distances from the ESA *Gaia* mission, allowed the radii of ~1000 of these planets to be constrained to a precision of 5% (Fulton et al. 2017, Fulton & Petigura 2018). Statistical analysis of the resulting high precision exoplanet sample revealed a bimodal radius distribution, with a population of rocky super-Earths and a population of gaseous sub-Neptunes separated by a gap in radius spanning approximately 1.5 - 2 $R_\oplus$ (see Figure 19). The observed radius gap is striking and indicative of the formation and evolution history of the planets. The shape of the distribution can be attributed to exoplanet atmosphere evolution under the influence of host star irradiation. Two important parameters that sculpt the distribution are host star mass and orbital separation. Both of these parameters are related to the X-ray and EUV irradiation history of the planet and are supported by the observed trend of a shift in the bimodal distribution toward smaller planets as host star mass decreases (and stellar activity increases, see Figure 19). This trend, the slope, and the width of the observed gap, are consistent with photoevaporation of atmospheres driven by host star irradiation as the dominant factor determining the radius distribution of small exoplanets. The observed radius trends are also in line with theoretical predictions of photoevaporative atmosphere loss (Lopez and Fortney 2013, Owen & Wu 2013, Fulton et al. 2017).

During its K2 mission, *Kepler* observed multiple fields around the Ecliptic plane that included populations of young stars. These included the Sco-Cen OB association (5 - 15 Myr), classic young open clusters like the Pleiades (125 Myr), the Hyades (600 - 800 Myr), and Praesepe (600 - 800 Myr), and isolated young stars in the Galactic field population

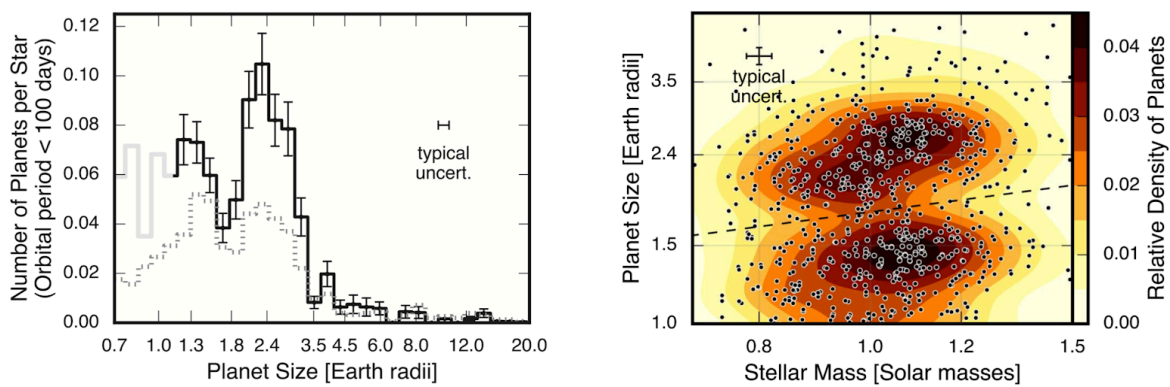

*Figure 19. Left: Radius distribution of short-period transiting exoplanets from the Kepler prime mission. The solid histogram shows the radii of planets in a completeness corrected sample and reveals two populations and a significant gap. The gray curve is planets in regions of poor completeness and the dotted line shows an arbitrarily scaled planet radius distribution prior to completeness corrections. Right: Two-dimensional representation of the planet radius distribution revealing that the observed bimodal population and radius gap tend toward smaller planets as host star mass decreases. These observed trends are consistent with irradiation driven photoevaporative mass-loss being the dominant factor in determining the radius distribution of small, close-in planets. Figure adapted from Fulton & Petigura (2018).*

(David et al. 2018a and references therein). These observations yielded the first young transiting exoplanets and this small population revealed an interesting emerging trend: young planets appear anomalously large compared to other planets on short period orbits transiting older stars of similar mass (David et al. 2018b, see Figure 20). This trend is most pronounced for young planets transiting low-mass stars which are more active for longer periods of time than their higher mass counterparts. This may be indicative of ongoing, space weather driven radius evolution. In this scenario, photoevaporative mass-loss from host star irradiation is still sculpting the planet's atmosphere and it has not yet reached its final size. At the same time, these young planets may also still be undergoing radius evolution due to core cooling and contraction (Vazan et al. 2017). Further observations of young stars spread across the sky by the NASA TESS mission will provide more examples of young transiting planets and add to the currently small sample. Detailed study of individual systems and statistical analysis of a larger sample will shed light on this interesting emerging feature in the exoplanet population.

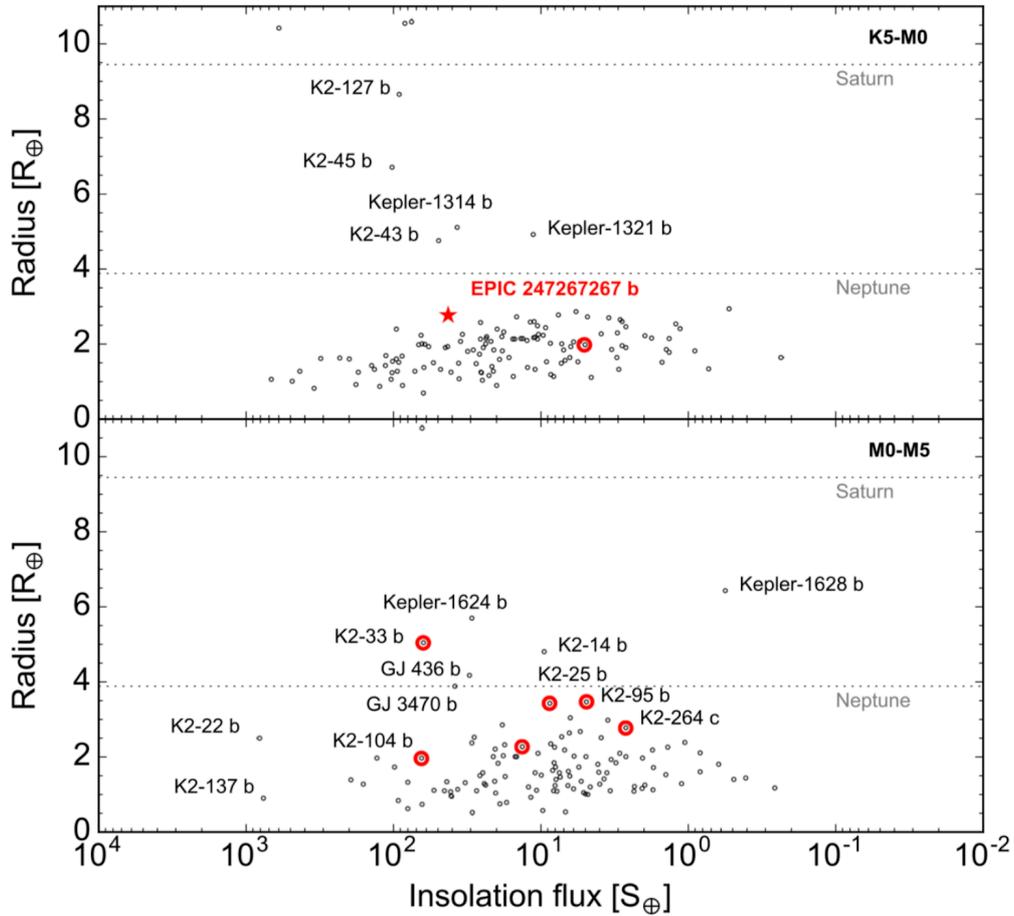

*Figure 20. Insolation flux vs. planet radius for known exoplanets transiting sample K5 - M5 type host stars (M = 0.7 - 0.1 M$_\odot$). The flux received by the Earth from the Sun is 1. Young (< 1 Gyr) transiting planets discovered by the K2 mission are circled in red. These planets tend to have radii that are larger than planets transiting old stars of similar mass and receiving similar insolation flux. This emerging trend may indicate these young planets are currently undergoing photoevaporative radius evolution. Figure adapted from David et al. (2018b).*

## 6.0 Impact of Space Weather on (Exo)planetary Atmospheric Chemistry
### 6.1 Effects of SEPs on Prebiotic Chemistry

As discussed in Section 3.3, during the first 500 Myr since its birth, our Sun was a magnetically active star generating a rich variety of high-energy eruptive events that generate ionizing radiation. This radiation in the form of XUV fluxes, CME driven energetic electrons and protons penetrate various layers of Earth's and Martian atmosphere: from magnetosphere to troposphere (see Figure 1). Ionizing radiation sources provide three major impacts on atmospheric molecules: photo and particle impact dissociation, excitation and ionization. These processes cause major changes in atmospheric chemistry that in turn control the dosage of ionizing radiation reaching the planetary surfaces, and thus directly affect habitability conditions on rocky planets.

While XUV radiation mostly ionizes the upper atmosphere and contributes to the formation of the Earth's ionosphere and thermosphere (at up to 90 km from the ground), auroral electrons can reach the altitudes of ~ 80 km, while relativistic electrons penetrate to the mesosphere at 40 km. Solar energetic protons (SEPs) with energies up to a few GeV can precipitate through the troposphere and reach the ground causing Ground Level Enhancement events as measured by ground monitors of neutrons, the byproducts of interactions of high-energy protons with atmospheric species (GLEs; see Figure 21). Impact electrons formed as a result of photo and collisional ionization can then interact with molecular nitrogen, carbon dioxide, methane and water vapor igniting a reactive chemistry in the Earth's lower atmosphere.

What would be the effects of such ionizing sources on the atmospheres of early Earth and Mars? Could these sources had been beneficial in starting life on Earth and possibly on Mars? The first clues on the relation between the sources of ionizing radiation and the production of biological molecules were obtained in revolutionary experiments by Miller (Miller 1953). In this experiment, a highly reducing gas mixture containing a water vapor, ammonia, $NO_3$, and methane, $CH_4$ and $H_2$ was exposed to a spark discharge. The chemical reactions initiated by the discharge promoted the formation of simple organic molecules including hydrogen cyanide, HCN, formaldehyde, $CH_2O$. These molecules and methane react via Strecker synthesis forming amino acids, the building blocks of proteins (amino acids) and other macromolecules.

However, later studies suggested that the early Earth's atmosphere was weakly reducing consisting of $N_2$, $CO_2$, CO and $H_2O$ with only minor abundance of $H_2$, $CH_4$ and $H_2S$ (Kasting et al. 1993; Lammer 2018) or neutral (Schaefer and Fegley 2017). Follow up experiments showed that non-thermal energy input in the form of lightning (spark discharge between centers of positive and negative charge) in a weakly reducing atmosphere does not efficiently produce abundant amino acids as reported in experiments in highly reducing environments (Cleaves et al. 2008 and references therein). Later, Patel et al. (2015) presented photochemically driven chemical networks that produce abundant amino acids, nucleosides, the building blocks of RNA and DNA molecules and lipids. Lipids are complex biomolecules that are used to store energy and serve as structural units of cell membranes. Recent experiments suggest that near-UV (NUV at $\lambda$=2000-2800Å) irradiation of the gas mixture is beneficial factor for formation of building blocks of life (Ranjan and Sasselov 2016; Rimmer et al. 2018). However, while UV emission promotes biochemical reactions with participation of HCN, it cannot break triple bonds of molecular nitrogen, $N_2$, that require 10 eV to create odd nitrogen (N), which is required to produce abundant HCN, the requirement for the chemical network by Patel et al. (2015).

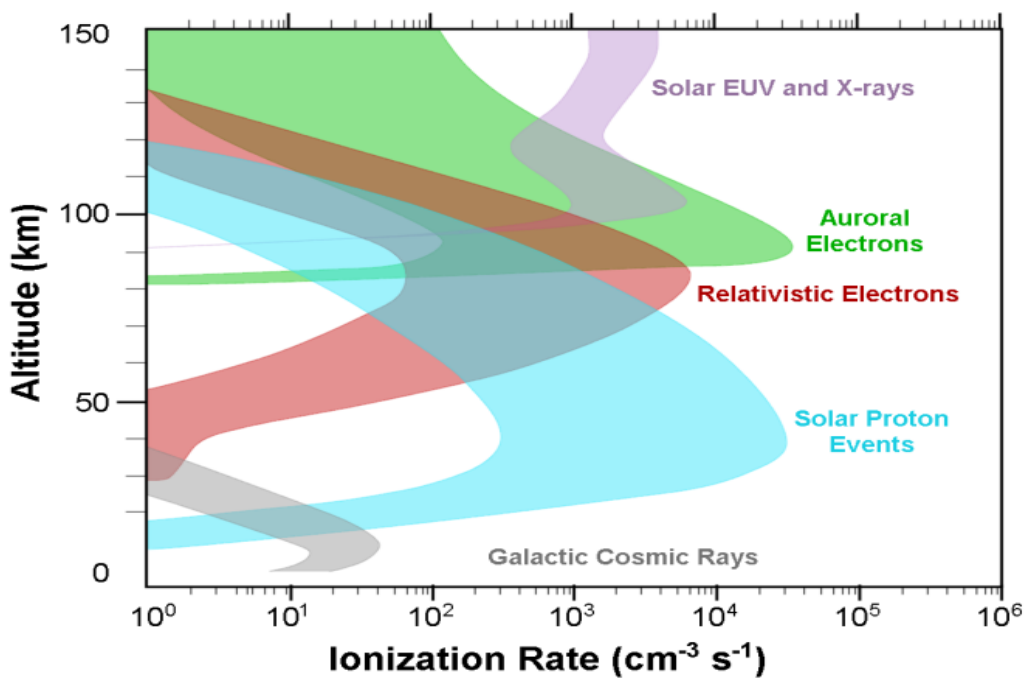

*Figure 21. Ionization of the Earth's atmosphere due to various factors of space weather including XUV emission, auroral electrons, SEPs and GCRs.*

HCN is the most important feedstock molecule for prebiotic chemistry that is the crucial for formation of amino acids, nucleobases and complex sugars. In order to produce HCN and other biologically important molecules, molecular nitrogen needs to be converted into odd nitrogen and follow up $NO_x$ ($NO$, $NO_2$) molecules, the process known as the nitrogen fixation. In order to efficiently precipitate these molecules down to the ground for further polymerization, the nitrogen fixation and subsequent production of HCN should occur in the lower atmosphere (stratosphere-troposphere layers) of Earth (Airapetian et al. 2016). This is an important requirement for the efficient delivery of HCN to the ground for subsequent synthesis into complex molecules (Airapetian et al. 2016).

The nitrogen fixation process can be achieved in the lower atmosphere either by high energy processes including extremely high temperatures (a few tens of thousands of Kelvins) that can be reached by atmospheric shock heating from impact processes, due to lightning processes that produce high energy electrons or solar EUV radiation with wavelengths shorter than 100 nm (Summers et al. 2012; Kasting 1990). Asteroid and comet impacts with Earth involve collisions with impact velocities from 20 km/s to 70 km/s. Such impact events heat the atmosphere and the ground to a few times of $10^4$ Kelvin. An impactor can create the favorable conditions for synthesis of reducing gas mixtures and form biologically relevant molecules (Chyba and Sagan 1992; Bernstein et al. 2006). Also, large molecules contained in the impactors can survive high velocity impacts and be delivered to the planetary surface (Pierazzo & Chyba 1999). These macromolecules can be dissolved into the "primodial" soup to provide further polymerization. However, the resulted concentration required to produce peptides, ribose, fatty acids, nucleobases, ribose and aligonucleotides through polymerization has not been discussed in details. EUV emission is another factor for nitrogen fixation, but it is absorbed at altitudes above 90 km above the ground (see Figure 21), and thus can provide significant amount of atomic hydrogen and other dissociated molecules required for the production of HCN in the upper atmosphere (Tian et al. 2011). However, the problem of efficient delivery of HCN to the lower atmosphere is problematic because of a slow vertical diffusion throughout the atmosphere. The alternative energy source, energetic particles from Galactic Cosmic Rays or SEP events can provide a mechanism to deliver the particles directly to the lower atmosphere. The penetration depth depends on the frequency of collisions of protons with ambient molecules and for the Earth's atmosphere is scaled with the proton's energy as $E^{-1.72}$ (Jackman et al. 1980). This suggests that the protons with energy of 300 MeV can penetrate to the heights of 4.5 km above the ground. The collisions with molecules produce enhanced ionization of the atmosphere and form a broad energy distribution of secondary electrons at > 35 eV. These electrons then thermalize to lower energies in the atmosphere and as soon their energy reaches 10 eV, they become very efficient in breaking $N_2$ into odd nitrogen with subsequent formation of $NO_x$. To study the pathways to complex organic molecules driven by energetic protons, Kobayashi et al. (1990); (1998); (2018); Miyakawa et al. (2002) have performed laboratory experiments exposing gas mixtures of $CO/CO_2$, CO, $N_2$ and $H_2O$ to 2.5 MeV protons. They reported production of amino acids precursors and nucleic acid bases as the results of secondary electron driven reactive chemistry. These experiments may have a direct relevance to the early Earth's atmospheric chemistry as high energy protons from SEP events precipitated through the Archean Earth's atmosphere causing showers as a result of impact ionization.

As discussed in Section 3.2, the young Sun was a source of frequent and energetic flares and associated CMEs. The CME driven shocks are the sites of efficient accelerations of SEPs that can accelerate particles to high energies (Fu et al. 2019). To study the role of SEPs from the young Sun as the source of high energy protons in gas phase prebiotic chemistry in the atmosphere of early Earth, Airapetian et al. (2016; 2019c) have applied a photo-collisional atmospheric chemistry model driven by the XUV flux and frequent SEP events. SEP driven protons with energies > 0.3 GeV (at 0.5 bar atmosphere) precipitate into the middle and lower atmosphere (stratosphere and troposphere) and produce enhanced ionization, dissociation, dissociative ionization, and excitation of atmospheric species. The destruction of $N_2$ into ground state atomic nitrogen, $N(^4S)$, and the excited state of atomic nitrogen, $N(^2D)$, as the first key step toward production of bio-relevant molecules. Reactions of these species with the products of subsequent dissociation of $CO_2$, $CH_4$ and $H_2O$ produces nitrogen oxides, NOx, CO and NH in the polar regions of the atmosphere. NOx then converts in the stratosphere into $HNO_2$, $HNO_3$ and its products including nitrates and ammonia.

The atmospheric model also predicts an efficient production of nitrous oxide, $N_2O$, driven primarily through $N(^4S) + NO_2 \rightarrow N_2O + O$; $NO + NH \rightarrow N_2O + H$. The recent model updated with the hard energy spectrum of protons, vertical diffusion and Rayleigh scattering outputs the $N_2O$ production in the lower atmosphere by a factor > 300 greater than the earlier model output of Airapetian et al. (2016). This is due to the fact that the CME driven shocks in the young Sun's corona, the source of CMEs, was at least a factor of 10 denser compared to the current Sun. Denser corona provided correspondingly larger concentrations of seed particles, while stronger shocks (larger compression ratio) produce SEPs with higher maximum energy and harder proton energy slopes. The particle acceleration via diffusive shock acceleration mechanisms on quasi-parallel strong shocks produce mostly SEPs with harder spectra (Fu et al. 2019). This is consistent with the recent statistical study of CME driven SEPs with hard energy spectra by Gopalswamy et al. (2016) and suggests the particle flux at 0.5 - 1 GeV is over 2 orders of magnitude greater than that assumed in the atmospheric chemistry model by Airapetian (2017) (see Figure 18).

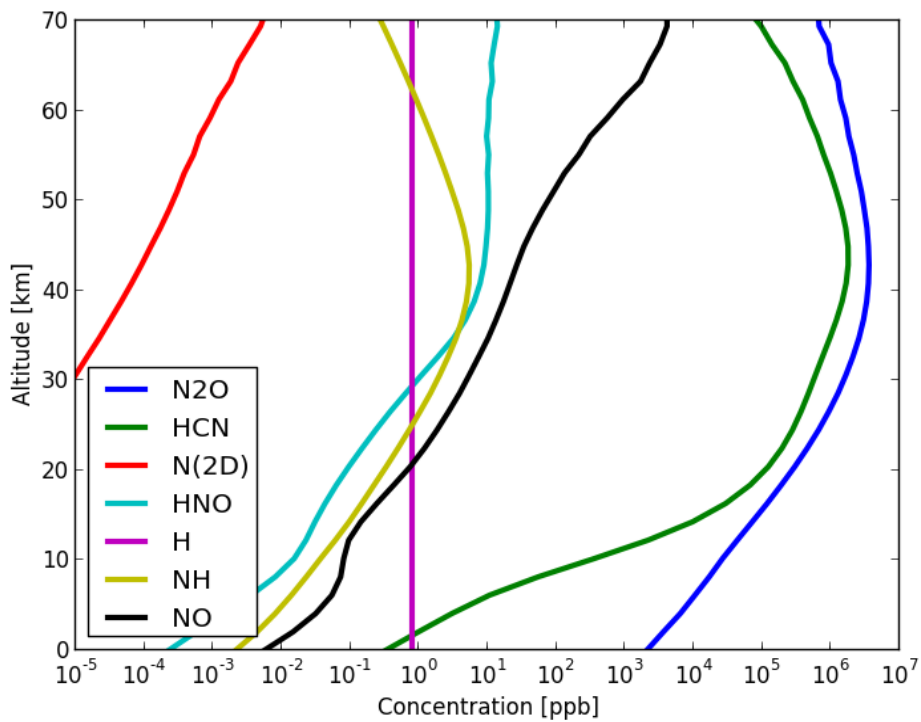

*Figure 22. Steady-state concentrations of molecules produced via photo-collisional chemistry from precipitating high-energy protons in the atmosphere of early Earth (Airapetian 2017).*

Laboratory experiments suggest that SEP driven chemistry predicts much more efficient production of HCN than that produced by lightning events for a weakly reducing ($N_2$ – $CO/CO_2$ – $CH_4$ – $H_2O$) atmosphere on early Earth (Kobayashi et al. 1998; Takano et al. 2004; Kobayashi 2018). HCN is formed primarily due to neutral reactions including $NO + CH \rightarrow HCN + O$, $CH_2 + N(^4S) \rightarrow HCN + H$, $CH_3 + N(^4S) \rightarrow HCN + H + H$ and $CH + CN \rightarrow HCN + H$. As it forms in the stratosphere - troposphere region, HCN may subsequently rain out into surface reservoirs and initiate higher order chemistry producing more complex organics. For example, the hydrolysis of HCN through reactions with water cloud droplets produces formamide, $HCONH_2$, that can rain out to the surface. Formamide serves as an important precursor of complex biomolecules that are capable of producing amino acids, the building blocks of proteins and nucleobases, sugars and nucleotides, the constituents of RNA and DNA molecules (Saladino et al. 2015; Hud 2018).

This scenario has recently been studied under lab conditions by Kobayashi et al. (2017; 2018), who performed a series of prebiotic chemistry experiments to study the formation of amino acids by irradiating weakly reducing gas mixtures ($N_2$-$CO_2$-$CH_4$-$H_2O$) that resemble a weakly reducing early Archean Earth's atmosphere with ionizing sources to simulate the energy flux from galactic and SEPs, UV emission and spark discharge (lightning) radiation. In these experiments, alanine and glycine were detected when gas mixtures with $CH_4$ molar ratio ($r_{CH4}$) as low as 0.5 % was irradiated by energetic protons with an energy range of 2.5-4 MeV generated from a van de Graaff accelerator (Tokyo Institute of Technology). The maximum G-value (the production rate in units of number of amino acid molecules per eV) for glycine is reached at $r_{CH4}$=5%. However, when the same mixture was exposed to irradiation by the spark discharge (accelerated electrons) or UV irradiation, amino acids were not detected for $r_{CH4}$ lower than 15 %. Considering fluxes of various energies on the primitive Earth (Kobayashi et al. 1998), energetic protons appear to be more efficient factor to produce N-containing organics than any other conventional energy sources like lightning or solar UV emission which irradiated the early Earth atmosphere.

Nitrous oxide, $N_2O$, is another abundant molecule produced by the chemical reaction network driven by SEP in the early Earth's atmosphere (see Figure 22). This molecule is a potent greenhouse gas, 300 times stronger in greenhouse power than $CO_2$ and is currently present in Earth's atmosphere produced at ~ 400 ppbv via biological processes. While its atmospheric concentration 1000 times less than $CO_2$, it contributes ~ 6% to the global warming of our planet. Airapetian et al. (2019) have recently applied the results of Figure 22 as input for climate model to show that the annually mean temperature of late Hadean Earth can be ~ $5.2^0C$. This is important to resolve a longstanding problem known as the Faint Young Sun (FYS) paradox (see a review by Feulner 2012). The young Sun was under late Hadean Earth was ~ 25% fainter that it is today, which will be insufficient to support liquid water on the early Earth contrary to geological evidence of its presence that time (Sagan and Mullen 1972; Feulner 2012). Further theoretical and experimental efforts are needed to study the pathaways and the products of the SEP & XUV modified gas phase chemistry to form "exotic" greenhouse molecules. This molecules may provide strong spectral signatures in the upcoming observational efforts to study signatures of life and also be instrumental in resolving the FYS for early Earth and Mars.

**6.2 Impact of SEPs on Surface Dosages of Ionizing Radiation**

SEPs induced by strong stellar CMEs may have a significant impact not only on the atmospheric chemistry, but also on the surface dosage of ionizing radiation that may affect surface habitability of terrestrial type exoplanets. Recent modeling of SEP-driven ionizing radiation in the atmospheres of exoplanets around M dwarfs of different chemical compositions and atmospheric pressures including Proxima-b and TRAPPIST-1 suggest that impact on life forms at 1 bar atmospheric pressure is relatively small compared to the critical dose at 10 Sv (Atri 2017, Yamashiki et al. 2019 and also shown in Figure 23). Considering that exoplanetary atmospheres may suffer severe atmospheric escape in proportion to the XUV flux from their host stars (Airapetian 2017b, see Fig. 17), the surface environment for the majority of exoplanets in the CHZs around M dwarfs may become inhospitable at least for complex terrestrial-type life forms (see Figure 23).

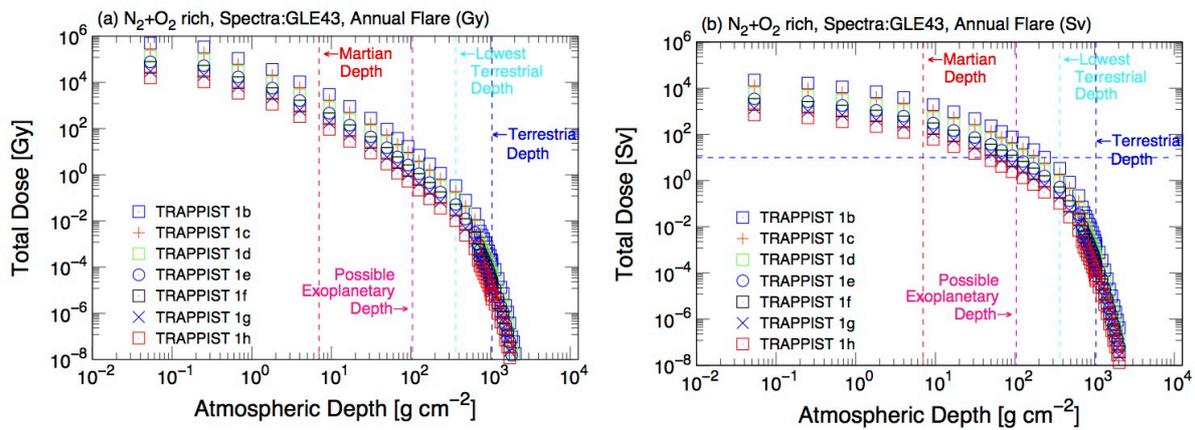

*Figure 23. Left: Vertical profile of radiation dose (in Gy) Right: (in Sv), caused by hard proton spectrum imitating GLE 43 penetrating $N_2+H_2$ rich terrestrial type atmosphere of TRAPPIST-1b (blue square), c(red cross), d(green square), e(blue circle), f(black square), g(blue cross) and h(red square) in logarithmic scale under Annual Maximum flare energy, calculating using PHITS (Sato et al.2015) and ExoKyoto. Vertical red dotted line represents Martian equivalent atmospheric depth, pink dotted line represents the depth 0.1 bar atmosphere, blue dotted line represents lowest terrestrial atmospheric depth observed at the summit of Himalaya, and blue dotted line as terrestrial atmospheric depth. Horizontal blue dotted line represents 10 Sv, which may be considered as critical dose for complex terrestrial-type lifeform. Figure adapted from Yamashiki et al. (2019).*

## 7.0   Space Weather, Habitability and Biosignatures

'Habitability' broadly refers to conditions which can sustain life. An unequivocable definition of life itself remains elusive although often-quoted is the "self-sustaining chemical system capable of Darwinian evolution" (*Joyce et al., 1994*). Life as we know it has three essential requirements, namely (i) an energy source, (ii) liquid water, and (iii) the chemical elements CHNOPS (see e.g. *Cockell et al., 2016*). The classical Circumstellar Habitable Zone (CHZ) refers to the circular region around a star where a terrestrial planet can support liquid water (*Huang 1960; Kasting et al., 1993*). An empirically-defined "conservative habitable zone" has also been introduced, with limits based on evidence that recent Venus and early Mars may have had conditions permitting liquid water on their surfaces (*Kasting et al., 2014*). The Continuous HZ (CHZ) (e.g. *Hart, 1979*) refers to the circular region around a star, which remains potentially habitable changes in stellar luminosity over geological time. Its inner and outer boundaries depend on stellar mass and metallicity, because they strongly affect the stellar luminosity and effective temperature (Gallet et al. 2013).

The non-classical HZ extends beyond the snowline and includes objects proposed to have subsurface oceans (*Lammer et al., 2009*). Other phenomena proposed to affect habitability include e.g. the output of the central star (*Beech, 2011*); the planetary atmospheric mass and composition (e.g. *Grenfell et al., 2010*); the presence of stabilizing climate feedbacks driven by e.g. plate tectonics (Korenaga et al., 2013); the presence of a large moon (*Laskar et al., 1993*); planetary orbital parameters such as the rotational rate (e.g. *Del Genio and Zhou, 1996; Edson et al., 2011; Yang et al., 2014*) and the role of the planetary magnetosphere (e.g. *Shizgal and Arkos, 1996*).

There are numerous ways in which space weather from planet hosts could influence habitability conditions within CHZs - for example lower mass main sequence stars have stronger magnetic fields and lager XUV fluxes which may impact exoplanetary magnetospheric and atmospheric erosion, two critical factors affecting surface climate and radiation as well as surface radiation doses as discussed in Section 6.

### 7.1 Atmospheric Models of Habitable Environments

A major goal of this Section is to discuss the factors necessary for habitability and the extent of habitable environments in the Universe. These clearly have repercussions for the distribution of potential life and biosignatures which drives the design of next generation exoplanetary missions. A central task is to constrain the HZ for a given planetary system by applying numerical models. Earlier 1D model studies (e.g. *Hart, 1979*) suggested a thin HZ for Sun-like stars extending from (0.95-1.01) Astronomical Units (AU). Subsequent works (e.g. *Kasting et al., 1988*) however, noted that including atmospheric climate feedbacks such as the carbonate-silicate cycle (*Walker et al., 1981*) could considerably widen the HZ.

Regarding the habitability of the early Earth, there exists a discrepancy between ancient soil (paleosol) data and the output of certain atmospheric climate-column models which suggests a frozen surface. Briefly, proposed solutions for the FYS paradox include e.g. (i) an increased abundance of greenhouse gases such as methane ($CH_4(g)$) (*Pavlov et al., 2003*) or/and nitrous oxide ($N_2O(g)$) (*Grenfell et al., 2011; Roberson et al., 2011*), (ii) enhanced line-broadening by molecular nitrogen ($N_2$) (*Goldblatt et al., 2009*), (iii) increased planetary albedo (*Rosing et al., 2010*) due to a lowering in biogenic cloud precursors, or (iv) updated carbon dioxide ($CO_2$) thermal absorption coefficients (*von Paris et al., 2008*). 3D model studies addressing the FYS problem (*Charnaey et al., 2013; Wolf & Toon 2013; Kunze et al., 2014*) noted that tropical latitudes could remain habitable even if the global mean surface temperature lies below the freezing point. Organic photochemical hazes that can form in reducing atmospheres with sufficiently high $CH_4$ concentrations had been thought to be an obstacle to keeping Archean Earth warm enough to sustain liquid water (*Sagan and Chyba, 1997*). More recent work suggests however that the fractal nature of such hazes (*Wolf and Toon, 2010*) and self-shielding effects (*Arney et al., 2016*) limit the surface cooling that can occur in the visible portion of the FYS spectrum while blocking harmful radiation in the UV.

For the inner HZ there are two boundaries. Firstly, the "water-loss" (or "moist greenhouse" limit") (*Kasting, 1988*) occurs where the planet's entire ocean reservoir is lost via diffusion-limited escape within the lifetime of the planet (originally estimated to occur at =0.95AU for Earth; (*Kasting et al., 1993*) and more recently at 0.99 AU (*Kopparapu et al., 2013*) for 1-dimensional cloud-free models and main-sequence stars). Secondly, the "runaway greenhouse limit" occurs where the surface temperature reaches the critical point of water ($T_c$=647K) (=0.84AU for Earth; (*Kasting, 1988*); see also *Wolf and Toon, 2015*). Recent 1D model studies (e.g. *Kasting et al., 2014*) noted the challenge of performing accurate climate transfer calculations in thick steam atmospheres near the inner HZ boundary. The 1D study of *Zsom et al. (2013)* suggested a very close (=0.38AU) inner HZ boundary for a desert world with high surface albedo (A=0.8) and low relative humidity (RH=1%) orbiting a solar-mass main sequence star. 3D model studies near the HZ boundary suggest e.g. climate-cloud feedbacks (*Yang et al., 2013*) and massive atmospheric circulation (e.g. Hadley cell) responses (*LecConte et al., 2013*) which suggest an expansion of the inner HZ boundary towards the star. *Yang et al. (2014a)* suggest a strong dependence of the inner HZ position upon planetary rotation rate associated with an atmospheric dynamical response driven by changes in the Coriolis force. A recent 3D model study (*Kopparapu et al., 2016*) suggested a modest revision of the inner HZ boundary inwards on including consistent planetary rotation due to tidal-locking and associated cloud effects.

For the outer HZ, the "maximum greenhouse limit" (=1.67AU for Earth; (*Kasting et al., 1993*) occurs where greenhouse warming of the planetary surface from carbon dioxide

($CO_2(g)$) is strongest. Increasing the abundance of atmospheric $CO_2(g)$ above this limit leads to a net cooling of the surface due to enhanced Rayleigh scattering. The $CO_2(g)$ abundance and its ability to stabilise climate depends upon outgassing rates from the interior and whether the planet features plate tectonics (see also section 4). Pierrehumbert and Gaidos (2011) suggested that collision-induced absorption from hydrogen ($H_2$) atmospheres could extend the outer HZ to as far as 10AU for planets orbiting a sunlike star, even though planets would lose their $H_2$ inventories within a few million years. Ramirez and Kaltenegger (2017) have recently proposed that the $CO_2$-$H_2O$ CHZ (e.g. Kasting et al. 1993) could be significantly widened on planets with volcanoes that outgassed significant amounts of $H_2$, which moves the outer edge limit outward (from 1.67 AU to ~2.4 AU in our Solar system). A recent 3D model study (Wolf, 2017) suggests that the 1D outer edge maximum greenhouse limit (1.67 AU) may not actually be realizable because of the high cloud and surface albedos of cold planets and subsaturation of water vapor. The extra warming from $H_2$ could partially offset this problem though (Ramirez and Kaltenegger, 2017). On the other hand, even for an Earth-like atmospheric composition, habitability may be possible for colder surface temperatures if the ocean is more saline than Earth's (*Cullum et al., 2016; Del Genio et al., 2017*).

**7.2 Habitability of Earth-like planets orbiting M-dwarf stars.**

Rocky planets orbiting in the HZ of M-dwarf stars are on the one hand favored objects for next-generation mission surveys since such stars are numerous in the solar neighborhood, their planets have high (planet/star) contrast ratios and the close proximity of HZ planets to the central star means more frequent transit events and hence faster data collection over a given observing interval. On the other hand, such planets have potential drawbacks with regards to their potential habitability as reviewed in *Scalo et al. (2007)* and *Shields et al. (2016)*. Firstly, the close proximity of the HZ to the central star could most likely lead to HZ planets becoming tidally-locked i.e. where strong gravitational interaction leads to the possibility of synchronous rotation (the planet having a constant day and night side). Whether the planet's surface remains habitable then depends on the ability of its atmosphere to transport heat from the dayside to the nightside. In practice, numerous 3D model studies have indicated that this is not an obstacle to habitability of the dayside region as long as (1) there is sufficient water to create an optically thick dayside convective cloud deck that shields the surface from direct starlight (e.g., *Yang et al., 2013; Kopparapu et al., 2016; Turbet et al., 2016; Boutle et al., 2017; Wolf, 2017*) and (2) the land-ocean configuration prevents water from being completely trapped as ice on the nightside (*Yang et al., 2014b*). Nonetheless, the "moist greenhouse" habitability limit of diffusion-limited escape occurs at lower instellation values for cooler stars because of their reducing Rayleigh scattering and predominantly near-infrared incident stellar flux, the latter which is strongly absorbed by water vapor, driving a stratospheric circulation that transports significant $H_2O$ to high altitudes where it is readily dissociated (e.g. Kasting et al. 1993; *Fujii et al., 2017; Kopparapu et al., 2017*).

Other factors that influence the habitability of M-star planets involve the evolutionary phase of bolometric luminosity and magnetic activity of host stars, their impact on a planet and internal planetary dynamics. For some planets, the luminous pre-main sequence phase of the host star may have driven the planet into a runaway greenhouse state before it had a chance to become habitable (*Ramirez and Kaltenegger, 2014; Luger and Barnes, 2015; Bolmont et al., 2017*) although worlds located in the pre-main-sequence HZ may be habitable (Ramirez and Kaltenegger, 2014). For others, conditions in the protoplanetary disk may have worked in favor of or against habitability, e.g., variations in water delivery to planets (*Ribas et al., 2016; Tian and Ida, 2015*) and an $H_2$-poor vs. $H_2$-rich nebula that may have provided initial shielding planetary envelopes of different thicknesses to protect Earth-size planetary cores and their atmospheres from complete photoevaporation (*Luger et al., 2015*). The California-Kepler survey suggests at least a small number of planets of a size that would be consistent with a remnant thin $H_2$ envelope (*Fulton et al., 2017*). The composition of the planet, to the extent that it is related to elemental abundances for its host star, may determine its interior structure (*Dorn et al., 2017*) and influence the tectonic history of the planet and thus the coupling of its climate to the interior (*Lenardic et al., 2016*). Finally, due (a) to the close proximity of the planet to the star and (b) (possible) weakening of the magnetospheric field (if present) due to tidal-locking - the stellar environment could lead to strong bombardment of the atmosphere with high energy particles and UV that either strip the planet of its atmosphere or leave its surface inhospitable to life (see also Section 5).

### 7.3 Atmospheric Biosignatures

The term biosignature refers here to evidence which suggests the presence of biological forms of life. The subject is broad, spanning a range of disciplines (e.g. biology, geology, atmospheric science) and physical criteria (e.g. fossil morphology, isotope signatures, chiral signals, departures from thermodynamic and redox equilibria, presence of gas-phase species etc.). Broadly speaking, an "ideal" biosignature should be (i) easily detectable with a large signal-to-noise ratio, (ii) unequivocally biological i.e. without abiotic sources and (iii) easily retrievable from the data without e.g. data degeneracies, numerical issues etc.

The topic may be broadly split into two parts, namely in-situ and remote biosignatures. In this paper, we focus on remote biosignatures via spectroscopic detection of atmospheric species in an exoplanet context. We discuss thereby possible abiotic production mechanisms (in order to rule out false signals from non-life), biosignature detectability and the role of space weather. A series of five comprehensive review papers in biosignature science (Schwieterman et al., 2018; Meadows et al., 2018; Catling et al., 2018; Walker et al., 2017; Fujii et al., 2017) organized by the NASA research network NExSS (Nexus for Exoplanet System Science) emphasized the following key points: 1. remote biosignatures need to be interpreted in their full environmental context i.e. combined with information on stellar input, planetary evolution etc (Airapetian et al. 2018a,b; Meadows et al. 2018); 2. exo-atmospheric biosignature science has recently developed a much more mature understanding of potential abiotic sources, especially for $O_2$(g) (see Grenfell et al. (2017) for a review of the response of atmospheric biosignatures in an exoplanet context); 3. given the observational challenges for biosignature detection and the open-ended nature of potential theories for false positives, a probabilistic approach to claims of biosignature detection will be necessary.

### 7.3.1. Impact of Stellar UV Emission on Biosignatures

The seminal study by Des Marais et al. (2002) investigated spectral detection of atmospheric biosignatures such as $O_2$ and $O_3$ on Earth-like worlds. Several works (e.g. Segura et al. 2003; Grenfell et al. 2007; 2014; Kaltenegger et al. 2007; Rugheimer et al. 2013, 2015) subsequently modeled the effect of the incoming stellar radiation (in particular UV) upon Earth-like planets orbiting in the HZ of different spectral classes. They found that electromagnetic radiation from the central star could play an important role for atmospheric climate, photochemistry and planetary spectral signals.

Regarding planets in the HZ of M-dwarfs, Segura et al. (2005) modelled the effect of UV suggesting that biosignatures such as $N_2O$(g) and $CH_3Cl$(g) could build-up by 2-3 orders of magnitude compared with the modern Earth. Segura et al (2010) modeled the UV (and the high-energy particle) effect of a stellar flare upon the atmospheric habitability for a hypothetical Earth-like planet orbiting in the HZ of the M-dwarf star AD Leo. Results suggested that flares may not present a direct hazard to potential life. Rauer et al. (2011); Grenfell et al. (2013) and Rugheimer et al. (2015) noted that changing the spectral class and activity (hence UV output) of the central M-dwarf star could have a potentially large effect upon atmospheric biosignature abundance and spectral signal. Because XUV flux may cause atmospheric erosion especially on exoplanets in HZs around M dwarfs, atmospheric pressure should be significantly lower than that assumed in these models unless a planet hosting star is magnetically quiet and planet is volcanically active to replenish high rates of escape (Airapetian et al. 2017a; Garcia-Sage et al. 2017).

While oxygen ($O_2$), its photochemical product, ozone ($O_3$); methane ($CH_4$), nitrous oxide ($N_2O$) have been proposed as gas-phase biosignatures, they cannot be considered robust in isolation due to the known abiotic production mechanisms for $O_2$, $O_3$, $CH_4$, and $N_2O$. A combination of biosignatures and an understanding of the planetary context will be required to claim a planet has a high potential of having life. Chloromethane ($CH_3Cl$) is another proposed biosignature that has no known false positive mechanisms and it is a highly specific molecule unlikely to be produced in large quantities on an exoplanet if at all (see Seager et al., 2013a for a discussion on types of biosignatures). Seager et al. (2013b) also suggested $NH_3$ as a potential biosignature in a temperate, $H_2$ dominated world. $NH_3$ has a short atmospheric lifetime on Earth as it is easily photolyzed by UV radiation. Ozone plays an additional role because its fate could be intermingled with those of the other species. This is because loss of the protective ozone layer on an Earth-like planet would likely result in a flooding of UV photons into the lower atmosphere which would strongly increase the photochemical removal of the other biosignatures, $CH_4$, $N_2O$, and $CH_3Cl$.

Regarding the effect of UV, ozone is on the one hand *produced* via molecular oxygen photolysis in the Herzberg region ($\lambda < 242$ nm) - producing oxygen atoms (O) which quickly react with $O_2$ to form $O_3$. On the other hand, ozone is photolytically *destroyed* at wavelengths ranging from ~ (242-320) nm. This implies a dependence of ozone abundance to UV with a "switchover" regime from ozone loss to ozone production with decreasing (UV) wavelength. The response of ozone to UV is further complicated by so-called catalytic cycles involving $HO_x$, $NO_x$ and $ClO_x$ (as discussed in section 2) which remove ozone in the middle atmosphere. The effect of increasing UV generally favors stronger catalytic loss, because it favors the release of $HO_x$, $NO_x$, and $ClO_x$ from their so-called reservoir (inert) species. For example, nitric acid ($HNO_3$) is a reservoir of both $HO_x$ and $NO_x$ because it can photolyze in the UV to release $HO_x$ and $NO_x$ species via the reaction: $HNO_3 + h\nu \rightarrow OH + NO_2$. Regarding the effect of high energy particles, the resulting air shower events generate secondary particles which destroy $N_2(g)$ and $O_2(g)$, leading to enhancements of NOx and HOx which can destroy ozone (as already discussed in section 2; see also references below).

If atmospheric ozone is removed, UV increases in the lower atmosphere, which can strongly affect other atmospheric biosignatures. $N_2O$, for example is mainly removed either directly via (i) UV photolysis or/and (ii) reaction with $O(^1D)$ (the O atom in an excited singlet state), which is mainly generated in the presence of UV via ozone photolysis. $CH_3Cl$ and $CH_4$ are also mainly removed via (i) UV photolysis or/and (ii) reaction with the hydroxyl (OH) radical. OH is favored by the presence of UV due to its central source reaction: $H_2O + O(^1D) \rightarrow 2OH$ (since $O^*$ is mainly generated in the presence of UV, as mentioned). Additionally, all of the known false positive mechanisms rely in some way on the UV environment through the photolysis of $H_2O$ or $CO_2$ (see section 5.3). Reviews of the photochemical responses of atmospheric biosignatures can be found in Brasseur and Solomon (2005), Holloway and Wayne (2010) and Grenfell et al. (2017).

**7.3.2 Impact of SEPs and GCR on Biosignatures**

The chemistry of exoplanets around active stars including active G, K and M dwarfs can be modified by space weather impacts that include SEP events formed during large flares and coronal mass ejections. These effects may represent a vastly underestimated factor regarding the chemical impacts of energetic particles including SEPs and the relative contribution from GCR. The Earth's middle atmospheric chemistry during large flares and associated SEP events is represented in production of nitrogen oxides, $NO_x$, $HO_x$ constituents and destruction of ozone (Jackman et al. 2005; 2008).

**The effects of SEPs on Earth-like atmospheric composition exoplanets.** To study the effects of SEPs on atmospheres of Earth-like exoplanets (with the current Earth's chemistry) around M dwarfs, Grenfell et al. (2012) applied an air shower approach in a coupled climate-photochemical column model considering formation of nitrogen oxides (NOx). Results suggested strong removal of the atmospheric biosignature ozone due to catalytic removal involving NOx cycles. Tabataba-Vakili et al., (2016) applied an updated version of the same model including hydrogen oxides (HOx) generated from stellar particle events. Results suggested that including the HOx effect led to some NOx removal into its unreactive reservoirs hence weaker loss of ozone. However, the assumptions about expected slopes and fluences of SEPs from M dwarfs as well as the atmospheric thickness of exoplanetary atmospheres need to be justified in the future first principle models of SEP initiation from active M dwarfs.

Airapetian et al. (2017b) have recently applied 2D GSFC atmospheric model to study the chemical response from strong SEP events on nitrogen-rich Earth-like planets and found that NOx and OHx are efficiently produced in the thermosphere as a result of photo-dissociation (X-ray and EUV) and collisional dissociation (via secondary electrons) of molecular nitrogen and water vapor. The abundance of NO and OH molecules increases by a factor of 100 during strong magnetic storms as compared to the quite time. The drastic enhancement of NO caused the depletion of ozone. This suggests that storms initiate time-varying emissions from broad - band molecular bands of NO at 5.3-μm, OH at 1.6 and 2 μm and $O_2$ ($^1\Delta$), the lowest electronically excited state of the $O_2$ molecule, at 1.27 μm, $N_2O$ at 3.7, 4.5, 7.8, 8.6 μm and $CO_2$ at 16 μm. Thus, the detection of time-varying (at the time scale of ~ 2 days) absorption bands at corresponding wavelengths in future direct imaging observations in the mid-IR band (see Section 9.4) would specify the fundamental ingredients of biologically compatible atmospheric conditions. These chemical ingredients of abiotically enhanced concentrations of these "beacons of life" would provide information about the presence of thick nitrogen atmosphere and atmospheric water from terrestrial type exoplanets around active G, K and early M dwarfs. Future simulations of SEP impacts on sulfur-based chemistry and methanogen and nitrogen rich chemistry would provide new set of molecules and their spectral signatures to be searched for in the upcoming space missions.

### 7.4 Detectability of Biosignatures via (Spectro)photometry

The detectability of biosignatures in an exoplanet atmosphere will depend on the abundance of the gas (in the VIS and IR) and the temperature vs pressure profile (in the IR) as well as the type of detection method used. Transmission spectra observations, such as with JWST will be best suited to detect biosignatures in planets orbiting M dwarfs where habitable planets have a higher transiting probability and transit more frequently for the same effective temperature. Planets orbiting FGK stars will be better targets for direct detection missions like LUVIOR/Hight Definition Space Telescope (HDST), where the inner working angle of the telescope will preclude close-in planets whereas planets orbiting M stars will be better targets for transmission spectra mission.

Because no single candidate gas is sufficient to be considered a biosignature, we will need to characterize a combination of gases combined with data that indicates the planetary context to claim a robust biosignature detection. Classically this means detecting an oxidizing gas like $O_2/O_3$ in combination with $CH_4$. Additionally, we will need data indicating habitable temperatures, water and $CO_2$ to indicate life as we know it. $N_2O$, $CH_3Cl$, dimethyl sulfide (DMS) are also useful biosignatures to consider with the proper planetary context.

Detecting biosignatures with the next generation of missions will be difficult though possible. It is estimated that JWST will characterize a handful of terrestrial exoplanets, and will have the possibility to detect some biosignatures like $O_3$ and $CH_4$. However, it is estimated to take several hundred hours of JWST time to observe these features for even nearby planets (Kaltenegger & Traub, 2009; Deming et al., 2009). Upcoming ELTs may also be able to detect biosignatures from the ground (Snellen et al., 2013). Missions still in the design concept phase such as LUVOIR/HDST should be able to overcome these current limitations and characterize hundreds of terrestrial exoplanets (Stark et al. 2014, 2015).

Clouds and hazes can be a strong limiting factor for the depth to which transmission spectra can sense and are already impacting exoplanet observations (Sing et al.,2015). For reflectance spectra, clouds can block access to the lower atmosphere, decreasing the depth of features, however clouds also increase reflectivity, and therefore our signal, due to their high albedo.

## 8.0   Internal Dynamics of Rocky Exoplanets and the Influence on Habitability

As terrestrial exoplanets evolve over billions of years, the cooling of their interiors drives outgassing, volcanism, surface recycling, dynamo generation, and might be a critical element in climate stabilization feedbacks. Those geodynamic processes depend strongly on planet mass, structure, composition, age, initial conditions, and maybe even the specific path of how the planet evolves. Therefore, space weather dynamics and interior planet evolution must be studied as a co-evolving complex system in order to be able to fully understand how the habitability of a terrestrial planet—the Earth or an alien planet—has evolved over time. In the following sections, we will discuss essential results on exogeophysics—geophysics as applied to rocky exoplanets.

### 8.1. Long-Term Thermal Evolution of Exoplanets.

The long-term thermal evolution of the interiors of rocky exoplanets over billions of years is driven by the decay of long-lived radiogenic nuclides such as Thorium, Potassium-40, and Uranium (e.g., McDonough & Sun, 1995 for the Earth's composition of radiogenic elements) and the release of primordial heat generated during the planet's formation and core differentiation processes.

The terrestrial planets in our own Solar system are differentiated into a rocky silicate mantle and a metallic core. It has been questioned whether super-Earths are also differentiated like the Earth (e.g., Elkins-Tanton and Seager 2008): rapid core formation for terrestrial planets (e.g., Kleine et al., 2002) requires widespread melting of the upper mantle leading to iron diapirs (a type of geologic intrusion in which a more mobile and ductily deformable material is forced into brittle overlying rocks) that sink towards the center via Stokes instabilities (Stevenson, 1990; Rubie 2007). The time scale of core formation depends on the viscosity of the lower mantle – the higher the viscosity the slower is core formation.

Assuming mantle and core separation, the main heat transport mechanism within a planet's silicate mantle is generally mantle convection due to large spatial and thermal contrasts in planetary mantles. In this situation, hot upwellings emerge at the core-mantle boundary and cold downwellings sink from the upper mantle (e.g, for an overview, see Schubert et al., 1969, 1979; Christensen, 1984; Turcotte & Schubert, 2002).

The thermal evolution of planetary interiors is commonly modeled with 2D or 3D mantle convection codes solving standard hydrodynamic partial differential equations of mantle flow using, e.g., codes like GAIA (e.g., Hüttig & Stemmer, 2008), StagYY (Tackley 2000), and Citcom/CitcomS (e.g., Tan et al., 2006; Zhong et al., 2000) amongst many others. Such calculations give detailed insight into specific aspects of mantle convection and are used to derive scaling laws applicable to the thermal state of planet interiors as a function of Rayleigh number—a non-dimensional property that indicates convective vigor (for a review of various Rayleigh numbers, see, e.g., Stamenković et al., 2012). Such computations are computationally expensive and not suited when hundreds to thousands of calculations are needed. For exoplanets large uncertainties in the planetary composition, structure, and initial conditions will inevitably remain. In combination with current-day uncertainties on how mantle rock behaves under high temperatures and especially high pressures, it is therefore critical to extend numerical convection model with simplified 1D thermal evolution models that are based on boundary instability theory or mixing length theory and use scaling laws derived from full mantle convection codes (e.g., Stevenson et al., 1983; or Stamenković et al., 2012 extended to super-Earths and pressure-dependent viscosity). Used together, both models, numerical convection and parameterized, can significantly expand our knowledge on the evolution of exoplanet interiors (e.g., Van Heck & Tackley, 2011; Foley et al., 2012; Stamenković et al., 2012; Stamenković et al., 2016).

In the last decade, various groups have applied such models to study the interior thermal evolution of exoplanets. Studied specifically were already discovered exoplanets (e.g., Cancri 55e (e.g., Demory et al., 2016), planets in the Trappist-1 system (e.g., Bourrier et al., 2017) amongst others), water worlds containing a significant mass-fraction of water (e.g., Sotin et al.,

2007; Fu et al., 2010), and Carbon rich planets (Stamenković & Seager, 2016) in order to understand in what way heat transport within exoplanets varies from the way heat is being transported within the Earth's interior.

At first driven by the discovery of super-Earths—considerable effort has been undertaken to study how planet mass, an essential exoplanet observable, affects heat transport within planetary mantles (e.g., amongst others, Gaidos et al., 2010; Van den Berg et al., 2010; Van Heck & Tackley, 2011; Stamenković et al., 2011, 2012, 2016). The cooling efficiency within rocky planets depends on the effectiveness of mantle convection, which strongly depends on a temperature- and pressure-dependent (amongst other factors) rock viscosity, with terrestrial mantle viscosity values in the order of ~$10^{20}$ Pas. Low viscosities are needed to effectively cool a planet's interior. In super-Earths, however, pressures within planetary mantles can reach up to 1 TPa. At such pressures, we lack experimental data to validate models simulating how the viscosity of mantle rock behaves. Early models studying the thermal evolution of exoplanets extrapolated a parameterized version for the Earth's viscosity as a function of temperature alone, and neglected any potential pressure effects. This critical assumption resulted in vigorous mantle convection for super-Earths due to their higher internal temperatures and hence smaller mantle viscosities (e.g., Papuc & Davies, 2007; Gaidos et al., 2010; Korenaga, 2010; Kite et al., 2009; O'Neill & Lenardic, 2007; Sotin et al., 2007; Valencia et al., 2006, 2007; Van den Berg et al., 2010; Van Heck & Tackley, 2011). However, as shown by Stamenković et al. (2011), for mantle rock minerals like MgO, perovskites, and post-perovskites, viscosity could also strongly depend on pressure. In such a case, numerical mantle convection models have shown that heat transport within the mantles of super-Earths might be sluggish (see, e.g., Stamenković et al., 2012). Other authors used steady state models (e.g., Tackley et al., 2013) to suggest that self-regulation of mantle viscosity, which generally works for smaller planets like the Earth and is known as the Tozer effect (Tozer, 1967), would take care of pressure effects by allowing the mantle to "quickly" heat up and reduce the viscosity to a level that facilitates vigorous convection and fast cooling (see Korenaga (2016) for a description of the Tozer effect). However, as shown in, e.g., Stamenković et al. (2012), the time scales needed for such self-regulation could be in the orders of many billions of years and are hence not compatible with an evolving planet. In this case "quickly" might not be fast enough.

Next to super-Earths, attention has been given to the thermal evolution of, yet hypothetical, carbon planets, which contain large amounts of SiC instead of SiO and might form in planetary systems that emerge in disks with large C/O ratios (Madhusudhan et al. 2012). The interior thermal evolution of carbon planets might significantly differ from that of silicate planets due to the very different thermal and transport properties of SiC in relation to SiO (Stamenković & Seager 2016). Specifically, theoretical models as well as experiments (e.g., Ghoshtagore & Coble 1966; Koga et al. 2005; Kröger et al. 2003, and Rüschenschmidt et al. 2004) suggest that the rheology of SiC could be much stiffer, resulting in a lower vigor of convection in carbon-rich planets in comparison to silicate counterparts (e.g., Stamenković & Seager 2016).

These examples illustrate how, when studying more massive rocky planets like super-Earths or planets with different mantle composition, planet evolution can start to significantly diverge from what we know from the Earth and our own Solar system. The discussion of impact of mantle composition on convection and interior structures can be found in Unterborn etbal. (2017); Dorn et al. (2018). Applying only scaling laws without questioning how fundamental principles might be different on exoplanets could constitute misleading approach.

**8.2 Evolution of Plate Tectonics on Exoplanets**

The heat flow through planetary mantles is strongly affected by the efficiency of crustal and lithospheric recycling and various surface boundary conditions. In this context, plate tectonics and stagnant lid convection reflect two extremes of how the surface exchanges with the planet interior. Plate tectonics occurs today only on one rocky planet in the Solar system, Earth, and allows for lithospheric and surface rocks to be recycled within the deeper mantle, with strong implications for the Earth's climate and its biogeochemical cycles as we shall explore in greater detail later. In stagnant lid convection, on the other hand, the lithosphere thermally insulates the bulk mantle and does not take part in the convective processes – this strongly affects the rate and type of volcanism. Stagnant lid convection is the prevailing mode on modern-day Mars (e.g., Spohn, 1991).

There have been great debates in the last ten years on the possibility of plate tectonics on exoplanets with planet masses, surface temperatures, structures, and compositions different from the Earth. Specifically, models initially explored whether super-Earths, hence rocky planets more massive than the Earth, were likely to facilitate a form of plate tectonics (e.g., amongst others Valencia et al., 2007; Van Heck & Tackley, 2011; Tackley et al., 2013; O'Neill & Lenardic, 2007; O'Neill et al., 2007; Stein et al., 2011, 2013; Noack & Breuer, 2014; Stamenković et al., 2012, 2016; Stamenković & Breuer, 2014; Lenardic & Crowley, 2012; Korenaga, 2010). Those studies focused mainly on the ability of super-Earth lithospheres to support mobile-lid convection (see for a summary of the various nuances of plate tectonics like convection, such as subduction, subduction geometry, plate failure, mobility, etc., e.g., Moresi & Solomatov 1998; Stein et al. 2004).

Stamenković & Breuer (2014) and Stamenković & Seager (2016) showed that the differences between the different groups could be traced back to different model assumptions—indicating how sensitive plate failure and subduction are affected by the assumed boundary conditions, mode of heating, rheology, and temperature profiles. Stamenković & Breuer (2014) and Stamenković & Seager (2016) showed however as well, that based on current day constraints including an uncertainty analysis of potential errors, generally an increasing planet mass (for planets more massive than the Earth) reduces the ability of plate failure due to decreasing shear stresses at the lithospheric plate base generated by a hotter upper mantle for more massive planets.

For the range of planetary conditions for silicate and carbon planets, the ideal candidate that would maximize the efficiency of plate yielding is an Earth-mass silicate super-Mercury planet with small concentration of radiogenic heat sources (~0.1 times the Earth's value) and as little iron as possible within its mantle (Noack et al. 2014; Stamenković & Seager (2016). As differentiation is highly likely to occur for Earth-sized planets (Breuer & Moore 2015), the characteristics of such "ideal candidate planets" would be a larger average rocky body density of ~7000 kg/m$^3$ (hence ~30% denser than an Earth-like analog)—which could be confirmed with simultaneous radial velocity and transit measurements. Carbon planets on the other hand do not seem to be ideal candidates for plate tectonics because of the slower creep rates and higher thermal conductivity for SiC in comparison to silicate planets (see Stamenković & Seager 2016).

The initiation and the maintenance of plate tectonics are affected not just by the planet mass, but by various factors from initial conditions, structure and composition (volatiles and radiogenic heat sources), surface temperature to evolutionary trajectory.The studies performed by different research groups suggest that the difficulty in understanding the plate tectonics comes from the attempts to image the multidimensional parameters space of plate tectonics into 2D – plate tectonics initiation versus planet mass.

## 8.3. Planet Habitability and Interior Evolution

### 8.3.1. Outgassing

The surface habitability of exoplanets will likely be significantly affected by interior processes. Volcanic outgassing is a strong function of planetary composition, upper mantle redox state (e.g., oxygen fugacity), planetary mass, and age as well as many other planetary properties, and will modulate a planetary atmosphere's composition and mass.

For planets with Earth-like compositions, outgassed material is mainly water and carbon dioxide but can vary widely for planets with more reducing planetary mantles (such as that on Mars which could lead to methane instead of carbon dioxide outgassing) or planetary interiors with a radically different composition (e.g., SiC rich mantles).

Outgassing processes vary strongly over the course of planetary evolution (for first 100 Myr see, e.g., Solomatov 2007). The early outgassing stage is characterized by magma oceans, where a large fraction of the planet's surface and mantle can be molten, followed by a later stage of mainly subsolidus mantle convection.

Specifically for water, during the early stages of magma ocean evolution in super Earths, outgassed fractions of water can range from less than 1mass% to up 20mass% of a super-Earth (1-30 $M_{Earth}$) terrestrial-like planet mass based on typical chondritic material compositions (see for a review, e.g., Elkins-Tanton & Seager 2008). Such early stage outgassing numbers are upper limits, as planets could have lost large amounts of water during the initial planet formation process. Moreover, as suggested by Hamano et al. (2013) planets with steam atmospheres at orbits with a stellar influx larger than about 300 $Wm^{-2}$ could have much longer magma ocean phases, possibly outgassing much more of their initial water in the first 10–100 Myr, and could thus result in planets that are much drier than those at greater distances from their host star.

For the post magma ocean phase, rocky planets cool mainly through subsolidus convection (see Section 8.1). Melt formation in the crust and upper mantle and subsequent transport to the surface through volcanic activity driving outgassing. As an example of a specific rocky exoplanet system with water outgassing, Bourrier et al. (2017) find for the TRAPPIST-1 system that, after the magma ocean phase, especially planets with more massive refractory parent bodies could be capable of outgassing water for much longer. In combination with the result that planets within the orbits of TRAPPIST-1d and TRAPPIST-1h could have entered the Habitable Zone within about 100 Myr to a few hundred million years, and considering that the largest atmospheric loss processes should have occured before entering the Habitable Zone, suggests that planets at more distant orbits and more massive planets could outgas up to one or two ocean masses of water after they entered the Habitable Zone. This emphasizes how late-stage geophysical outgassing might be a critical component for sustaining habitable environments on rocky planets as found in the TRAPPIST-1 system and beyond.

### 8.3.2 Space Weather and Geodynamics: approach

The surface habitability of exoplanets is affected by two major types of processes: from inside out via geodynamic processes and from outside in by exoplanetary space weather. These two aspects of exoplanetary habitability are both time-dependent and intertwined and have to be studied together for an evolving planet. It is highly questionable as to how much information we can gain solely from steady-state approaches as shown by Stamenković et al. (2012, 2016). In this context, work such as Bourrier et al. (2017) exemplify a reasonable way forward, uniting exogeodynamics with space weather by accounting for evolving planets and assuming for all geophysical aspects a probabilistic approach, which self-consistently takes the known geophysical uncertainties (thermal and transport properties; planet composition, structure, and initial conditions; as well as model uncertainties) into account and seeks for robust results (following Stamenković & Seager 2016).

### 8.3.3. Potential Impact of Plate Tectonics on Planet Habitability

Plate tectonics impacts both climate and surface redox state on the Earth, and thus our own planet's habitability, by 1) bringing oxidized material back into the interior and reduced rock to the surface (impacting the delivery of fresh nutrients), 2) by (partially) returning buried atmospheric carbon to the atmosphere via volcanism caused by subduction (on geological time scales rather continuously), and 3) by causing continental uplift at subduction zones, which actually moves oceanic carbon to long-term storage on massive cratons and also leads to formation of more exposed land (depending on amount of surface water ), which affects climate and also cools the mantle. Hence, the latter process will lead to a higher heat flux at CMB, and thus can help to maintain magnetic dynamo and associated planetary magnetosphere (Stamenković et al. 2012).

The latter process might aid burying carbon and building up oxygen levels (e.g., Wilson cycle, see, e.g., Falkowski & Isozaki, 2008) and might therefore be possibly necessary to satisfy the large oxygen demand for mammals (e.g., Catling et al., 2005). Up to now, the exact role of plate tectonics impacts on atmospheric composition and climate is a critical field to be explored.

Although plate tectonics impacts a planet's climate and biogeochemical cycles, it is not clear whether it is really necessary for complex surface habitability over billions of years on an Earth-sized planet – especially because stagnant lid convection on such planets is also associated with long lasting strong volcanism and offers many mechanisms that could mimic the processes used to legitimate the necessity of plate tectonics for life on Earth (such as long-time volcanic $CO_2$ outgassing and carbon burial). The link between plate tectonics and habitability on super-Earths (rocky planets 1-10 times more massive than Earth) is, however, strengthened by previous studies (Stamenković et al., 2012), which indicate that for a variety of initial conditions, volcanism without plate tectonics might be limited to smaller planets due to melt extraction issues on more massive rocky planets (Kite et al. 2009; Noack et al. 2017; Dorn et al. 2018).

Although it is not clear to what degree plate tectonics is needed for creating habitable environments, it remains certain that we must deepen our understanding of the planetary conditions that impact the tectonic mode of a rocky planet in order to: 1) understand the Earth's biogeochemical cycles and long-term climate variability (especially the carbon cycle and the rise of oxygen), 2) to understand how common Earth like planets are in the Milky Way, and 3) to reduce false-positive biosignature gases caused by geophysical activity on an alien rocky exoplanet—by being capable to assess the tectonic mode (and hence tectonic type related volcanic fingerprints) of an exoplanet (see Stamenković & Seager, 2016). Tosi et al. (2017) and Foley (2019) suggested that Earth like planets could maintain a habitable period without plate tectonics. Therefore, studying plate tectonics for a diversity of planets unites solid Earth, Solar system, and exoplanet research in one field.

### 8.4. Effects of Tidal Heating on Volcanism and Habitability of Close-In Exoplanets

The classical habitable zone (Kasting et al., 1993) around low luminous M dwarf stars occurs at sufficiently short orbital periods for tidal effects between the star and planet to become relevant for a planetary energy budget which affect climate conditions. Terrestrial planet interior processes outside this region are generally dominated by long-term radionuclide decay. In radiogenic-only situations, the heat flux from the interior, as well as corresponding metrics such as the rate of outgassing, volcanism, and convective vigor, are primarily determined by the size and age of the planet, along with the metallicity of the planet host that determines the radiogenic heat sources in an exoplanetyary system. If an exoplanet has an Earth-like plate-tectonics, then carbonate-silicate cycle and recycling of other atmospheric components chemically trapped at the surface become important in regulating an exoplanetary climate (Walker et al. 1981). When tidal heating is relevant, primary control of the internal heat flux, and interior-atmosphere interactions, switches to the present value of the orbital eccentricity, planetary spin rate, and planetary obliquity.

Each of these terms may lead to heating rates far higher than Earth-like radiogenic isotope decay would ever provide (e.g., Ferraz-Mello et al., 2008; Jackson et al., 2008; Barnes et al., 2009; Henning et al., 2009; Barnes et al., 2008; Henning & Hurford, 2014). Tidal heating generally leads to damping of eccentricity, spin rate, and obliquity, at rates inversely proportional to the intensity of heating. Control of the overall intensity of this process is dominated by semi-major axis (a) to the power of -7.5. Obliquity tides damp rapidly and are often weak. Tides induced by non-synchronous spin are generally intense, and can either be short lived or long term depending on initial spin rate or presence of a moon(s). Further issues relating to spin-synchronization are discussed in Section 8.5. Tidal heating due to nonzero eccentricity, as a primary driver of activity on Io, Europa, and Enceladus, may be very long-lived and intense, if an orbital resonance exists. The strongest and most stable such resonance, a mean-motion resonance, exists when there is an integer ratio between orbital periods of co-orbiting objects, such as Io and Europa. Tides may also strongly alter habitability if a large exomoon or binary-planet system exists (Heller & Barnes, 2013).

Several outcomes are worth special attention. First, consider worlds similar to TRAPPIST-1g and TRAPPIST-1h (Gillon et al., 2017). These planets lie outside the classical surface-temperature-based habitable zone based on instellation alone. Multiple parameters however make them ideal candidates for sufficient internal heating to sustain extensive liquid water oceans underneath ice shells: they are large enough that radionuclide-containing cores are expected to supply non-negligible levels of internal heating, even though the exact rates are highly uncertain. Furthermore, they exist in a crowded orbital neighborhood, were perturbations are strong; with multiple bodies in the TRAPPIST-1 system close to mean motion resonance (Luger et al, 2017), allowing for sustained non-zero eccentricity maintained in opposition to tidal damping. Finally, stellar distances, sizes, and suspected compositions imply tidal heating rates in a range appropriate to provide liquid water if the surfaces are water rich (Barr et al., 2017; Turbet et al., 2017).

A scenario such as this is important because it circumvents stellar radiation issues for habitability, by providing conditions for life's origin and evolution, which comes with a thick shield: in the form of a few kilometers of surface ice. Both tides, and tides in combination with radionuclides, can lead to underground liquid water for worlds otherwise outside the instellation-only ice line. Such a world would have shielding from radiation for organisms at depth within the ocean, and perhaps not-insurmountable barriers to chemical exchange between the atmosphere and liquid habitat. Long-term escape of modest internal heat rates can lead to ice shell convective motion, surface cracks, cryovolcanism, and diapirs, and thus some degree of chemical communication between the ice surface and subsurface ocean (e.g., Kargel et al. 2000). Such chemical transport will generally be reduced for cases with thicker ice shells, and low internal heat flux, such as cases of conductive heat transport only. Tidal flexure and the focusing of tidal dissipation into viscoelastic ice shells, as occurs for Europa and Enceladus, additionally aids in making ice shells chemically permeable (in both directions, up and down) for improved transport of biologically important molecules.

It is not yet known how common close-packed terrestrial systems at short periods around M dwarf stars are, however the discovery of the TRAPPIST-1 system itself is suggestive.

It is important to consider that internal heating is not always favorable to habitability, and may also cause worlds to be too hot (Henning et al. 2009). In the extreme limit, planets with large-scale surface magma oceans, with temperatures maintained by combinations of stellar heating, radiogenic heating, and tidal heating, are possible. The short period M dwarf planet 55 Cancri e is one possible example of such a world, although this possibility is far from confirmed (Demory et al., 2015).

Moderate levels of eccentricity and subsequent tidal activity, even when not playing a strong role in controlling surface temperature, can still be a factor worth investigating with regards to the biochemistry of habitability, via outgassing and volcanism. However, this is true mainly in cases when the XUV driven escape rate is still less than any such outgassing rates. High internal heat rates imply both high volcanic outgassing, and high overturn or depletion rates for volatile components within a planet's mantle. $CO_2$ and $H_2O$ are the two dominant gases in all volcanic eruptions on Earth. This is expected to remain true for exoplanets with the same volatile composition and redox state (Kaltenegger et al., 2010). We should note that in the case of a reduced mantle, outgassing will include primarily $H_2$, $H_2O$ and $CO$ (Schaefer and Fegley 2017). $SO_2$ and $H_2S$ are typical second tier volcanic gas components. High release rates of $CO_2$ in particular may have a critical role on climate, and must be considered carefully in conjunction with the presence/absence of a surface ocean or exposed silicate surfaces for carbon cycle feedback. Transient $SO_2$ and $H_2S$ spectral signatures provide possible means to identify worlds with active and extreme volcanism (Kaltenegger et al. 2010). Note that plate tectonics is not necessary for volcanism to assist biochemical habitability, as even volcanic sources on Earth that are sourced from the deep mantle still provide volatile elements to the surface for very long timescales, including non-zero $CO_2$ that can accumulate and break planets free from snowball-Earth (Kirschvink, 1992) style episodes.

Exoplanets may also experience modest tidal dissipation rates, especially near M-dwarf hosts, due to less-studied dynamical processes such as tides due to librations, shell adjustments, or due to ongoing forced small-angle obliquity as occurs for an object in a dynamical Cassini state (Ward, 1975).

For planets around stars more luminous than M dwarfs, star-planet tides play less of a role for habitability, as the larger host distance of the classical habitable zone does not support significant tidal dissipation. For planets or moons at arbitrary distances from a luminous host (including "free-floating" worlds not bound gravitationally to any host star (Stevenson, 1999; Osorio et al., 2000; Abbot & Switzer 2011; Badescu 2011)), many configurations of planet and moon can lead to subsurface liquid water. Observational constraints for the occurrence rates of "free-floating" systems (Debes & Sigurdsson 2007) at the terrestrial-planet mass level are lacking, and remain limited even for Neptune and Jupiter mass primaries, however early microlensing statistics (Sumi et al., 2011; Schild et al., 2012; Mróz et al., 2017), imply they are sufficiently numerous to contribute significantly to overall galactic habitability. A large exomoon around Kepler-1625b was recently discovered, but the distribution of exomoon properties remains poorly known (Teachey et al., 2018; Teachey & Kipping 2018). Still, if Europa and Enceladus alone are reasonable examples, the volume of liquid water in the galaxy may be dominated by such subsurface environments more so than by exposed surface oceans such as on Earth. Thin-ice shell ocean worlds, allowing chemical communication with the silicate interior and surface, but shielded from stellar radiation, may be highly favorable for life analogous to the Precambrian ocean-only ecosystem of the early Earth. The presence of high pressure ices however, may prevent an ocean from having direct contact with a rocky core below, for objects approximately of the mass of Ganymede or higher (Noack et al. 2016).

**Terrestrial Planet Spin Synchronization**. When tidal dissipation induced due to non-synchronous spin is of sufficient magnitude, a planetary body will evolve toward a state of spin synchronization, in the same manner that one face of Earth's moon is synchronized to face the Earth, in what is called a 1:1 spin-orbit resonance. Mercury however is locked in a similarly stable 3:2 spin orbit resonance (Orbital period 87.969 days, spin period 58.646 days).

Short period exoplanets, especially near M dwarf stars, are often assumed to exist in 1:1 spin-orbit resonance states leading to strong dayside-nightside differences in instellation and space weather exposure. This is based on the typically short (1-10 Myr) timescales for the damping of plausible initial spin rates. Strong temperature dichotomies and evidence for hypersonic dayside-to-nightside circulation on Hot Jupiters such as HD 189733b (Majeau et al., 2002; Knutson et al., 2007) support this expectation, but cannot yet discern between exact-1:1 and near-1:1 (pseudo-synchronous) spin-orbit states. For worlds that have synchronized, internal heating may allow for habitability on nightsides that are naturally shielded from the majority of stellar radiation. Heat transport via a thick atmosphere may also help to maintain small temperature differences between daysides and nightsides.

Recent studies based on advances in modeling dynamical outcomes that may arise from more realistic models of viscoelastic silicate and icy mantle materials suggest capture into higher-order spin orbit resonances may be more common than previously expected (Makarov, 2012; Makarov et al., 2012; Makarov & Efroimsky, 2014). In particular, use of the Andrade rheology, a model of the frequency dependent response of silicate or ice to stress that shows unique success in the laboratory, suggests that second-order effects in the despin process may prevent widespread 1:1 spin orbit resonance. Similar effects may also alter the prevalence of the phenomenon of pseudo-synchronization, whereby the equilibrium spin rate is greater than the orbital period by magnitudes in the range of 3% (Murray & Dermott, 1999). Pseudo-synchronous rotation is no longer expected to be a stable outcome for fully solid worlds (Makarov & Efroimsky, 2013), but be more likely than 1:1 resonance for worlds with a liquid, semi-liquid, or partial melt layer (Makarov, 2015), including silicate asthenospheres and water oceans. Exact 1:1 spin synchronization is a time-averaged phenomenon, and the degree of dayside-nightside libration or drift is predicted to depend strongly on internal layer structure and the existence of any surface or internal liquid layers. Slightly inexact spin-orbit resonances are also found as solutions when advanced material models are tested. Significant planetary asymmetry or triaxiality (e.g., Olympus Mons or the lunar crustal dichotomy) increases the probability of dynamical capture into a strong exact 1:1 spin-orbit resonance (the current state of Earth's Moon), whereas lack of an asymmetric (quadrupole) term for torque has generally been seen to favor pseudo-synchronization. Atmospheric torque has historically been regarded as a key control in the spin history for Venus. Crucially, Leconte et al. (2015) find that even a thin atmosphere can be sufficient to disturb the gravitational torques maintaining 1:1 spin-orbit synchronization. Other factors such as rapid atmospheric circulation induced by instellation, and regular shifts in the mass distribution of the planet (e.g., true polar wander), may also act to upset strict dayside-nightside divisions.

Resolving such issues will be critical for the assessment of the habitability of exoplanets of interest. In the meantime, climate models and star-planet interaction models for worlds at M dwarf stars should consider both 1:1 synchronous cases, pseudo-synchronous cases, and higher-order spin-orbit cases such as a 3:2 resonance.

### 8.5. Generation of Intrinsic Magnetic Fields of Rocky Exoplanets

An intrinsic (of the internal origin) magnetic field of a terrestrial planet is vital for protecting life and habitable environment from harmful electromagnetic radiation from sources external to the planet, such as the host star. The existence and the evolution of the magnetic field are among the key science questions for understanding planetary habitability. In the remainder of this section, the intrinsic planetary magnetic fields are simply referred to as the "planetary magnetic fields".

Generation of the planetary magnetic fields is governed by the "dynamo theory" (Larmor 1919), which can be summarized as that a planetary field is generated and maintained by convection of electrically conducting rotating fluid within the planet, as illustrated in Figure 24. Thus, the planetary field requires two fundamental physical conditions: a planetary-scale metallic fluid core, e.g. the Earth's molten outer core, and a strong driving force to stir up the fluid motion, such as the thermal and compositional buoyancy forces from planetary secular cooling and differentiation (Braginsky and Roberts 1996), and driving forces from planetary rotation variations (Tilgner 2005). The secular cooling and the differentiation of a planet, and their consequences on e.g. solidification of the solid inner core at the center, depend ultimately on the dynamical processes in the overlaying mantle (e.g. Buffett *et al.*, 1996; Nakagawa and Tackley, 2013). Therefore, generation of a planetary magnetic field depends on the core dynamics, and on various interactions within the planet and with astronomical systems external to the planet.

The working of the dynamo theory is very simple, but the resulting planetary fields can be very complex, as observed from our own Solar system: Earth has a strong (slightly tilted and dipole dominant) magnetic field over its accessible history (e.g. Biggin *et al*, 2015); Mercury has a very weak (off-centered dipole) field at present (Anderson *et al,* 2012); Mars and Venus are currently non-magnetic. However, Mars had a strong magnetic field in its early history, as shown from the strongly magnetized ancient surface in the southern hemisphere detected by NASA Mars Global Surveyor (MGS) mission (Mitchell *et al*, 2007). But no similar conclusion could be made to Venus simply because of its young and hot surface. These observations suggest that a planetary magnetic field could change significantly over geological time scales, and its life-span can even be much shorter than that of the planet.

Substantial progress has been made in our understanding of the generation and the evolution of the planetary magnetic fields with large-scale numerical dynamo simulations over the past two decades (e.g., Roberts and King, 2013; Stanley and Glatzmaier 2010). However, there are many fundamental questions to be answered, such as the amount of heat transport across the core-mantle boundary (e.g., Karato, 2011; Stamenković et al., 2011; Tackley et al., 2013), and on the thermo-chemical properties of the metallic cores (e.g. Valencia *et al.* 2006; Stamenković *et al.* 2011, 2012). Their impacts on the magnetic field generation can be far more complex than what we have understood at present time. For example, recent results of Kuang et al. (2019) indicate that the generation of planetary magnetic fields can be significantly weakened or even terminated if the inner core grows to beyond 50% of the outer core's radius, even though the inner core growth is vital for producing the necessary buoyancy force to drive the core dynamo. Moreover, simulations suggest that our planet could lose its magnetic field, and thus the planetary magnetospheric protection from space weather effects including solar wind and energetic particles, in as short as 700 Myr with the current growth rate of the inner core (Tarduno et al. 2015; Bono et al 2019). Obviously, the age and thus the growth rate of the inner core depends on the geochemical properties in the Earth's core, e.g. smaller thermal conductivities can result in lower inner core growth rates (Labrosse 2015; Hirose *et al* 2017). However, these may not necessarily lead to longer geodynamo life: slower inner core growth rates imply lower amount of energy available for the geodynamo which, in turn, implies smaller inner core radii necessary to maintain the geodynamo. These results have far reaching implications for habitability of Earth-like planets around active stars including M dwarfs and imply that the planet's age should be an important habitability factor.

More observational, experimental and theoretical studies are needed for understanding the generation and evolution of the planetary magnetic fields, and for addressing the fundamental science questions directly relevant to habitable worlds, such as potential causes of destabilizing the field (e.g. magnetic polarity reversals), impacts of planetary interior structures (e.g. geometry and boundary changes) and their chemical composition (metallicity) and star-planet-moon interactions (e.g. tidal forcing on planetary fluid systems) on the planetary magnetic fields.

The magnetized planets in our solar system provide valuable diverse dynamic systems for us to research, test and validate scientific hypothesis and models, and thus provide pathways to extrapolate results to the diversity of magnetized exoplanets. Their magnetospheres can be detected through the radio signatures introduced by the effects of the interactions between close-in exoplanets with astrospheres of their host stars. A unique Jupiter-Io interacting system, in which Io orbits within the Jupiter's magnetosphere and produces the UV-bright aurora (Mura et al. 2018), can serve as a good example of the planet-moon system that possibly could be detected in the closest exoplanets including TRAPPIST-1 and Proxima Centauri systems using upcoming Low Frequency Array (LOFAR) radio observations (Turnpenney et al. 2018).

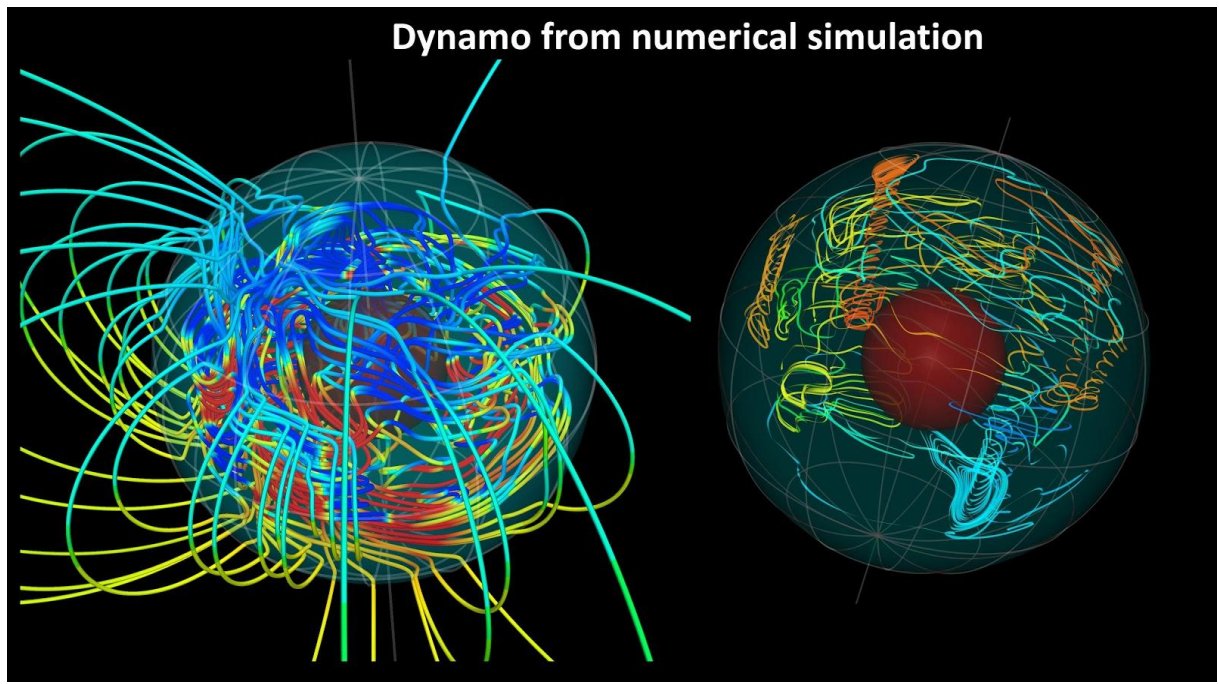

*Figure 24. Simulated planetary dynamo in a fluid outer core (between the transparent green surface at the top and the opaque red surface near the center). The helical chaotic convective flow (on the right) generates and maintains a very complicated magnetic field (on the left) spreading out from the core to the exterior of the planet (Kuang et al. 2019).*

**9.0   Observational Methods and Strategies for the Detection of Habitable Planets**

The observational picture for the detection of potentially habitable Earth-sized exoplanets is bright due to continued improvements in the methods that have been used in the past 22 years after the initial discoveries of exoplanets around solar-type stars by Mayor and Queloz (1995) and by Marcy and Butler (1996). These researchers used the precision radial velocity method, which measures the very small Doppler shift in the velocity of the host star along the line of site to the observer, caused by the motion of the host star about the center of mass of the two objects. In this section we focus on the strengths and limitations of each method, particularly their observational biases.

The incredible progress made in the last two decades has relied strongly on utilizing results from multiple techniques and combining them so that important physical attributes of an exoplanet can be determined, including its orbital parameters, size, effective temperature or insolation, and where enough information is available, its bulk density. Once the density is known then it is possible to determine, within appropriate observational and theoretical constraints, whether or not a given detected exoplanet is, for example, a rocky super-Earth with a thin atmosphere and the potential for a magnetic field that protects its atmosphere, or a mini-Neptune with a thick hydrogen atmosphere. Historically, most exoplanets were discovered using the radial velocity method until the photometric precision of the transit method improved to the point that it was sufficient to detect exoplanets. Both of these methods have a bias towards more massive or larger exoplanets that are relatively close to the host stars. The reason for the bias towards close-in planets for radial velocity is the relatively larger and easier to measure amplitude of the doppler shift for a given planet mass and observational precision. The bias towards close-in planets for transit observations is related to the probability of the orbital plane of the exoplanets being close to edge-on so that the transit can be viewed along the line of sight to the observer. The likelihood of successfully observing a transit is proportional to the ratio of the diameter of the host star to the semi-major axis of the orbit of the planet. Secondly, it is also dependent on the area of the exoplanet itself compared to that of the host star, which determines the size of the signal, while the instrumental photometric precision determines the noise level.

The radial velocity method is discussed in more detail in Section 9.1 and the transit method in Section 9.2. Both methods also favor geometries that are either edge-on (transits) or within a range of angles with greatest sensitivity to mostly edge-on geometries, but limited to measuring the mass times the sine of the inclination angle (precision radial velocity).

In recent years, the gravitational lensing method has been employed to search for exoplanets beyond the snow line. This requires larger separations between the host star and planet than is accessible with the first two, on the order of several AU from the host star, typically around 4 AU or greater around a G-type star. Thus, it fills in an important observational gap. Much more will be known about these exoplanets in the future when the WFIRST mission (Spergel et al. 2015) completes its gravitational lensing observations in the 2020s. The gravitational lensing method also allows searches for exoplanets that are not bound the host star, the so-called "free floating" exoplanets (Penny et al. 2014). These may be exoplanets that have been ejected from protoplanetary disks during the planet formation epoch, or at other stages of planetary system evolution. However, because the focus of this review is on exoplanets in the habitable zones of their host stars, we only mention it briefly for completeness.

Section 9.3 discusses the fundamentals of direct imaging techniques, which are further developed in Sections 9.4 and 9.5 discussing direct imaging methods in the visible and infrared, respectively. Direct imaging methods hold the most promise for the future, because they allow for all geometries, from face-on to edge-on, and with them, spectroscopic methods are employed that allow for detection of biosignature gases for exoplanets within the habitable zones of the host stars. A possible future ground-based observatory for the detection of exoplanets around nearby host stars is discussed in Section 9.6.

### 9.1. Radial Velocity Method

The radial velocity method was used to discover the first exoplanets as noted in the Introduction to this section. It is beyond the scope of this paper to discuss this technique in great detail. The paper by Lovis and Fischer (2010) provides a good introduction and overview. Here we provide a brief history and a discussion of the essential physics of the method. The radial velocity method has been extremely successful with >775 confirmed exoplanets as of November 2018 (http://exoplanet.eu/catalog).

The quantity measured by the radial velocity method is the semi-amplitude, $K_1$, which can be written in terms of quantities scaled to a Jupiter mass, the star's mass scaled to the Sun's mass, and the orbital period in years (adapted from Eq. 14 of Lovis and Fischer 2010),

$$K_1 \sim (28.4 / (1-e^2)^{1/2}) \, (m_p/M_{Jup}) \, \sin(i) \, (M_*/M_{Sun})^{2/3} \, (P/ 1 \text{ yr})^{1/3} \text{ m/sec},$$

where $m_p$ is the mass of the exoplanet, $M_{Jup}$ is Jupiter's mass, $M_*$ is the stellar mass, $M_{Sun}$ is the mass of the Sun, where $e$ is the eccentricity. If $K_1$ is observed over one or more complete orbital periods it is possible to determine $m_p \sin i$, the eccentricity, $e$, and the period, P. It is also necessary to independently estimate the stellar mass, $M_*$, which can be done by a variety of methods, in order to determine *$m_p$ sin i*.

A Jupiter mass planet at 1 AU has a semi-amplitude of 28.4 m/sec, whereas an Earth mass planet at 1 AU has a semi-amplitude of only 0.09 m/sec. A super-Earth of 5 Earth masses would have a semi-amplitude of 0.45 m/sec at 1 AU, and 1.4 m/sec at 0.1 AU. Thus, for a given RMS semi-amplitude precision and stellar spectral type, radial velocity detections are limited to a region of exoplanetary masses and periods that provide signals roughly of this magnitude. Repeated observations can improve the sensitivity somewhat below this naïve limit, particularly for shorter period exoplanets, where folding the radial velocity time series into a single-phase curve and averaging is possible.

The major limitation of the radial velocity technique for an exoplanet observed with this method is that only the minimum mass is measured, i.e., *$m_p$ sin i*, so a measurement of the inclination angle is necessary to determine the true mass. This can be done through astrometric means in which the orbit of the host star around the center of mass of the system is measured. This can also be accomplished through direct detection methods, if the angular separation between star and the exoplanet is larger than the inner working angle of the coronagraph or nulling interferometer, and sufficient star-planet contrast is achieved. For statistical studies, however, this limitation may not matter as much, since the majority of observed inclination angles for randomly distributed inclinations will be between 45 and 135 degrees. Other method to measure inclination angles relies on the observations of dust belts (Anglada et al.2018).

Over the past several years, improvements to the precision of such measurements have been a major focus of the radial velocity community with the goal of achieving a precision of ~ 0.1 m/sec, sufficient to detect Earth-mass exoplanets. Besides improving the measurement precision to measure smaller exoplanet masse the community is also extending this technique to the near-infrared in order to increase the sample to include more low-mass cool stars.

Currently there are many spectrometers around the world that are being used for such exoplanet searches; essentially all are echelle spectrometers with very high resolution, of the order of R~80,000 to 100,000. Unfortunately, it is beyond the scope of this paper to discuss these in detail. Listings of current and planned spectrographs can be obtained in the CARMENES website (http://carmenes.caha.es/ext/spectrographs/index.html), and also the web page (http://pendientedemigracion.ucm.es/info/Astrof/invest/actividad/echelle.html#spectrographs). An excellent reference discussing progress and challenges in the radial velocity technique is in the whitepaper from the Exoplanet Program Analysis Group (EXOPAG), Science Analysis Group (SAG), SAG-8, see Plavchan et al. (2015).

CARMENES, short for Calar Alto high- Resolution search for M dwarfs with Exo-earths with Near- infrared and optical Échelle Spectrographs, is an important instrument as it is one of the first to operate in the near-infrared at the Y, J, and H bands, from 950-1700 nm, with a resolution of about 85,000 and is in operation at the Calar Alto Observatory in Spain on the 3.5-m telescope. This instrument also has a visible wavelength channel, with a resolution of about 80,000, and is designed for a precision of $\leq 1$ m/sec. It is optimized for searches around nearby late M-dwarf stars, thus opening up discovery space around this important class of low-mass stars. (see Quirrenbach et al. 2014 for a more detailed description of this instrument and its performance).

Another instrument that is currently under development is the Extreme Precision Doppler Spectrometer (EPDS), led by Dr. Suvrath Mahadevan at Penn State University. This instrument is being funded through NASA in a joint program with the NSF to be installed on the 3.5-m Wisconsin-Indiana-Yale-NOAO (WIYN) telescope at the Kitt Peak Observatory. This instrument has a minimum precision of 0.5 m/sec with the ultimate goal of reaching $\leq 0.1$ m/sec precision, sufficient to detect Earth-mass planets around nearby stars. Note that even if such an instrumental precision is achieved, stellar effects from convection and granulation, for example, can obscure the signal from the motion of the star about the center of mass of the system, since these effects can be of comparable size. This instrument will also be able to provide the radial velocity measurements needed for mass and orbital determination for the large number of exoplanets expected to be discovered by the TESS satellite, which was launched in 2018.

**9.2 The Transit Method and Transit Spectroscopy**

The transit method has been the most successful method of all of those employed so far, principally due to the incredible success of the Kepler mission. As of this writing there are over 3000 confirmed exoplanets detected with this method. In principal, the transit method is very simple: a planet passes in front of the host star viewed along the line of site to the observer, and a small dimming of the brightness of the host star can be observed. This is called the primary transit. When the exoplanet passes behind the host star, then a secondary transit occurs, which usually called the eclipse or occultation. There are many good references discussing the transit method and transit spectroscopy. An overview is given in the review paper by Winn (2010).

Fundamentally, the observational problem is that the amount of dimming of the host star is very small, as can be seen from a simplified formula for the transit depth, assuming $R_p$ is the radius of the planet, and $R_*$ is the stellar radius, then the transit depth $\delta_{tra}$ is $\delta_{tra} \approx (R_p/R_*)^2$, where the night side emission from the exoplanet is assumed to be negligible (simplified from Eq. 22 of Winn (2010)).

For a Jupiter-sized planet around a star like the Sun, the transit depth is $\sim 10^{-2}$, whereas the transit depth for an Earth-radius planet is $\sim 10^{-4}$. This means the photometry should be substantially better than 100 parts per million (ppm), at least during each measurement. Systematic drifts in the photometry can be overcome by careful fitting, and repeated observations of a transit can be overlapped in a phase curve, increasing the signal-to-noise for a given detection as is done for radial velocity observations. A similar approximate formula for the depth of the secondary eclipse is, e.g., Winn (2010), Eq. 23: $\delta_{ecl} \approx (R_p/R_*)^2 \, I_p(t_{ecl}) / I_*$, where, $\delta_{ecl}$, is the eclipse depth, and $I_p(t_{ecl})$ is the emission or scattered light from the planet's dayside just before the planet begins its ingress behind the star, and $I_*$ is the stellar intensity.

The probability of a transit varies, largely depending orbital distance of the planet to the star. In the case when the planet is much smaller than the star, and the orbit is circular, a simple formula can be derived that helps one understand how many stars are needed to be observed in a transit survey like Kepler to achieve the desired number of detections for stars of any given spectral class and exoplanets in various size regimes, as in Eq. (11) in Winn (2010). In this equation, the symbols $p_{tra}$ and $p_{ecl}$ are the probabilities of a transit or eclipse respectively, a is the semi-major axis, $p_{tra} = p_{ecl} \approx 0.005 \, (R_* / R_{Sun}) \, (a / 1 \, AU)^{-1}$, giving a probability of transit of 0.5% or 1/200 for an exoplanet at 1 AU for a star of the same radius as the Sun.

The transit probability is only 1/20 or 5% for an exoplanet at 0.1 AU. This equation makes it clear that an observer needs to have a large sample of stars in order to obtain a statistically meaningful number of detections of exoplanets with this method. For Kepler, in order to determine the probability of a solar-type star to have an Earth-sized exoplanet in the habitable zone required a sample of roughly 100,000 stars. If the entire sample consisted only of G-type stars like the Sun, then there would be 500 such detections. However, G-type stars are less than 10% of all stars in the galaxy, giving a sample size of roughly 50 stars, if the survey lasted sufficiently long (on the order of 3-5 years) in order to make all possible detections. Also note that there is also a distribution in eccentricities that we have neglected.

Thus, sample size is one issue for transiting planet observations. Another issue is that the detection system noise should be low enough to be close to the photometric limits. A third limitation is that transits last for a few hours at most. For the case of relatively low signal to noise, the characteristic timescale of a transit, $T_C$, adapted from Eq. 19 of Winn (2010), is $T_C = R_* P / (\pi a) \approx 13 \, hr \, (P / 1 \, yr)^{1/3} \, (\rho_* / \rho_{Sun})^{-1/3}$, where $\rho_*$ and $\rho_{Sun}$ is the mean density of the star and the Sun, respectively. From this equation, for a star of the same density as the Sun and a planet with a 1-year orbital period, the characteristic time scale is only a few hours. This limits the observations to have a characteristic cadence, hence integration time per measurement, which much be significantly less than $T_C$ or important details in the transit light curves would be lost.

The final tool in the transit method toolkit is transit spectroscopy. As a planet with an extended atmosphere transits a star the effective radius of the planet changes as a function of wavelength due to the varying optical depth of the molecules and atoms in its atmosphere. This effect can be strong when the wavelength is at a particularly strong molecular or atomic transition. The size of the atmosphere will change by a few scale heights, H, which is given by $H = k_B T/(\mu_m g)$, where $k_B$, T, $\mu_m$, and g, are Boltzmann's constant, atmospheric temperature, mean molecular weight, and planet's surface gravity, respectively.

If $R_\tau$ is radius at which the optical depth is greater than unity at all wavelengths, then the fractional change in the transit depth is given by $\Delta\delta / \delta \approx 2 N_H (H/ R_\tau)$, where $N_H$ is the number of scale heights, and the other symbols were previously defined. For an Earth-like planet (i.e., temperature, surface gravity, and mean molecular weight of its atmosphere) around a star similar to the Sun, the effect is of the order of $10^{-2}$, assuming only a single scale height (see Eq. 36 in Winn (2010) for more details). The result is that the photometer must have an overall precision better than $10^{-6}$ to measure these very small changes in transit depth to characterize the atmospheres of Earth-sized planets.

The eclipse depth gives the ratio of the star to planet disk-averaged intensities as discussed previously, however, at wavelengths where the Rayleigh-Jeans approximation can be used for the Planck function, the eclipse depth as a function of wavelength reduces to $\delta_{ecl,them}(\lambda) \approx \delta_{tra} (T_b / T_*)$, where $T_b$ is the brightness temperature of the planet (disk averaged), and $\delta_{tra}$, and $T_*$ are as defined previously. This equation shows that if the primary transit is measured and the stellar temperature is well-known, then it is possible to determine the brightness temperature as a function of wavelength. For the case where scattering dominates, the eclipse depth depends on the wavelength dependent geometric albedo, $A_\lambda$, and the amount light scattered or reflected towards the observer, which depends on the planet radius, $R_p$, and the semi-major axis, a, such that $\delta_{ecl,scat}(\lambda) \approx A_\lambda (R_p / a)$. For a Jupiter sized planet at 1 AU the eclipse depth is of the order of $10^{-4}$, and very small for an Earth-sized planet at 1 AU, of the order of one part per billion (see Winn (2010) Eqs. 37-39 for further details).

Details of how transit and eclipse broad-band and spectroscopic observations are reduced and fitted to models of the exoplanets and their atmospheres are beyond the scope of this paper. However, there are other more recent reviews focusing on the observational literature and on potential future observations with JWST, in particular Seager and Deming (2010). For future observations, particularly with upcoming telescopes like JWST, we refer the reader to e.g., Beichman et al. (2014), Cowan et al. (2015), Greene et al. (2016).

### 9.3 Direct Imaging of Rocky Exoplanets

We now turn to direct imaging of rocky exoplanets, which here means exoplanets ranging in size from that of Mars with a radius of ~0.5 $R_E$ to Super Earths with radii of up to ~2 $R_E$. The requirements to detect and characterize rocky exoplanets are daunting, since the light reflected by the Earth is ~ $10^{-10}$ that of the Sun at visible wavelengths near 0.5 and ~$10^{-7}$ emitted by the Earth at 10 μm. Thus, technologies that can achieve such extremely high contrast ratios are required. Besides this high contrast requirement, direct imaging systems also require high angular resolution. For example, if the Solar system is viewed from a distance of 10 parsecs, the Earth would have an angular distance of ~ 0.1 arcsec from the Sun.

The search for life in exoplanets around nearby stars is fundamentally a search for molecular constituents in the atmosphere, or biosignature gases (as discussed in Section 7), e.g., which are out of chemical equilibrium due to the presence of life. This means that at a minimum a low-resolution spectrograph is needed in addition to broadband filters that are needed for the detections. This requires a minimum size for the telescope so as to detect enough light in a narrow bandwidth of each spectral resolution element within a given exposure time.

Clearly, the orbital parameters of the exoplanets will also have to be measured, not just the semi-major axis, but also the eccentricity, inclination, and line-of-nodes are necessary to characterize an exoplanet, and to determine its potential habitability. This is particularly important for direct imaging missions, in which the minimum angular resolution required depends on the maximum separation between host star and exoplanet, which occurs during maximum elongation.

These considerations also imply that if an exoplanet is not detected by an indirect method in which at least some orbital parameters can be determined as described in the previous sections of this section, a revisit strategy is necessary in order to obtain a sufficient number of observations to adequately determine the orbital parameters.

Modern visible and near-infrared (wavelengths shorter than the thermal infrared) coronagraphs create a dark hole in the region around the location of the star, and the contrast varies from the Inner Working Angle (IWA) with the worst performance to an Outer Working Angle (OWA) where it is best. The dark hole geometry is dependent on details of the coronagraph design. The inner work angle of most coronagraphs varies from $\sim$ 2-4 $\lambda/D$, where $\lambda$ is the wavelength, and D is the diameter of the telescope. Such telescopes need to have diameters of the order of 4 to 15 meters to achieve the necessary IWA.

At thermal infrared wavelengths, defined here as those greater than about 3 μm, two main considerations lead to substantial differences in the instrumentation needed for exoplanet detection and characterization from that in visible wavelengths. First, because the wavelength is 10 to 20 times longer, any telescope also must be that much larger in effective diameter to obtain an equivalent inner working angle. Thus, in order to search a volume of space of the order of 10-30 parsecs from the Sun, a very large telescope is needed, with a diameter of roughly 20-100 m. Second, the background in the thermal infrared is extremely large compared to the expected emission from the planet and even the host star itself. For example, at 10 μm, the thermal emission for a ground-based telescope operating at ~ $0^0$C is ~ $10^9$ photons $s^{-1}$ $arcsec^{-2}$. Thus, in order to achieve the necessary sensitivity, such a telescope must be cooled to temperatures of the order of 40-50 K, like the James Webb Space Telescope. The solution to the requirement of a very large telescope is to use a dilute pupil rather than a filled one, hence an interferometer with at least two telescopes is required. For an interferometer, the angular resolution is $\lambda/(2B)$, where B is the baseline, i.e., the center-to-center separation of the telescopes.

**9 .4 Visible wavelength direct imaging**

In the past 20 years, great progress has been made in the development of coronagraphs for exoplanet detection. Modern coronagraphs all have heritage to the original one of Lyot (Lyot 1939), who developed it in order to see the coronae of the Sun. The original concept was to take the light in the image plane of the telescope and block it with a black spot or mask. The light diffracted around the spot is scattered to the edges of the pupil plane downstream, and a mask is inserted at the pupil plane to block the diffracted light. This light is further reimaged to create a new one with as much of the original starlight removed as possible, which depends on the design of both the spot or mask at the first image plane and the mask at the pupil plane. The image at the second focal plane after the pupil mask retains residual scattered light called speckles. The field of coronagraphy for exoplanet studies has exploded over the last few years, and is beyond the scope of this paper to cover in much detail.

Here we capture some of the essential features and discuss the status of next generation instruments.

Currently, there are a number of systems in ground-based astronomy that are in operation, and which rely on advanced adaptive optics system to correct the wavefront coming into the coronagraph to the level necessary for the design contrast ratio. The largest contrast achievable by any ground-based coronagraph is limited in principal by the minimum number of photons that can be detected in a given fraction of the pupil (depends on the number of modes corrected in the Adaptive Optics or AO system) at the rate at which the AO control loop is closed, typically of the order of 1 kHz. Thus, the amount of correction to the wavefront is limited, which also means that the achievable contrast ratio is limited. Present limits for ground-based coronagraphs are contrast ratios of the order of $10^{-5}$ to $10^{-6}$ at a few $\lambda/D$, i.e., for the Gemini Planet Imager (GPI) (MacIntosh et al. 2014) at the Gemini South telescope, SPHERE (Langlois et al. 2014) at the Very Large Telescope, and SCExAO (Jovanovic et al. 2015) at the Subaru telescope.

Techniques have been developed using deformable mirror technologies to cancel out the residual speckles and create the darkest holes possible (e.g., Oppenheimer and Hinkley 2009, Traub and Oppenheimer 2010, Lawson et al. 2013). The best achieved contrast in fairly recent laboratory experiments is raw contrast of $10^{-8}$, bandwidth 10%, at a central wavelength of 550 nm (Traub et al. 2016).

At the present time, space-based coronagraphs are being developed as part of the technology development plan from the 2010 Astrophysics Decadal Survey. The Coronagraph Instrument (CGI) is under development and it will have the best contrast ratio ever achieved in practice, expected to be of the order of $10^{-9}$ (Spergel et al. 2015). Other technologies are needed for a complete system, and as previously mentioned, a type of low-resolution spectrograph is necessary. WFIRST is developing a relatively new type of spectrograph that takes a low-resolution spectrum of every pixel in the image plane, i.e., an Integral Field Spectrograph (IFS). In order to reduce background and diffracted light (cross-talk) each pixel is imaged onto a mask at the pupil using a lenslet array, and the light from the pupil mask is sent through a prism or grism and reimaged to a focal plane in such a way that each pixel in the image plane creates a spectrum that does not overlap with the spectrum of any adjacent pixels.

With the science results achieved with the planned WFIRST CGI and the ground-based coronagraphs discussed here so far, the field is moving rapidly, and new ever improved systems are on the horizon, including instruments on the upcoming 30-m class telescopes on the ground, such as the Thirty Meter Telescope (TMT) and Giant Magellan Telescope (GMT) being developed in the US, the European Extremely Large Telescope (E-ELT) developed by the European Southern Observatory, and the >20 m diameter Exo-Life Finder (ELF) (Kuhn et al. 2018), see also Section 9.6. For space astronomy, the Large UltraViolet Optical InfraRed (LUVOIR) and Habitable Exoplanet (HabEx) telescopes being developed by NASA.

**9.5 Infrared Direct Imaging**

In the infrared, due to the long wavelengths, in order to obtain the necessary angular resolution required, relatively large structures are needed, and interferometry has been the preferred solution. Historically, one of the first exoplanet detection concepts was that of Bracewell (1978) who proposed a rotating two-telescope cooled space interferometer operating in the far-infrared in order to detect large cold exoplanets of the size of Neptune to Jupiter or larger. The physical principle that this concept relies on is called "nulling" and it is achieved by an achromatic 180 degree or $\pi$ radian phase-shift of the light between the two beams. The beams can be combined in the image plane or the pupil plane, and the requirement for a deep null or high contrast relies primarily on equalizing the paths from each telescope to a high precision, and on having a very small wavefront error on the beam coming from the each of the telescopes. Visualized in the plane of the sky, the result is a pattern of dark and light stripes projected on the star and its planetary system. The "broad-band" dark stripe at which the path lengths are equal is centered on the star, and the maximum response to an exoplanet or other emitting material is on the first bright stripes on either side of the dark stripe. The star is much smaller than the width of the stripes, but even so, the small amount of light leakage from the star itself limits the achievable contrast. If the interferometer is rotated 180 degrees about the line of sight to the source, a dark hole can be synthesized, with rings surrounding the dark hole where an exoplanet or other material can be detected, separated by $\lambda/(2B)$.

Over the past twenty years Bracewell's concept has been realized in both ground-based interferometers and laboratory testbeds. Two ground-based nulling interferometers have been funded by NASA. One is the Keck Interferometer with the Keck Interferometer Nuller (KIN) (Colavita et al. 2009, Serabyn et al. 2012) which is no longer in operation due to funding constraints. More recently, NASA funded the Large Binocular Telescope Interferometer (LBTI) (Hinz et al. 2016). Both interferometers have key science programs focused on obtaining measurements of the luminosity function of debris disk material (dust) in the habitable zones of nearby solar-type stars as well as the amount of such material in nearby stars that are likely targets for direct imaging missions.

Detailed discussions of the science of infrared interferometry for direct imaging of exoplanets and the state of the relevant technologies are beyond the scope of this paper. However, a review of the state of the art of this field was published in the Chapter on "Infrared Direct Imaging" (Danchi et al. 2009) in the 2009 Exoplanet Community Report (Lawson, Traub, and Unwin 2009). Progress in the last few years was reviewed recently by Defrère et al. (2018). Briefly, laboratory testbeds developed at the end of the last decade for the Terrestrial Planet Finder – Interferometer (TPF-I) mission concept in the US had reached contrast ratios of 1.2 × 10$^{-5}$ in a 32% bandwidth at 10 $\mu$m (Peters et al. 2008), and made tests of nulling detection of Earth-sized exoplanets, including simulations of rotational modulation and chopping schemes with contrast ratios of ~10$^{-8}$ for narrow bandwidths (~1%) at 10 $\mu$m (Martin et al. 2012). In France the PERSEE testbed, developed for the Darwin mission concept, achieved its goals at 9 × 10$^{-8}$ in the 1.65-2.45 μm spectral band (37% bandwidth) during 100 s. This result was extended to a 7h duration with an automatic calibration process (Le Duigou 2012). These testbeds achieved the desired contrast and showed that the detection strategies worked at the level required for the mission concepts. To progress further, what remains to be done is for cryogenic testbeds to be developed with realistic signal and noise levels, appropriate for the actual astrophysical measurements.

The Large Binocular Telescope Interferometer (LBTI) Hunt for Observable Signatures of Terrestrial Planets (HOSTS) study (Hinz et al. 2016, Danchi et al. 2016, 2018, Ertel et al. 2018a) has recently set new limits on exozodiacal emission for solar-type stars (Ertel et al. 2018b). With roughly half of the original sample observed (~38 stars), the detection rate is comparable for both early and solar-type stars, ranging from 71+11-20 % for stars with cold dust previously detected and 8% for stars without such an excess. The upper limits on Habitable Zone (HZ) dust is 11+9-4 times the Solar system value (95% confidence limit) for all stars without cold dust, and 16 times the Solar system value for Sun-like stars not having currently detected cold dust. Upper limits for stars without an excess detected by the LBTI, the limits are approximately a factor of two lower.

The recent constraints on exo-zodiacal emission demonstrate the power of LBTI for vetting potential targets for future direct imaging missions such as LUVOIR (Fischer et al. 2018) or HabEx (Gaudi et al. 2018). They also demonstrate the importance of completing and enlarging the study in the next few years. The WFIRST CoronaGraph Instrument (CGI) may provide additional information about the properties of the dust detected by LBTI for some of the sample, as the CGI instrument will observe scattered light from dust rather than thermal emission detected by the LBTI. The two measurements taken together allow for analyses of the physical properties of the dust grains including morphology and chemistry.

**9.6 Extremely Large Telescopes (ELTs) Beyond the 30-m Class Telescopes**

Currently under construction are new 20+ meter class extremely large telescopes being developed by teams in Europe and the United States, as mentioned in Section 9.4. Direct imaging campaigns with the sensitivity to measure exoplanet reflected light against "background" light will likely depend on background- and diffraction-limited optical systems. For example, it is well-known that in this case there is a very large advantage to large-aperture optics (for telescopes of given scattered light performance). Here the integration time to achieve a given signal-to-noise ratio (S/N) scales like $D^{-4}$, where D is the optical system's pupil diameter. For the exoplanet problem, where the background light is dominated by the scattered-light PSF of the optical system, the advantage of a larger aperture is even greater. In this case the exoplanet in larger optics is seen at larger angles in units of $\lambda/D$ from the bright stellar background source, and smaller backgrounds (Kuhn et al. 2012).

Unfortunately, the Keck-era ground-based telescopes scale in cost like their mass which grows at least as fast as $D^2$. The scaling function is anchored by the Keck telescope at about $100M. This curve applies to the predicted cost of all of the planned "World's Largest Telescopes" -- The Giant Magellan Telescope (GMT), the Thirty Meter Telescope (TMT) and the European Extremely Large Telescope (EELT). Accounting for a fixed (in units of $\lambda/D$) telescope scattering function allows an estimate of how the achievable signal-to-noise of, for example, a one-hour Proxima-b measurement depends on wavelength pass-band and telescope aperture (Kuhn et al. 2014, 2018, Berdyugina et al. 2018).

One concept for an extremely large 100-m diameter class ground-based telescope is the ExoLife Finder Telescope (ELF), that is a hybrid design with multiple sub-apertures arranged on a steerable ring structure ~ 100-m in diameter, with a total collecting area equivalent to a single 20-m primary mirror. Each of the sub-apertures is an independent telescope, and active and adaptive optical systems can be used to co-phase all of the individual sub-apertures into a coherent image with a field of view of at least a few arcseconds in diameter. Such a ground-based system could potentially be built on a cost scale of substantially lower than conventional filled aperture telescopes since this design would not obey conventional scaling relations such as apply to the ELTs currently under construction (Kuhn et al 2018).

## 10.0     Conclusions: Future Prospects and Recommendations

In this paper we described the current research and promising key research areas and goals in the field of the exoplanetary space weather, its roots, signatures and effects on (exo)planetary atmospheric dynamics, climates, habitability conditions and observational strategies to detect habitable planets. Here we will discuss the major directions and priorities in this emerging aspect of astrobiology – exoplanetary space weather.

Current research efforts described in this paper suggests that exoplanetary space weather is an emerging and vibrant field of exoplanetary science and astrobiology addressing fundamental questions at the intersection of astrophysics, heliophysics, planetary science, Earth science, chemistry and biochemistry of life.

We find that exoplanetary habitability, the critical aspect of astrobiology, is dynamic in nature, and thus should be expanded to include the impact of evolving space weather from planet hosting stars. The process of evolving habitability relates to necessity of characterization of the signatures of space weather from cool (G-K-M spectral type) main-sequence stars of different ages, its evolving impact on atmospheric erosion and chemistry of terrestrial exoplanets, the evolution of planetary internal dynamics, the conditions initiating prebiotic chemistry, and thus the planetary abiogenesis. The observational and theoretical characterization of these factors is a key in understanding the prerequisites for developing exobiogenic zones and atmospheric biosignatures to be observed with upcoming missions. To address these ambitious issues, a broad range of sophisticated multi-dimensional multispecies physico - chemical models that are validated for the extreme conditions in our Solar system along with laboratory experiments and new observational facilities and strategies are required. Also, the models of paleo space weather and its impact on early Earth, Mars and Venus should provide better constraints for the evolving habitability in our Solar System (Airapetian 2018). These studies will provide a framework for detection of biosignatures of life with emerging technologies in the coming decades.

The key research goals in the characterization of exoplanetary habitability and for signs of life in the next 10-20 years should include the following:

### A. Characterization of Space Weather from Planet Hosting Stars

Understanding the drivers and fluxes of ionizing radiation from F-M type stars hosting exoplanets is critical for characterization of their impact on exoplanetary atmospheres, will be an organic component of exoplanetary science over the next 10-20 years. Given the current progress in developing theoretical modeling and observational tools required for such characterization, we urge the committee to consider the following recommendations:

1. Recognize that modeling of extended astrospheres of stars hosting planets is one of the major components for characterizing habitable exoplanets, especially impacts on atmospheric erosion, chemistry and surface radiation dosages.

2. Promote the development of coordinated multi-wavelength multi-observatory programs to derive critical inputs for theoretical and empirical tools using X-ray, FUV, NUV, radio observations and surface magnetic field maps (magnetograms). The community should strongly encourage the development of new X-ray space telescope (see Wolk et al. 2019), FUV/NUV space telescope (The Star-Planet Activity Research CubeSat (SPARCS), see Scowen et al. 2018; UV space probe mission, CETUS, Danchi et al. 2018, Colorado Ultraviolet Transit Experiment (CUTE), Fleming + 2018) missions, and radio telescope facilities (LOFAR and Very large Array (ngVLA) (Osten et al. 2018) to open new windows into the nature of nearby planet hosting stars. Characterization of stellar magnetic activity signatures include characterization of dynamics of magnetic starspots and their association with (super)flares using data from Kepler, TESS and the upcoming missions including CHEOPS, JWST, PLATO 2.0 and ARIEL. A comprehensive description of flaring properties (frequency, maximum energy, energy partition) is a crucial component of exoplanetary atmospheric evolution models.

3. Perform direct, detailed characterization of stellar magnetic fields, in particular those of key exoplanet host stars, through high-resolution spectroscopy, spectropolarimetry and interferometric polarimetry using data obtained with ground-based optical and infrared facilities including the upcoming instrument SPIRou for infrared spectropolarimetry at the CFHT. Characterize the evolution of stellar magnetic structures in large samples of stars using indirect proxies, including spots and their association with flares using data from Kepler, TESS, and upcoming missions including CHEOPS, JWST, PLATO 2.0 and ARIEL. Develop dynamo models of F-M stars.

4. Refine the characterization of the ages of planet hosting stars using Li, rotation rates, CaII H&K, and patterns of magnetic activity in the form of frequency distribution of stellar flares, their duration and maximum energy and maximal sizes of starspots. Thus, dedicated observations and comprehensive characterization of flares at different phases of evolution of F-M stars are required along with flare frequency.

5. Encourage the extension of heliospheric MHD, kinetic and hybrid models to be used for reconstruction of coronal properties, XUV fluxes and stellar wind properties from G-K-M dwarfs using empirical and theoretical models that incorporate coordinated X-ray, FUV, NUV, optical, radio (with next generation Very large Array (ngVLA)) and mm emission and surface magnetic field observations (magnetograms). This requires a multi-observatory network of coordinated observations using ground-based and space telescopes including TESS, HST, NICER, XMM-Newton.

6. Develop MHD models for initiation and development of stellar flares & CMEs. These models can also be used to characterize the magnetic connectivity between a star and a planet. This is important for space weather impact on habitability as it will determine (for instance) the trajectories of stellar energetic particles. These models will should also characterize the impact of activity transmission spectra of exoplanets, which is an unexplored area.

7. Search for stellar CMEs from active stars based on an array of X-ray, UV and radio observations as predicted by current and future multidimensional magnetohydrodynamic models.

8. Characterization of stellar ages of planet hosting stars based on a set of observables including Li, rotation rates, CaII H&K, patterns of magnetic activity. Dedicated observations of flares on G-K-M stars at different phases of evolution are required along with characterization of flare frequency.

### B. Characterization of Star-Planet Interactions

1. Development of multi-dimensional multi-fluid MHD, hydrodynamic and kinetic models that describe the coupling of energy flows between star and planet as well as energy transfer and dissipation in exoplanetary magnetosphere - ionosphere - thermosphere - mesosphere environments. This requires a well-coordinated interdisciplinary effort from the cross-disciplinary community including heliophysics, astrophysics, planetary and Earth science.

2. Derive thresholds of impact of astrospheric space weather on factors of habitability including atmospheric escape (thermal and non-thermal escape rates) from exoplanets around F-M dwarfs.

3. Characterize chemistry changes due to: steady state and transient FUV, XUV fluxes, stellar winds, CMEs & energetic particles.

4. Search for radio and optical stellar CME signatures by performing extended long-term observations at lower frequencies (< 10 MHz) with space or lunar radio missions.

5. Search for planetary outflows in spectral lines of H (hot Jupiters) and nitrogen and metals (terrestrial planets) driven by powerful stellar flares from active G-K-M dwarfs.

6. Explore when M dwarf habitable cases actually shift beyond the ice line due to severe SW, when combined with ameliorating internal heating, including radiogenic sources as well as tidal heating within compact multi-body TRAPPIST-1 analog systems.

### C. Characterization of Exoplanetary Environments

1. Understanding of mechanisms of ionosphere-thermosphere-mesosphere (MIT) system response to extreme space weather on the early Earth, Mars and exoplanets. This includes the development of multi-dimensional modeling efforts of MIT system to characterization of the rates of atmospheric erosion resulting from various modes of star-planet interactions and resulting atmospheric thermodynamics and chemistry.

2. Search for signatures of nitrogen-rich atmospheres through UV and mid-IR bands using transmission and direct imaging observations, as necessary to determine how common they are in terrestrial type exoplanets in the solar neighborhood.

3. Detection and characterization of atmospheric signatures of hydrogen-rich (primary atmospheres) of terrestrial-type exoplanets around very young planet hosting stars.

4. Detection and characterization of $N_2 - CO_2 - H_2O$ rich atmospheres through identification of spectral atmospheric signatures of young terrestrial-type exoplanets at a pre-life phase: signatures of prebiotic chemistry.

5. Extension of 3D Global Climate modeling efforts to include diverse chemistry of exoplanetary atmospheres impacted by volcanic activity and space weather factors.

6. Modeling and laboratory experiments of atmospheric biosignatures of life guided by prebiotic chemistry and observations of Earth's upper atmosphere in response to current space weather. This includes theoretical and laboratory studies of the initiation of prebiotic chemistry and biochemical pathways to the building blocks of life as we know it.

7. Characterization of exoplanetary magnetic dynamos, mantle activity, and the interplay between volcanic/tectonic activity and the role exomoons in the generation and maintaining exoplanetary magnetic dynamos. Magnetospheric low-frequency radio observations provide crucial information about exoplanetary magnetic field, the factor of habitability (Lazio et al. 2018).

8. Instrument development in support of direct imaging exoplanetary missions to detect atmospheric biosignatures in the mid-infrared band including characterization of host stars in the FUV and X-ray bands.

**Acknowledgements**. The authors thank the referee, Prof. Lena Noack, for her constructive suggestions that improved this paper. V. Airapetian was supported by NASA Exobiology grant #80NSSC17K0463 and TESS Cycle 1. V. Airapetian, A. Glocer, K. Garcia-Sage, W. Danchi, G. Gronoff, M. Way were partially supported by Sellers Exoplanetary Environments Collaboration (SEEC) Internal Scientist Funding Model (ISFM). The part of O. Verkhoglya-dova's work was performed at the Jet Propulsion Laboratory, California Institute of Technology, under a contract with NASA.**References**

Abbot, D. S., & Switzer, E. R. (2011). The Steppenwolf: a proposal for a habitable planet in interstellar space. *The Astrophysical Journal Letters*, *735*(2), L27.

Acuña, M. H., J. E. P. Connerney, N. F. Ness, R. P. Lin, D. Mitchell, C. W. Carlson J. McFadden, K. A. Anderson, H. Reme, C. Mazelle, D. Vignes, P. Wasilewski, P. Cloutier (1999). Global distribution of crustal magnetization discovered by the Mars Global Surveyor MAG/ER Experiment. *Science, 284,* 790-793.

Airapetian, V. S., Adibekyan, V., Ansdell, M. and 83 co-authors (2019a). Reconstructing Extreme Space Weather from Planet Hosting Stars. White Paper submitted to the Astronomy and Astrophysics Decadal Survey (Astro2020), eprint arXiv:1903.06853.

Airapetian, V. S., Jin, M., Lugtinger, T., Danchi, W., van der Holst, B., Manchester, W. B. (2019b) One Year in the Life of The Variable Young Sun: Dynamic Paleo Space Weather. submitted to *Nature Astronomy*.

Airapetian, V. S. (2018) Terrestrial planets under the young Sun, *Nature Astronomy*, 2, 448-449.

Airapetian, V. S., Danchi, W. C., Dong, C. F. and 33 co-authors (2018a) Life Beyond the Solar System: Space Weather and Its Impact on Habitable Worlds, the white paper was submitted to the National Academy of Sciences in support of the *Astrobiology Science Strategy for the Search for Life in the Universe*, eprint arXiv:1801.07333.

Airapetian, V. S., Adibekyan, V., Ansdell, M. and 46 co-authors (2018b) Exploring Extreme Space Weather Factors of Exoplanetary Habitability, the white paper submitted to the US *National Academy of Sciences call on Exoplanet Science Strategy*, eprint arXiv:1803.03751

Bell, J. M. (2008) The dynamics in the upper atmospheres of Mars and Titan, PhD Thesis, University of Michigan; 2008. Publication Number: AAT 3328763.

Bell, J. M. (2008) The dynamics in the upper atmospheres of Mars and Titan, PhD Thesis, University of Michigan; 2008. Publication Number: AAT 3328763.

Biggin, A.J., E.J. Piispa, L.J. Pesonen, R. Holme, G.A. Paterson, T. Veikkolainen and L. Tauxe (2015), Paleomagnetic field intensity variations suggest Mesoproterozoic inner-core nucleation", *Nature, 526*, doi:10.1038/nature15523

Birkmann S, Ferruit P, Giardino G, Kempton E, Carey S, Krick J, Deroo PD, Mandell A, Ressler ME, Shporer A, Swain M, Vasisht G, Ricker G, Bouwman J, Cross_eld I, Greene T, Howell S, Christiansen J, Ciardi D, Clampin M, Greenhouse M, Sozzetti A, Goudfrooij P, Hines D, Keyes T, Lee J, McCullough P, Robberto M, Stansberry J, Valenti J, Rieke M, Rieke G, Fortney J, Bean J, Kreidberg L, Ehrenreich D, Deming D, Albert L, Doyon R, Sing D (2014) Observations of Transiting Exoplanets with the James Webb Space Telescope (JWST). *PASP, 126*:1134, DOI 10.1086/679566.

Bolmont, E., et al. (2017). Water loss from terrestrial planets orbiting ultracool dwarfs: implications for the planets of TRAPPIST-1. *Monthly Notices of the Royal Astronomical Society, Volume 464, Issue 3,* p.3728-3741.

Bono, R.K., J.A. Tarduno, F. Nimmo and R.D. Cottrell (2019), Young inner core inferred from Edicarran ultra-low geomagnetic field intensity, *Nature Geoscience,* doi: 10.1038/s41561-018-0288-0.

Boro Saikia, S., Lueftinger, T., Jeffers, S. V., Folsom, C. P., See, V., Petit, P., Marsden, S. C., Vidotto, A. A., Morin, J., Reiners, A. and 2 coauthors (2018) Direct evidence of full dipole flip during the magnetic cycle of a sun-like star. *A&A, 620*, L11-L23.

Borucki, W. J., et al. (1982). Comparison of Venusian lightning observations. *Icarus, 52,* 302-321.